\documentclass[a4paper,10pt]{article}

\usepackage{authblk}
\usepackage{ifthen,ifpdf}

\ifpdf
  \usepackage{hyperref}
  \hypersetup{colorlinks=false,allcolors=blue}
  \usepackage{hypcap}
  \pdfpagewidth=\paperwidth
  \pdfpageheight=\paperheight
\fi

\usepackage[utf8]{inputenc}
\usepackage[T1]{fontenc}
\usepackage{amsmath, amsthm, amssymb}
\usepackage{mathtools}
\usepackage{enumerate}
\usepackage{afterpage}
\usepackage{tabularx}
\usepackage{dsfont}
\usepackage[table]{xcolor}

\usepackage{graphicx}

\usepackage{algorithm}
\usepackage{algpseudocode}
\usepackage{caption}
\DeclareCaptionLabelFormat{andtable}{#1~#2  \&  \tablename~\thetable}
\usepackage[labelformat=simple]{subcaption}

\usepackage{color}
\usepackage{units}
\usepackage{array, multirow}
\usepackage{booktabs}

\newcolumntype{M}[1]{>{\vspace{3pt}\raggedleft\arraybackslash}m{#1}}

\usepackage[binary-units]{siunitx}
\usepackage{tikz}
\usepackage{pgfplots}
\usepackage{pgfplotstable}
\usepgfplotslibrary{units}
\pgfplotsset{compat=1.15}
\usepackage{enumitem}
\usepackage{cite}
\usepackage{numprint}

\usepackage{anysize} 
\marginsize{2.5cm}{2.5cm}{2cm}{2cm}
\usepackage{wrapfig}
\usepackage{xspace}
\usepackage{scalerel,stackengine}

\newcommand{\review}[1]{#1}

\theoremstyle{plain}

\theoremstyle{remark}

\numberwithin{equation}{section}

\newcommand{\set}[1]{\left\{#1\right\}} 

\newcommand{\abq}{ABAQUS\xspace}

\renewcommand{\d}{{\,\rm  d}}

\newcommand{\T}[1]{{#1}^{\sf  T}} 

\newcommand{\pd}[2]{\displaystyle\frac{\partial #1}{\partial #2}}

\newcommand{\pdn}[3]{\displaystyle\frac{\partial^{#3} #1}{\partial #2^{#3}}}

\newcommand{\pdmixed}[3]{\displaystyle\frac{\partial^2 #1}{\partial #2 \partial #3}}

\newcommand{\norm}[1]{\Vert #1 \Vert}
\newcommand{\normg}[1]{\left\Vert \, #1 \, \right\Vert}

\renewcommand{\div}[1]{{\rm div } #1 }
\newcommand{\tr}[1]{{\rm tr  }\left( #1 \right)}
\newcommand{\diag}[1]{{\rm diag  }\left( #1 \right)}

\newcommand{\nablas}{\nabla^\textrm{s}}
\renewcommand{\unit}[1]{~\rm #1}

\newcommand{\Sym}[1]{{\rm Sym }( #1 )}
\newcommand{\Dev}[1]{{\rm Dev }( #1 )}
\newcommand{\Sph}[1]{{\rm Sph }( #1 )}

\newcommand{\sty}[1]{\mbox{\boldmath $#1$}}
\newcommand{\styy}[1]{{\mathbb{#1}}}

\definecolorset{rgb}{KIT}{}{%
   green, 0.0, 0.58, 0.49;%
   blue, 0.0, 0.34, 0.62;%
   lead, 0.44, 0.44, 0.45;%
  darkgray, 0.53, 0.53, 0.54;%
  gray, 0.89, 0.89, 0.90;%
  lightgray, 0.93, 0.93, 0.93}
\definecolor{KITgreen}  {RGB}{  0,150,130} 
\definecolor{KITgreen70}{RGB}{ 76,181,167} 
\definecolor{KITgreen50}{RGB}{127,202,192} 
\definecolor{KITgreen30}{RGB}{178,223,217} 
\definecolor{KITgreen15}{RGB}{217,239,236} 
\definecolor{KITblue}  {RGB}{ 70,100,170} 
\definecolor{KITblue70}{RGB}{125,146,195} 
\definecolor{KITblue50}{RGB}{162,177,212} 
\definecolor{KITblue30}{RGB}{199,208,229} 
\definecolor{KITblue15}{RGB}{227,232,242} 
\definecolor{KITblack}  {RGB}{  0,  0,  0} 
\definecolor{KITblack70}{RGB}{ 77, 77, 77} 
\definecolor{KITblack50}{RGB}{128,128,128} 
\definecolor{KITblack30}{RGB}{179,179,179} 
\definecolor{KITblack15}{RGB}{217,217,217} 
\definecolor{KITpalegreen}{RGB}{130,190,60}
\colorlet{KITpalegreen70}{KITpalegreen!70}
\colorlet{KITpalegreen50}{KITpalegreen!50}
\colorlet{KITpalegreen30}{KITpalegreen!30}
\colorlet{KITpalegreen15}{KITpalegreen!15}
\definecolor{KITyellow}{RGB}{250,230,20}
\colorlet{KITyellow70}{KITyellow!70}
\colorlet{KITyellow50}{KITyellow!50}
\colorlet{KITyellow30}{KITyellow!30}
\colorlet{KITyellow15}{KITyellow!15}
\definecolor{KITorange}{RGB}{220,160,30}
\colorlet{KITorange70}{KITorange!70}
\colorlet{KITorange50}{KITorange!50}
\colorlet{KITorange30}{KITorange!30}
\colorlet{KITorange15}{KITorange!15}
\definecolor{KITbrown}{RGB}{160,130,50}
\colorlet{KITbrown70}{KITbrown!70}
\colorlet{KITbrown50}{KITbrown!50}
\colorlet{KITbrown30}{KITbrown!30}
\colorlet{KITbrown15}{KITbrown!15}
\definecolor{KITred}{RGB}{160,30,40}
\colorlet{KITred70}{KITred!70}
\colorlet{KITred50}{KITred!50}
\colorlet{KITred30}{KITred!30}
\colorlet{KITred15}{KITred!15}
\definecolor{KITlilac}{RGB}{160,0,120}
\colorlet{KITlilac70}{KITlilac!70}
\colorlet{KITlilac50}{KITlilac!50}
\colorlet{KITlilac30}{KITlilac!30}
\colorlet{KITlilac15}{KITlilac!15}
\definecolor{KITcyan}{RGB}{80,170,230}
\colorlet{KITcyan70}{KITcyan!70}
\colorlet{KITcyan50}{KITcyan!50}
\colorlet{KITcyan30}{KITcyan!30}
\colorlet{KITcyan15}{KITcyan!15}

\definecolor{boxrahmen}{gray}{0.0}
\definecolor{boxhintergrund}{gray}{0.999}
\newsavebox{\tmpbox}

\newcommand{\fa}{\sty{ a}}
\newcommand{\fb}{\sty{ b}}

\newcommand{\fe}{\sty{ e}}

\newcommand{\fn}{\sty{ n}}

\newcommand{\fq}{\sty{ q}}

\newcommand{\fu}{\sty{ u}}

\newcommand{\fx}{\sty{ x}}

\newcommand{\fz}{\sty{ z}}
\newcommand{\fA}{\sty{ A}}

\newcommand{\fC}{\sty{ C}}

\newcommand{\fN}{\sty{ N}}

\newcommand{\fW}{\sty{ W}}

\newcommand{\ffC}{\styy{ C}}
\newcommand{\ffD}{\styy{ D}}

\newcommand{\ffF}{\styy{ F}}

\newcommand{\ffI}{\styy{ I}}

\newcommand{\ffN}{\styy{ N}}

\newcommand{\ffP}{\styy{ P}}

\newcommand{\ffR}{\styy{ R}}

\newcommand{\ffV}{\styy{ V}}

\newcommand{\falpha}{\mbox{\boldmath $\alpha$}}

\newcommand{\fsigma}{\mbox{\boldmath $\sigma$}}

\newcommand{\feps}{\mbox{\boldmath $\varepsilon $}}

\newcommand{\effective}[1]{\bar{#1}}
\newcommand{\fmicrostrain}{\feps}
\newcommand{\fmacrostrain}{\effective{\feps}}
\newcommand{\fmicrostress}{\fsigma}
\newcommand{\fmacrostress}{\effective{\fsigma}}

\newcommand{\microstrain}{\varepsilon}
\newcommand{\macrostrain}{\effective{\microstrain}}
\newcommand{\microstress}{\sigma}
\newcommand{\macrostress}{\effective{\sigma}}

\newcommand{\freeenergy}{\psi}
\newcommand{\dissipation}{\phi}

\newcommand{\microtemp}{\theta}
\newcommand{\macrotemp}{\effective{\microtemp}}

\newcommand{\microrpl}{D}
\newcommand{\macrorpl}{\effective{\microrpl}}
\newcommand{\microdissipation}{\mathcal{D}}
\newcommand{\macrodissipation}{\effective{\microdissipation}}

\newcommand*\dif{\mathop{}\!\mathrm{d}} 
\newcommand*{\eg}{e.g.,\@\xspace}
\newcommand*{\ie}{i.e.,\@\xspace}
\newcommand*{\wrt}{w.r.t.\@\xspace}
\newcommand*{\cf}{see\@\xspace}

\newcommand{\DMNLIN}[1]{\mathcal{DMN}^{\mathcal{L}}_{#1}}

\newcommand{\BB}{\mathcal{B}}

\newcommand{\condensedpotential}{\Psi}

\newcommand{\statev}{z}
\newcommand{\fstatev}{\fz}
\newcommand{\fzero}{\sty{ 0}}
\newcommand{\mean}[1]{\left\langle #1 \right\rangle_\textrm{Y}}

\newcommand{\IDTWO}{\mathbf{1}}
\newcommand{\IDFOURS}{\ffI_\textrm{s}}

\newcommand{\fkappa}{\sty{ \kappa}}

\newcommand\equalhat{\mathrel{\stackon[1.5pt]{=}{\stretchto{%
				\scalerel*[\widthof{=}]{\wedge}{\rule{1ex}{3ex}}}{0.5ex}}}}

\title{An FE-DMN method for the multiscale analysis of \review{thermomechanical} composites}

\author[1]{Sebastian Gajek}
\author[1]{Matti Schneider}
\author[1,*]{Thomas Böhlke}

\affil[1]{Karlsruhe Institute of Technology (KIT), Institute of Engineering Mechanics}
\affil[*]{correspondence to: \texttt{thomas.boehlke@kit.edu}}

\date{\today}

\begin{document}

\maketitle 
\begin{abstract}
	
	\noindent We extend the \textrm{FE}-\textrm{DMN} method to fully coupled thermomechanical two-scale simulations of composite materials. In particular, every Gauss point of the macroscopic finite element model is equipped with a deep material network (DMN). Such a DMN serves as a high-fidelity surrogate model for full-field solutions on the microscopic scale of inelastic, non-isothermal constituents.\\
	Building on the homogenization framework of Chatzigeorgiou et al. [Int. J. Plast, vol. 81, pp. 18--39, 2016], we extend the framework of DMNs to thermomechanical composites by incorporating the two-way thermomechanical coupling, \ie the coupling from the macroscopic onto the microscopic scale and vice versa, into the framework. We provide details on the efficient implementation of our approach as a user-material subroutine (UMAT). We validate our approach on the microscopic scale and show that DMNs predict the effective stress, the effective dissipation and the change of the macroscopic absolute temperature with high accuracy. After validation, we demonstrate the capabilities of our approach on a concurrent thermomechanical two-scale simulation on the macroscopic component scale. 
	 
\end{abstract}

\section{Introduction} \label{sec:introduction}

Many common engineering materials are characterized by a thermomechanically coupled mechanical behavior, \ie involving a coupling between temperate and deformation. In particular, variations in temperature may affect the mechanical response of a structural material. In addition, deformations may lead to changes in temperature as well, \eg via changes in internal entropy or in the form of internal energy dissipation. For instance, dissipation-induced self-heating is commonly observed for thermoplastic polymers subjected to cyclic loading. As such polymers are particularly sensitive to temperature fluctuations, especially in the vicinity of their glass transition temperature, deformation-induced self-heating effects may significantly influence the mechanical properties of such materials and even lead to premature failure~\cite{Katunin2019}.\\
To complicate matters, many structural materials consist of composite materials, \ie they feature a (spatially varying) complex microstructure. As the effective response of composite materials depends both on the constituent materials and the geometric composition of the microstructure, predicting the thermomechanical response of such materials is a challenging task, even for rather simple geometries.\\
For instance, a monolithic finite element (FE) model of a structural component, also resolving the microstructure heterogeneities, is typically not feasible with today's computational power. Alternatively and under a suitable separation of scales, asymptotic homogenization methods~\cite{homogenization1,homogenization2,homogenization3} may be used to derive so-called effective material models which account for the geometric composition of the microstructure and the material behavior of the constituents on the lower scale. Chatzigeorgiou et al.~\cite{Chatzigeorgiou2016} applied first-order asymptotic homogenization to composites of small-strain non-isothermal generalized standard materials (GSM)~\cite{HalphenNguyen} and deduced the governing equations for the macroscopic and microscopic scale. In particular, provided the force term varies only slowly on the macroscopic scale, Chatzigeorgiou et al.~\cite{Chatzigeorgiou2016} deduced that the balance of linear momentum on the microscopic scale, the thermomechanical cell problem, only depends on the macroscopic temperature, \ie temperature fluctuations on the microscopic scale constitute only a lower-order contribution to the effective stress. By solving the thermomechanical cell problem on a suitable microstructure, the effective, non-isothermal model of the composite emerges naturally. For linear  constituent materials, the effective material behavior can be pre-computed and cached for later use. The outlined strategy does not, however, extend to inelastic materials as the internal variables naturally live on the microscopic scale and cannot be homogenized to the macroscopic scale. For this reason, $\textrm{FE}^2$ methods~\cite{Renard1987, Smit1998,Feyel1999, Feyel2000, Feyel2003} were developed. In a $\textrm{FE}^2$ simulation, each Gauss point of the macroscopic finite element simulation is furnished with a finite element model of the microstructure on which the cell problem is solved. Thus, the evolution of the internal variables can be accounted for. The $\textrm{FE}^2$ method for \review{thermomechanical} composites was investigated, for instance, in the context of thermo-elastoplasticity~\cite{Ozdemir2008a, Ozdemir2008b}, phase transforming polycrystals under dynamic loading~\cite{Sengupta2012} or single-crystal thermo-elastoviscoplasticity~\cite{Li2019}. Recently, Tikarrouchine et al.~\cite{Tikarrouchine2019} investigated a short-fiber reinforced composite in a concurrent two-scale setting accounting for heat conduction and convection but temperature-independent material properties.\\ 
As an alternative to FE models on the microscale, FFT-based computational micromechanics~\cite{MoulinecSuquet1994,MoulinecSuquet1998,Schneider2021Review} may be used to solve the thermomechanical cell problem more efficiently giving rise to the so-called \textrm{FE}-\textrm{FFT} method~\cite{Spahn2014,Kochmann2016, Kochmann2018}. Recently, Wicht et al.~\cite{WichtThermo2020} proposed an efficient, fully implicit FFT-based solution scheme for thermomechanical composites.\\
To reduce the computational burden on the microscopic scale, model order reduction techniques (MOR) exploit that the cell problem is solved repeatedly, but with slightly different input parameters, in order to derive a reduced order model. MOR techniques include the transformation field analysis (TFA)~\cite{TFADvorakBenveniste1992, TFA2, TFA3, TFA4}, the self-consistent clustering analysis (SCA)~\cite{SCALiu2016,SCA_CVP_2017,SCA_softening_2018,SCA_convergence,SCAThermo2021} and the non-uniform transformation field analysis (NTFA)~\cite{NTFA1, NTFA_viscoelastic1, NTFA_viscoelastic2, NTFA_TSO1,NTFA_TSO2,NTFA_TSO3,FritzenLeuschner,FritzenHodappLeuschner}. These approaches can be incorporated into a concurrent two-scale framework giving rise to the $\textrm{FE}^{2\textrm{R}}$ (R for reduced) method~\cite{FE2R}. Furthermore, these approaches allow to incorporate thermomechanical loading, thermal eigenstrains and temperature-dependent material parameters. However, they typically do not consider the back-coupling of the mechanical deformation onto the temperature evolution.\\
In contrast to approximating the solution of the cell problem, alternative strategies seek to approximate the effective properties directly. Data driven approaches, \eg artificial neural networks (ANN), are predestined for such tasks as they effortless operate on a high-dimensional domain of interest. For instance, the regularity of the effective stress allows to approximate the stress-strain relationship directly. Being by no means exhaustive, we refer to the works of Jadid~\cite{NasserJadid1997}, Penumadu-Zhao~\cite{Penumadu1999} or Srinivasu et al.~\cite{Srinivasu2012} for different approaches. By considering the temperature as an additional degree of freedom of the feature space, ANNs can be extended to thermomechanical problems, see for example the works of Ji at al.~\cite{Ji2011} or Li et al.~\cite{Li2012}. Machine learning approaches were applied in a concurrent two-scale setting both for isothermal and non-isothermal problems, see, \eg Acuna et al.~\cite{Acuna2020} or Fritzen et al.~\cite{Fritzen2019}. Using ANNs comes with two significant drawbacks, however. For a start, the capabilities to extrapolate beyond the training domain is limited for ANNs, in general. Secondly, the underlying physical principles, \eg thermodynamic consistency or preservation of stress-strain monotonicity, may be violated unless specifically accounted for by the model. Recently, Masi and co-workers~\cite{Masi2021a,Masi2021b} proposed so-called thermodynamics-based artificial neural networks (TANN) which ensure thermodynamic consistency a priori. Their findings indicate that the predictive capabilities of TANNs outperform those of standard ANNs. Please note that the mentioned approaches only consider a one-way thermomechanical coupling, \ie from the temperature on the effective properties, and not vice versa.\\
Applying the concepts underlying deep learning in a more micromechanics-aware context, Liu and co-workers~\cite{Liu2018, Liu2019} proposed so-called deep material networks (DMN) as a surrogate model for micromechanical computations. To be more precise, for a $N$-phase microstructure, they consider a $N$-ary tree structure of $N$-phase laminates with intermittent rotations associated with the edges of the tree as their primary modeling approach. Instead of approximating the stress-strain relationship directly, DMNs approximate the effective stiffness of a fixed microstructure and variable constituents. For identifying the free parameters of the DMN, the so-called training process, Liu et al.~\cite{Liu2018, Liu2019} rely upon stochastic gradient descent and automatic differentiation. Once the training process is complete, DMNs can be applied to inelastic problems at finite and infinitesimal strains with impressive accuracy. Subsequently, direct DMNs were introduced by Gajek et al.~\cite{Gajek2020, Gajek2021} which allow for an efficient solution scheme in the inelastic setting as they do not involve additional rotations. Furthermore, Gajek et al.~\cite{Gajek2020} motivated the approximation capabilities of (direct) DMNs by showing that, to first-order in the strain rate, the effective inelastic behavior of composite materials is determined by linear elastic localization. In addition, Gajek et al.~\cite{Gajek2020} clarified that DMNs inherit thermodynamic consistency and stress-strain monotonicity from their phases. The former is crucial for stable and fast simulations, especially in a two-scale context, as it ensures that the effective model inherits stabilizing numerical properties, \eg strong convexity, from its phases. Recently, DMNs were augmented by cohesive zone models to account for interface damage~\cite{Liu2020a} or multiscale strain localization modeling~\cite{Liu2021}. Liu et al.~\cite{Liu2020b} and Gajek et al.~\cite{Gajek2021} extended DMNs to accelerate two-scale concurrent simulation giving rise to the \textrm{FE}-\textrm{DMN} method.\\
In this work, we extend the framework of direct DMNs~\cite{Gajek2020, Gajek2021} to composites with full thermomechanical coupling, effectively enabling thermomechanical two-scale simulations of industrial problems. As point of departure, we recapitulate the results of Chatzigeorgiou et al.~\cite{Chatzigeorgiou2016} in Section~\ref{sec:asymptotic_homogenization}, who introduced a framework for the first-order asymptotic homogenization of thermomechanical composites. Subsequently, we extend the framework of direct DMNs to thermomechanical composites, see Section~\ref{sec:direct_DMN_thermo}. We take special care in incorporating the coupling of microscopic mechanical deformation onto the macroscopic temperature and vice versa into our approach. For this purpose, we exploit the homogeneity of the absolute temperature on the microscopic scale to arrive at an efficient solution scheme for solving the balance of linear momentum of a direct DMN. To accelerate a component-scale simulation of industrial complexity, we discuss the efficient implementation of our approach as a user-material subroutine (UMAT) only relying on the provided interfaces.\\
To demonstrate the capabilities of the proposed approach, we consider a short-fiber reinforced polyamide featuring a pronounced thermomechanical coupling, see Section~\ref{sec:material_parameters}. In Section~\ref{sec:surrogate_model_validation}, we elaborate on the training and the validation of the identified DMN surrogate model separately. We show that the DMN is able to predict the effective stress, the effective dissipation as well as the deformation-induced change in temperature of the composite with sufficient accuracy for all investigated loading conditions and strain rates. Later on, we demonstrate the power of our approach in Section~\ref{sec:computational_example}, where we conduct a fully coupled thermomechanical two-scale simulation of a asymmetric notched specimen subjected to cyclic loading also considering heat conduction and convection on the macroscopic scale.

\section{First-order asymptotic homogenization of thermomechanical composites} \label{sec:asymptotic_homogenization}

In their work, Chatzigeorgiou et al.~\cite{Chatzigeorgiou2016} introduced a framework for the (first-order) asymptotic homogenization of thermomechanical composites at small strains. More precisely, they considered quasi-static, non-isothermal generalized standard materials (GSM)~\cite{HalphenNguyen} and derived governing equations for the microscopic and macroscopic scale.\\
Let $\Sym{d}$ denote the set of symmetric $d \times d$ matrices. Then, in $d \in \set{2, 3}$ spatial dimensions, we consider a small-strain, quasi-static, non-isothermal GSM to be a quadruple $\left(Z, \freeenergy, \dissipation, \fstatev_0 \right)$ comprising 
\begin{enumerate}[label=D\arabic*]
	\item a (sufficiently large) Banach vector space $Z$ of internal variables,
	\item a Helmholtz free energy density $\freeenergy : \Sym{d} \times \ffR_{> 0} \times Z \rightarrow \ffR$, which we assume to be differentiable \wrt all arguments,
	\item an extended-real-valued dissipation potential $\dissipation : \ffR_{> 0} \times Z \rightarrow \ffR\cup\{+\infty\}$, which we assume to be proper, convex, lower semicontinuous in its second argument, and to satisfy $\dissipation(\cdot, 0) = 0$ as well as $0\in \partial_{\dot{\statev}} \dissipation(\cdot, 0)$, where $\partial_{\dot{\statev}} \dissipation$ denotes the subdifferential of the convex function $\dissipation$ \wrt the second argument,
	\item and an element $\fstatev_0 \in Z$ serving as initial condition for the dynamics.
\end{enumerate}
For every strain path $\fmicrostrain: [0, T] \rightarrow \Sym{d}$, temperature path $\microtemp: [0, T] \rightarrow \ffR_{> 0}$ and internal variables $\fstatev: [0, T] \rightarrow Z$ with final time $T \in \left(0, \infty\right]$, the Cauchy stress $\fmicrostress: [0, T] \rightarrow \Sym{d}$ is expressed in terms of the potential relation 
\begin{equation}
	\fmicrostress(t) = \pd{\freeenergy}{\fmicrostrain}(\fmicrostrain(t), \microtemp(t), \fstatev(t)),
\end{equation}
and the evolution of the internal variables satisfies the initial value problem described by Biot's equation
\begin{equation}\label{eq:biots_euqation}
	\pd{\freeenergy}{\fstatev}(\fmicrostrain(t), \microtemp(t), \fstatev(t)) + \pd{\dissipation}{\dot{\fstatev}}(\microtemp(t), \dot{\fstatev}(t)) = 0 \quad \textrm{with} \quad \fstatev(0) = \fstatev_0.
\end{equation}
With these definitions at hand, we turn our attention to the first-order homogenization of non-isothermal GSMs. We refer to Chatzigeorgiou et al.~\cite{Chatzigeorgiou2016} for more details.\\
We consider a macroscopic body $\Omega \subseteq \ffR^d$ with macroscopic point $\effective{\fx} \in \Omega$. To every macroscopic point $\effective{\fx}$, we associate a (rectangular) two-phase periodic microstructure $Y \subseteq \ffR^d$.

\paragraph{The microscopic cell problem}  The microstructure $Y$ comprises two non-isothermal GSMs, \ie $\left(Z_1, \freeenergy_1, \dissipation_1, \fstatev_{1,0} \right)$ and $\left(Z_2, \freeenergy_2, \dissipation_2, \fstatev_{2,0} \right)$, with measurable characteristic functions $\chi_{1/2}: Y \rightarrow \set{0, 1}$ whose associated sets are mutually disjoint and cover all of $Y$, \ie the conditions
\begin{equation}
	\chi_1 \chi_2 = 0 \quad \textrm{and} \quad \chi_1 + \chi_2 = 1
\end{equation} 
hold. Then, on the microscopic level, the so-called thermomechanical cell problem of first-order homogenization, \ie the (quasi-static) microscopic balance of linear momentum, reads
\begin{equation}\label{eq:micro_cell_problem}
	\div_{\! x}\left(\sum_{i=1}^{2} \chi_i \pd{\freeenergy_i}{\fmicrostrain}(\fmacrostrain + \nablas_{\!\! x}\, \fu, \macrotemp, \fstatev_i) \right) = \fzero,
\end{equation}
where $\div_{\! x}$ and $\nablas_{\!\! x}$ refer to the divergence and the symmetrized gradient operator \wrt the microscopic point $\fx \in Y$, respectively. Furthermore, $\fmacrostrain: \Omega \times [0, T] \rightarrow \Sym{d}$ denotes the macrostrain, $\fu : \Omega \times Y \times [0, T] \rightarrow \ffR^d$ symbolizes the periodic displacement fluctuation with anti-periodic normal derivative and $\fstatev_{1/2}: \Omega \times Y \times [0, T] \rightarrow Z_{1/2}$ stands for the fields of internal variables. Chatzigeorgiou et al.~\cite{Chatzigeorgiou2016} established that, for first-order homogenization, the absolute temperature is a macroscopic quantity, \ie there is no temperature fluctuation on the microscopic level. Most importantly, there is no need to solve for the temperature on the microscopic level. Thus, the macroscopic absolute temperature $\macrotemp: \Omega \times [0, T] \rightarrow \ffR_{> 0}$ as well as the macrostrain $\fmacrostrain$ enter Equation \eqref{eq:micro_cell_problem} as inputs and constitute the one-way coupling between the macroscopic and the microscopic scale.

\paragraph{The macroscopic balance of linear momentum and the macroscopic heat equation} On the macroscopic level, two governing equations emerge. First, the quasi-static balance of linear momentum, governing the evolution of the macrostrain $\fmacrostrain$, reads
\begin{equation}\label{eq:macro_impuls}
	\div_{\! \effective{x}} \mean{\sum_{i=1}^{2} \chi_i \pd{\freeenergy_i}{\fmicrostrain}(\fmacrostrain + \nablas_{\!\! x}\, \fu, \macrotemp, \fstatev_i)} + \fb = \fzero,
\end{equation}
where $\mean{\cdot}$ denotes the volume average over $Y$
\begin{equation}
	\mean{\cdot} = \frac{1}{|Y|} \int_{Y} (\cdot) \d V.
\end{equation}
Furthermore, $\fb: \Omega \times [0, T] \rightarrow \ffR^d$ denotes the vector of volume forces and $\div_{\! \effective{x}}$ designates the divergence operator \wrt the macroscopic point $\effective{\fx} \in \Omega$. Secondly, the macroscopic heat equation reads
\begin{equation}\label{eq:macro_heat}
	\effective{c}_{\microstrain} \, \dot{\macrotemp}=  \effective{w} - \div_{\! \effective{x}}(\effective{\fq}) + \macrorpl,
\end{equation}
which governs the evolution of the macroscopic absolute temperature $\macrotemp$. Here, $\effective{w}: \Omega \times [0, T] \rightarrow \ffR$ denotes the macroscopic heat source and $\effective{\fq}: \Omega \times [0, T] \rightarrow \ffR^d$ stands for the macroscopic heat flux. The effective heat capacity at constant strain $\effective{c}_{\microstrain}$ is given explicitly by
\begin{equation}
	\effective{c}_{\microstrain} = -\macrotemp \mean{\sum_{i=1}^{2} \chi_i  \pdn{\freeenergy}{\microtemp}{2}(\fmacrostrain + \nablas_{\!\! x}\, \fu, \macrotemp, \fstatev_i)}.
\end{equation}
To keep the notation reasonable, we introduced the thermomechanical coupling term
\begin{equation}
	\begin{split}
		\macrorpl &= \macrotemp \mean{ \sum_{i=1}^{2} \chi_i \pdmixed{\freeenergy_i}{\microtemp}{\fmicrostrain}(\fmacrostrain + \nablas_{\!\! x}\, \fu, \macrotemp, \fstatev_i) : (\dot{\fmacrostrain} + \nablas_{\!\! x}\, \dot{\fu})}\\ 
		&+  \macrotemp \mean{ \sum_{i=1}^{2} \chi_i \pdmixed{\freeenergy_i}{\microtemp}{\fstatev}(\fmacrostrain + \nablas_{\!\! x}\, \fu, \macrotemp, \fstatev_i) \cdot \dot{\fstatev}_i} - \mean{\sum_{i=1}^{2} \chi_i \pd{\freeenergy_i}{\fstatev}(\fmacrostrain + \nablas_{\!\! x}\, \fu, \macrotemp, \fstatev_i) \cdot \dot{\fstatev}_i}
	\end{split}
\end{equation}
as an additional source term of the macroscopic heat equation. The former constitutes the back-coupling between the microscopic scale and the evolution of the macroscopic temperature. Please note that the coupling term $\macrorpl$ may be decomposed further. The first two terms are linked to changes in entropy, whereas the last summand is commonly referred to as the dissipation 
\begin{equation}
	\macrodissipation = - \mean{\sum_{i=1}^{2} \chi_i \pd{\freeenergy_i}{\fstatev}(\fmacrostrain + \nablas_{\!\! x}\, \fu, \macrotemp, \fstatev_i) \cdot \dot{\fstatev}_i}.
\end{equation}
The dissipation measures the dissipated energy of the composite due to the evolution of the internal variables, \eg the dissipated energy due to plastic flow, and is the primary cause for the self-heating of the material due to irreversible processes. \\
Typically, in a concurrent two-scale setting, the macroscopic balance of linear momentum~\eqref{eq:macro_impuls} and the macroscopic heat equation~\eqref{eq:macro_heat} are solved on the macroscopic scale while, in every Gauss point of the macroscopic model, the thermomechanical cell problem~\eqref{eq:micro_cell_problem} is solved as well. Here, the above-mentioned two-way thermomechanical coupling prevails. On the one hand, the macrostrain and the macroscopic absolute temperature influence the mechanical behavior at the microscopic scale. On the other hand, the evolution of the macroscopic absolute temperature is driven by the coupling term $\macrorpl$, which comprises deformation induced changes of entropy and dissipated energy on the microscopic level.\\ 
In the article at hand, we consider speeding up such a thermomechanical two-scale simulation by means of direct DMNs. In this context, a DMN might be regarded as a surrogate for the underlying microstructure for which the thermomechanical cell problem~\eqref{eq:micro_cell_problem} can be solved efficiently. However, to use a DMN to speed up such a fully coupled thermomechanical two-scale simulation, the aforementioned two-way thermomechanical coupling needs to be taken into account. This will be the topic of the following section.  


\section{Direct deep material networks for thermomechanical composites} \label{sec:direct_DMN_thermo}

\subsection{The framework of direct deep material networks}
\begin{figure}[h!]
	\centering
	\begin{subfigure}{0.33\textwidth}
		\includegraphics[width=\textwidth]{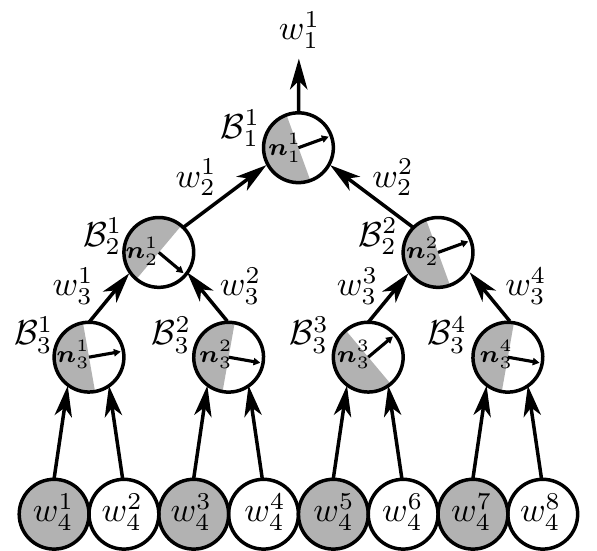}
		\caption{Weight propagation}
		\label{fig:DMN_weight_propagation}
	\end{subfigure}
	\begin{subfigure}{0.33\textwidth}
		\includegraphics[width=\textwidth]{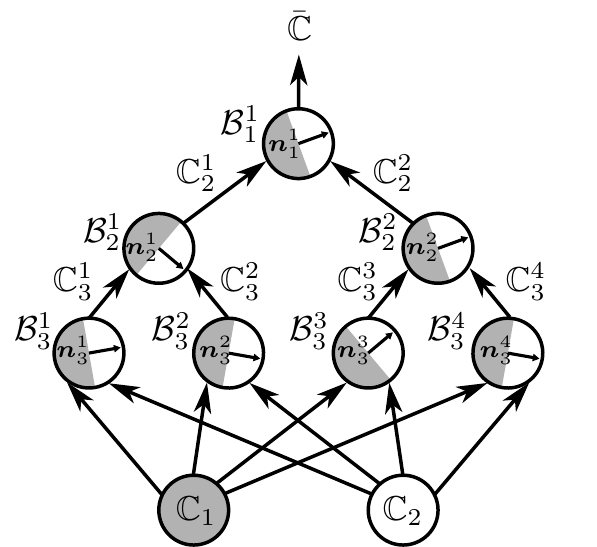}
		\caption{Stiffness propagation}
		\label{fig:DMN_stiffness_propagation}
	\end{subfigure}
	\caption{Weight and stiffness propagation (from the bottom to the top) in a two-phase direct DMN~\cite{Gajek2021} of depth $K=3$}
	\label{fig:DMN_schematics}
\end{figure}

We start with the formal definition of a direct DMN. For more detailed information, we refer to Gajek et al.~\cite{Gajek2020, Gajek2021}. We consider a two-phase direct DMN to be a perfect, ordered, rooted binary tree of depth $K$, see Fig.~\ref{fig:DMN_schematics} for an illustration.
Each node of the binary tree is given by a two-phase laminate $\BB^i_k$ with unknown direction of lamination $\fn^i_k$ and unknown volume fractions $c^i_{k,1}$ and $c^i_{k,2}$. Here, we denote the depth of a node by the letter $k = 1, \dots, K$ and consistently index the horizontal position by the letter $i = 1, \dots, 2^{k-1}$. Then, the DMN's free parameters are given by the directions of lamination, which we collect in the form of a (large) vector
\begin{equation}
	\vec{\fn} = \left[ \fn^1_K, \dots, \fn^{2^{K-1}}_K, \fn^1_{K-1}, \dots, \fn^{2^{K-2}}_{K-1}, \dots, \fn^1_1  \right] \in \left(\ffR^d\right)^{2^K - 1},
\end{equation}
and the volume fractions of all laminates $c^i_{k,1}$ and $c^i_{k,2}$. For reasons of numerical stability during the parameter identification, Liu et al.~\cite{Liu2018, Liu2019} proposed a change of coordinates when parameterizing the volume fractions. For this reason, the laminates' volume fractions are expressed in terms of the (input) weights $w^i_{K+1}$. These weights are assigned to the laminates at the bottom layer of the binary tree in pairs. By traversing the binary tree from the leaves to the root, the weights on level $k$ are inductively computed by a pairwise summation of the weights of the previous level, \ie
\begin{equation}
	w^i_k = w^{2i-1}_{k+1} + w^{2i}_{k+1}
\end{equation}
holds, see Fig.~\ref{fig:DMN_weight_propagation} for a schematic. Then, for every laminate, the volume fractions $c^i_{k,1}$ and $c^i_{k,2}$ are computed by normalization
\begin{equation}
	c^i_{k,1} = \frac{w^{2i-1}_{k+1}}{w^{2i-1}_{k+1} + w^{2i}_{k+1}} \quad \textrm{and} \quad c^i_{k,2} = 1 - c^i_{k,1}.
\end{equation}
For consistency, the weights $w^i_{K+1}$ need to be non-negative and sum to unity, \ie the conditions
\begin{equation}
	w^i_{K+1} \ge 0 \quad \textrm{and} \quad \sum_{i=1}^{2^K} w^i_{K+1} = 1
\end{equation}
hold. In the following, we collect the input weights $w^i_{K+1}$ into the vector
\begin{equation}
	\vec{w} = \left[w^1_{K+1}, \dots, w^{2^K}_{K+1} \right] \in \ffR^{2^K}_{\geq 0}.
\end{equation}
Thus, the network topology of a two-phase direct DMN of depth $K$ is uniquely determined by the vector $\vec{\fn}$, containing $2^K-1$ independent directions of lamination, and the vector $\vec{w}$ of weights comprising $2^K$ scalar parameters, for which $2^K-1$ parameters are independent. We call the process of identifying these free parameters the \textbf{offline training}. During the offline training, the DMN is fitted to the effective elastic response of a fixed microstructure $Y$ but varying stiffness parameters of the constituting phases. Afterwards, during the \textbf{online evaluation}, the free parameters $\vec{\fn}$ and $\vec{w}$ are fixed. Then, the DMN acts as a high-fidelity surrogate model for inelastic computations on the microscopic scale.

\subsection{Offline training} \label{sec:offline_training}

\review{For isothermal problems, DMNs are trained on linear elastic data alone, see Liu et al~\cite{Liu2018, Liu2019}. As we wish to predict the effective stress response of the composite for nonlinear and non-isothermal constituents, we assume that the linear elastic training still suffices.}  Thus, the following section serves as a brief summary of Gajek et al.~\cite{Gajek2020, Gajek2021}.\\ 
In the following, we treat the linear elastic training data as given, see Section~\ref{sec:sampling} for more information on the sampling of the training data. The training data is represented by a sequence of triples of stiffnesses $\set{\left( \effective{\ffC}^s, \ffC^s_1, \ffC^s_2 \right)}^{N_\textrm{s}}_{s=1}$ where $s$ enumerates the sample index and $N_\textrm{s}$ the number of samples. For a fixed microstructure $Y$, the training data is generated by sampling $N_\textrm{s}$ tuples of input stiffnesses $\left( \ffC^s_1, \ffC^s_2 \right)$ and computing the associated effective stiffness $\effective{\ffC}^s$ by means of computational homogenization.\\
The offline training, \ie the parameter identification, is expressed by solving the optimization problem
\begin{equation}\label{eq:optim_problem}
	J(\vec{\fn}, \review{\vec{w}}) \longrightarrow \min_{\vec{\fn}, \review{\vec{w}}} \quad \review{\textrm{s.t.} \quad w^i_{K+1} \geq 0}.
\end{equation}
Please note that there is some freedom in selecting a suitable objective function $J$, see, \eg Liu et al.~\cite{Liu2018, Liu2019} or Gajek et al.~\cite{Gajek2020, Gajek2021}. \review{In this work, we} follow Gajek et al.~\cite{Gajek2020, Gajek2021} and prescribe the following objective function 
\begin{equation} \label{eq:loss_function}
	J\left(\vec{\fn}, \vec{w}\right) = \frac{1}{N_b} \sqrt[q]{\sum_{s=1}^{N_b} \left(\frac{\normg{\effective{\ffC}^s - \DMNLIN{}\left(\ffC^s_{1}, \ffC^s_{2}, \vec{\fn}, \vec{w}\right)}_p}{\normg{ \effective{\ffC}^s }_p}\right)^q} + \lambda \left( \sum_{i=1}^{2^K} w^i_{K+1} - 1\right)^2.
\end{equation}
\review{Here, the $\norm{\cdot}_{p}$-norm on the stiffness tensors is defined via the $\ell^p$-norm of the components in (normalized) Voigt-Mandel notation and $p,q \geq 1$ and $\lambda \gg 0$ hold. }
\review{Furthermore}, $\DMNLIN{}$ denotes the DMN's linear elastic homogenization function
\begin{equation} \label{eq:lin_homogenization_function}
	\DMNLIN{}: \fC \times \fC \times \left(\ffR^d\right)^{2^K-1} \times \ffR^{2^K} \rightarrow \fC, \quad \left(\ffC_1, \ffC_2, \vec{\fn}, \vec{w}\right) \mapsto \effective{\ffC},
\end{equation}
which maps two input stiffnesses $\ffC_1$, $\ffC_2$ and the parameter vectors $\vec{\fn}$, $\vec{w}$ to the DMN's predicted effective stiffness. Efficiently evaluating the linear elastic homogenization function $\DMNLIN{}$ is paramount and involves computing a sequence of effective stiffnesses of two-phase laminates which are propagated from the bottom to the top of the binary tree. More formally, the effective stiffness 
\begin{equation}\label{eq:laminate_homogenization_linear}
	\ffC^i_k = \BB^i_k(\ffC^{2i-1}_{k+1}, \ffC^{2i}_{k+1})
\end{equation}
of a single laminate at level $k$ and position $i$ with direction of lamination $\fn^i_k$ and volume fractions $c^i_{k,1}$ and $c^i_{k,2}$ is computed by solving the equation
\begin{equation}
	\left(\ffP(\fn^i_k) + \lambda \left[ \ffC^i_k - \lambda \IDFOURS \right]^{-1}\right)^{-1}  = c^i_{k,1} \left(\ffP(\fn^i_k) + \lambda \left[ \ffC^{2i-1}_{k+1} - \lambda \IDFOURS \right]^{-1}\right)^{-1} + c^i_{k,2} \left(\ffP(\fn^i_k) + \lambda \left[ \ffC^{2i}_{k+1} - \lambda \IDFOURS \right]^{-1}\right)^{-1}
\end{equation}
for the effective stiffness $\ffC^i_k$, see Section 9.5 in Milton's book~\cite{Milton2002}. With $\IDFOURS: \Sym{d} \rightarrow \Sym{d}$, we denote the identity on $\Sym{d}$ and  $\ffP: \Sym{d} \rightarrow \Sym{d}$ stands for a projection operator which reads
\begin{equation}
	\left(\ffP(\fn)\right)_{mnop} =  \frac{1}{2}(n_m \delta_{no} n_p + n_n \delta_{mo} n_p + n_m \delta_{np} n_o + n_n \delta_{mp} n_o) - n_m n_n n_o n_p
\end{equation}
in Cartesian coordinates. Here, $\delta$ denotes the Kronecker symbol and $\lambda$ is a parameter which needs to be chosen either sufficiently large or suitably small, see Appendix in Kabel et al.~\cite{Kabel2015}. Starting at level $K+1$, the input stiffnesses $\ffC_1$, $\ffC_2$ are assigned pairwise to laminates at the $K$-th level and the respective effective stiffnesses are computed. These homogenized stiffnesses serve as the input for the next higher level, \ie the level $K-1$, until the effective stiffness of the DMN $\ffC^1_1 = \effective{\ffC}$ emerges on the highest level. We refer to Fig.~\ref{fig:DMN_stiffness_propagation} for a schematic of the stiffness propagation.\\ 
The objective function $J$ penalizes the difference of the DMN's predicted effective stiffness to the actual effective stiffness $\effective{\ffC}^s$ of microstructure $Y$ for all sampled input stiffnesses $\set{\left(\ffC^s_1, \ffC^s_2 \right)}^{N_s}_{s=1}$.  Additionally, the mixing constraint on the weights 
\begin{equation}
	\sum_{i=1}^{2^K} w^i_{K+1} = 1
\end{equation}
is encoded by the quadratic penalty term of Equation~\eqref{eq:loss_function}. To ensure that the non-negativity constraint on the weights $w^i_{K+1} \geq 0$ holds, we express the constrained weights $\vec{w} \in \ffR^{2^K}_{\geq 0}$  in terms of the unconstrained weights $\vec{v} = \left[v_1, \dots, v_{2^K}\right]  \in \ffR^{2^K}$ by projecting each element of $\vec{v}$ onto the positive real number line, \ie  
\begin{equation}
	\vec{w} = \langle \vec{v} \rangle_{+} \quad \textrm{with} \quad \langle \cdot \rangle_{+}: \ffR^{2^K} \rightarrow \ffR^{2^K}_{\geq 0}, \quad \vec{v} \mapsto \left[\max(0, v_1) , \dots, \max(0, v_{2^K})\right],
\end{equation}
holds. \review{In this way, the regression problem \eqref{eq:optim_problem} may be rewritten as
\begin{equation}\label{eq:optim_problem2}
J(\vec{\fn}, \langle\vec{v}\rangle_{+}) \longrightarrow \min_{\vec{\fn}, \vec{v}},
\end{equation}
which we solve} by means of accelerated stochastic gradient descent using mini batches of size $N_b$. A training epoch $j$ consists of the following steps: First, the loss function~\eqref{eq:loss_function} is evaluated for all stiffness samples in a batch. Then, the gradients ${\partial J}/{\partial \vec{\fn}}\left(\vec{\fn}_{j}, \review{\langle\vec{v}_j\rangle_{+}}\right)$, ${\partial J}/{\partial \review{\vec{v}}}\left(\vec{\fn}_{j}, \review{\langle\vec{v}_j\rangle_{+}}\right)$ are computed by means of automatic differentiation. Subsequently, the fitting parameters are updated by
\begin{equation}
	\vec{\fn}_{j+1} = \vec{\fn}_{j} - \alpha_{\vec{n}} \frac{\partial J}{\partial \vec{\fn}}\left(\vec{\fn}_{j}, \review{\langle\vec{v}_j\rangle_{+}}\right), \quad \review{\vec{v}}_{j+1} = \review{\vec{v}}_{j} - \alpha_{\review{\vec{v}}} \frac{\partial J}{\partial \review{\vec{v}}}\left(\vec{\fn}_{j}, \review{\langle\vec{v}_j\rangle_{+}}\right) \quad \textrm{and} \quad \review{\vec{w}_j = \langle\vec{v}_j\rangle_{+}}
\end{equation}
where $\alpha_{\vec{n}}, \alpha_{\review{\vec{v}}} \in \ffR_{>0}$ strictly larger than zero denote the learning rates. This procedure is repeated for all batches in the training set and for a \review{pre-defined} number of epochs. Upon convergence, the unknown fitting parameters of the DMN, \ie $\vec{\fn}$ and $\vec{w}$, are given.\\

\subsection{Online evaluation} \label{sec:online_evaluation}

For fixed fitting parameters $\vec{\fn}$ and $\vec{w}$, the goal of the online evaluation is to efficiently integrate a deep material networks implicitly at a single Gauss point of a macroscopic FE simulation. Indeed, direct DMNs are defined as a hierarchy of nested laminates. For this reason, they inherit \emph{thermodynamic consistency} and \emph{stress-strain monotonicity} from their phases, see Section 3.1 and Appendix C in Gajek et al.~\cite{Gajek2020} for a discussion. Thus, extending DMNs to non-isothermal problems does not infer any challenges from the point of view of thermodynamics. The governing equation, \ie the DMN's balance of linear momentum, emerges naturally by incorporating the homogeneity of the absolute temperature into the framework. Furthermore, considering the back-coupling from the microscopic onto the macroscopic scale is straightforward as well. Both will be explained in the following.\\
We consider a two-phase DMN of depth $K$ comprising two non-isothermal GSMs $\mathcal{G}_1 = (Z_1, \freeenergy_1, \dissipation_1, \fstatev_{0,1})$ and $\mathcal{G}_2 = (Z_2, \freeenergy_2, \dissipation_2, \fstatev_{0,2})$ as phases. We consider the former as a single laminate with a complex kinematics, comprising $2^K$ independent phases in total, see Gajek et al.~\cite{Gajek2020} for a schematic. We index these phases by the letter $i = 1, \dots, 2^K$ and assign to each phase the non-isothermal GSM $\mathcal{G}_i$ which alternates between $\mathcal{G}_1$ and $\mathcal{G}_2$, \ie
\begin{equation}
	\mathcal{G}_i = \left\{ \begin{array}{rl}
		\mathcal{G}_1 = (Z_1, \freeenergy_1, \dissipation_1, \fstatev_{0,1}), & i\textrm{ odd,}\\
		\mathcal{G}_2 = (Z_2, \freeenergy_2, \dissipation_2, \fstatev_{0,2}), & i\textrm{ even.}
	\end{array}
	\right.
\end{equation}
Let the superscript $n$ refer to the $n$-th time step at time $t^n$ and let $\triangle t = t^{n+1} - t^{n}$ denote the time increment. Then, for each phase $i = 1, \dots, 2^K$, discretizing Biot's equation~\eqref{eq:biots_euqation} in time with an implicit Euler method gives rise to the condensed free energy potential ${\condensedpotential_i: \Sym{d} \times \ffR_{> 0} \times Z_i \rightarrow \ffR}$,
\begin{equation}
	\condensedpotential_i\left(\fmicrostrain^{n+1}_i, \microtemp^{n+1}_i, \fstatev^n_i\right) = \inf_{\fstatev^{n+1}_i \in Z_i}\left(\freeenergy_i\left(\fmicrostrain^{n+1}_i, \microtemp^{n+1}_i, \fstatev^{n+1}_i\right) + \triangle t \, \dissipation_i\left(\microtemp^{n+1}_i, \frac{\fstatev^{n+1}_i - \fstatev^n_i}{\triangle t}\right) \right),
\end{equation}
solely depending on the strain $\fmicrostrain^{n+1}_i \in \Sym{d}$, the absolute temperature $\microtemp^{n+1}_i \in \ffR_{> 0}$ and the internal variables $\fstatev^{n}_i \in Z_i$ of the last converged time step. Then, for a fixed absolute temperature, the stress of phase $i$
\begin{equation}\label{eq:incremental_stress}
	\fmicrostress^{n+1}_i = \frac{\partial \condensedpotential_i}{\partial \fmicrostrain} \left(\fmicrostrain^{n+1}_i, \microtemp^{n+1}_i, \fstatev^{n}_i\right)
\end{equation}
is given by a nonlinear elastic law, \cf Lahellec-Suquet \cite{Lahellec2007_2} for more information. For the sake of exposition, we omit explicit reference to time step $n+1$ from here on.\\
First, we consider the kinematics of the DMN by collecting the phase strains into the vector $\vec{\fmicrostrain} = \left[\fmicrostrain_1,\ldots,\fmicrostrain_{2^K}\right] \in (\Sym{d})^{2^K}$, introducing the vector of macrostrains $\vec{\fmacrostrain} = \left[\fmacrostrain,\ldots,\fmacrostrain\right] \in (\Sym{d})^{2^K}$ and the vector of the unknown displacement jumps $\vec{\fa} \in (\ffR^d)^{2^K-1}$. The latter inherits its ordering from the vector of lamination directions $\vec{\fn}$. Then, the DMN's kinematics admits the representation
\begin{equation}\label{eq:DMN_kinematics}
	\vec{\fmicrostrain} = \vec{\fmacrostrain} + \fA \vec{\fa},
\end{equation}
where $\fA:(\ffR^d)^{2^K-1} \rightarrow (\Sym{d})^{2^K}$ is a gradient-type operator comprising the DMN's topology, \ie the tree structure, lamination directions and volume fractions, into a single linear mapping, see Gajek et al.~\cite{Gajek2020} for the derivation of the special structure of $\fA$. Secondly, the homogeneity of the absolute temperature on the microscopic scale, see Section~\ref{sec:asymptotic_homogenization}, infers that only the macroscopic absolute temperature $\macrotemp$ needs to be considered, \ie $\microtemp_i \equiv \macrotemp$ holds for all phases $i = 1, \dots, 2^K$. Here, the macroscopic absolute temperature $\macrotemp$ and the macrostrain $\fmacrostrain$ act as inputs to the DMN. Both are provided by the macroscopic finite element simulation for every Gauss point and for every increment of the global (Newton) solver. As outputs, the effective stress $\fmacrostress$, the thermomechanical coupling term $\macrorpl$ and algorithmic tangents, \ie the partial derivatives of the effective stress and thermomechanical coupling term \wrt the effective strain and macroscopic temperature, need to be returned.\\
We start with deriving the governing equation of a thermomechanically coupled direct DMN. Let $\effective{\condensedpotential}:(\Sym{d})^{2^K}\!\!\!\!\times \mathcal{Z} \rightarrow \ffR$ denote the averaged condensed free energy of the DMN
\begin{equation}
	\effective{\condensedpotential}(\vec{\fmicrostrain}, \macrotemp, \vec{\fstatev}^{\, n}) = \sum_{i=1}^{2^K} w^i_{K+1} \condensedpotential_{i}(\fmicrostrain_i, \macrotemp, \fstatev_i^n)
\end{equation}
where $\vec{\fstatev}^{\, n} = \left[\fstatev_1^n, \dots, \fstatev_{2^K}^n\right] \in \mathcal{Z} := Z_1 \oplus Z_2 \oplus \cdots \oplus Z_1 \oplus Z_2$ denotes the vector of internal variables of the last time step. Then, critical points of the optimization problem
\begin{equation}
	\effective{\condensedpotential}(\vec{\fmacrostrain} + \fA \vec{\fa}, \macrotemp, \vec{\fstatev}^{\, n}) \longrightarrow \min_{\vec{\fa} \in (\ffR^d)^{2^K-1}}
\end{equation}
encode the DMN's (microscopic) balance of linear momentum
\begin{equation}\label{eq:euler_lagrange}
	\T\fA \fW\vec{\fmicrostress}(\vec{\fmacrostrain} + \fA \vec{\fa}, \macrotemp, \vec{\fstatev}^{\, n}) = \fzero.
\end{equation}
 Here, $\vec{\fmicrostress} = \left[\fmicrostress_1, \dots, \fmicrostress_{2^K} \right] \in (\Sym{d})^{2^K}$ represents the vector of phase stresses for which Relation~\eqref{eq:incremental_stress} holds. Furthermore, the mass matrix $\fW:\Sym{d}^{2^K} \rightarrow \Sym{d}^{2^K}$ 
\begin{equation}\label{eq:weight_matrix}
	\fW = \diag{w^1_{K+1}, \dots, w^{2^K}_{K+1}}
\end{equation}
associates the weights $w^i_{K+1}$ to the corresponding phase stresses $\fmicrostress_i$ and may be represented by a diagonal matrix. Indeed, $\T\fA \fW: (\Sym{d})^{2^K} \rightarrow (\ffR^d)^{2^K-1}$ may be regarded as a divergence-type operator, such that the similarity of Relation~\eqref{eq:euler_lagrange} to the thermomechanical cell problem in general form~\eqref{eq:micro_cell_problem} is immediately revealed.\\
For solving the DMN's balance of linear momentum~\eqref{eq:euler_lagrange} for the unknown displacement jumps $\vec{\fa}$, we rely upon Newton's method. Let $j$ denote the $j$-th Newton increment. Then, for an initial guess ${\vec{\fa}_0 \in (\ffR^d)^{N-1}}$, the unknown displacement jump vector is iteratively updated, 
\begin{equation}
	\vec{\fa}_{j+1} = \vec{\fa}_{j} + s_j\,\triangle \vec{\fa}_j,
\end{equation}
for which the increment $\triangle\vec{\fa}_j \in \left(\ffR^d\right)^{2^K-1}$ solves the linear system
\begin{equation} \label{eq:newton_complex}
	\left[\T\fA \fW \frac{\partial \vec{\fmicrostress}}{\partial \vec{\fmicrostrain}}(\vec{\fmacrostrain} + \fA\vec{\fa_j}, \macrotemp, \vec{\fstatev}^{\, n}) \fA\right] \triangle \vec{\fa}_j = - \T\fA \fW \vec{\fmicrostress}(\vec{\fmacrostrain} + \fA\vec{\fa_j}, \macrotemp, \vec{\fstatev}^{\, n}).
\end{equation}
To ensure convergence, a step size $s_j\in(0,1]$ less than unity may arise from backtracking, and the Jacobian $\partial \vec{\fmicrostress} / \partial \vec{\fmicrostrain}$ may be represented by a block-diagonal matrix comprising the (stress-strain related) algorithmic tangents of the phase materials $\partial \fmicrostress_i / \partial \fmicrostrain \, (\fmicrostrain_i, \macrotemp, \statev^n_i)$, \ie
\begin{equation}\label{eq:block_diagonal_tangents_sress_strain}
	\frac{\partial \vec{\fmicrostress}}{\partial \vec{\fmicrostrain}}(\vec{\fmicrostrain}, \macrotemp, \vec{\fstatev}^{\, n}) = \textrm{block-diag}\left(\frac{\partial \fmicrostress_1}{\partial \fmicrostrain}(\fmicrostrain_1,  \macrotemp, \fstatev_1^n), \dots, \frac{\partial \fmicrostress_{2^K}}{\partial \fmicrostrain}(\fmicrostrain_{2^K},  \macrotemp, \fstatev_{2^K}^n) \right)
\end{equation} 
holds. Upon convergence, the DMN's effective stress $\fmacrostress$ is computed by averaging the phase stresses by
\begin{equation}\label{eq:effective_stress_new}
	\fmacrostress = \T{\left[\IDFOURS, \IDFOURS, \dots, \IDFOURS\right]} \fW \vec{\fmicrostress}(\vec{\fmacrostrain} + \fA \vec{\fa}, \macrotemp, \vec{\fstatev}^{\, n}),
\end{equation}
where $\left[\IDFOURS, \dots, \IDFOURS\right] \in \Sym{d}^{2^K}$ stand for a vector of the identity operators on $\Sym{d}$ and $\fW$ constitutes the weight matrix~\eqref{eq:weight_matrix}.\\
In Section~\ref{sec:asymptotic_homogenization}, we learned that the evolution of the macroscopic temperature $\macrotemp$ is coupled to the microscopic scale by the thermomechanical coupling term $\macrorpl$. For computing $\macrorpl$ efficiently, we introduce the phase-wise coupling terms 
\begin{equation}
	\microrpl_i(\fmicrostrain_i, \macrotemp, \fstatev_i^n) = \macrotemp \pdmixed{\condensedpotential_i}{\microtemp}{\fmicrostrain}(\fmicrostrain_i, \macrotemp, \fstatev_i^n) : \frac{\fmicrostrain_i - \fmicrostrain^n_i}{\triangle t} +  \left[\macrotemp \pdmixed{\condensedpotential_i}{\microtemp}{\fstatev}(\fmicrostrain_i, \macrotemp, \fstatev_i^n) - \pd{\condensedpotential_i}{\fstatev}(\fmicrostrain_i, \macrotemp, \fstatev_i^n)\right] \cdot \frac{\fstatev_i - \fstatev^n_i}{\triangle t}
\end{equation}
for every phase $i = 1, \dots, 2^K$, individually. With the vector of coupling terms $\vec{\microrpl} = \left[\microrpl_1, \dots, \microrpl_{2^K}\right] \in \ffR^{2^K}$ and the vector of ones, $\left[1, \dots, 1\right] \in \ffR^{2^K}$, we compute $\macrorpl$ by averaging, \ie
\begin{equation}\label{eq:effective_coupling_new}
	\macrorpl  = \T{\left[1, \dots, 1\right]} \fW \vec{\microrpl}(\vec{\fmacrostrain} + \fA \vec{\fa}, \macrotemp, \vec{\fstatev}^{\, n})
\end{equation}
holds.\\ 
To employ a DMN in a two-scale setting, four algorithmic tangents need to be computed  and provided to the macroscopic solver. We start with the algorithmic tangents related to the effective stress. Derivation of the effective stress $\fmacrostress$~\eqref{eq:effective_stress_new} \wrt the effective strain $\fmacrostrain$ and the absolute temperature $\macrotemp$ gives rise to the DMN's (stress-related) algorithmic tangents 
\begin{equation}\label{eq:algorithmic_tangent_ddsdde}
	\ffC^\textrm{algo}_{\macrostrain} := \frac{\partial \fmacrostress}{\partial \fmacrostrain} = \T{\left[\IDFOURS, \dots, \IDFOURS\right]} \fW \left[ \frac{\partial \vec{\fmicrostress}}{\partial \fmacrostrain}(\vec{\fmacrostrain} + \fA \vec{\fa}, \macrotemp, \vec{\fstatev}^{\, n}) + \frac{\partial \vec{\fmicrostress}}{\partial \vec{\fmicrostrain}}(\vec{\fmacrostrain} + \fA \vec{\fa}, \macrotemp, \vec{\fstatev}^{\, n}) \fA \frac{\partial \vec{\fa}}{\partial \fmacrostrain} \right]
\end{equation}
and
\begin{equation}\label{eq:algorithmic_tangent_ddsddt}
	\ffC^\textrm{algo}_{\macrotemp} := \frac{\partial \fmacrostress}{\partial \macrotemp} = \T{\left[\IDFOURS, \dots, \IDFOURS\right]} \fW \left[ \frac{\partial \vec{\fmicrostress}}{\partial \macrotemp}(\vec{\fmacrostrain} + \fA \vec{\fa}, \macrotemp, \vec{\fstatev}^{\, n}) + \frac{\partial \vec{\fmicrostress}}{\partial \vec{\fmicrostrain}}(\vec{\fmacrostrain} + \fA \vec{\fa}, \macrotemp, \vec{\fstatev}^{\, n}) \fA \frac{\partial \vec{\fa}}{\partial \macrotemp} \right].
\end{equation}
To get compact expressions, we introduced the vectors of algorithmic tangents
\begin{equation}
	\frac{\partial \vec{\fmicrostress}}{\partial \fmacrostrain}(\vec{\fmicrostrain}, \macrotemp, \vec{\fstatev}^{\, n}) = \left[\frac{\partial \fmicrostress_1}{\partial \fmicrostrain}(\fmicrostrain_1,  \macrotemp, \fstatev_1^n), \dots, \frac{\partial \fmicrostress_{2^K}}{\partial \fmicrostrain}(\fmicrostrain_{2^K},  \macrotemp, \fstatev_{2^K}^n) \right]
\end{equation}
and
\begin{equation}
	\frac{\partial \vec{\fmicrostress}}{\partial \macrotemp}(\vec{\fmicrostrain},  \macrotemp, \vec{\fstatev}^{\, n}) = \left[\frac{\partial \fmicrostress_1}{\partial \microtemp}(\fmicrostrain_1,  \macrotemp, \fstatev_1^n), \dots, \frac{\partial \fmicrostress_{2^K}}{\partial \microtemp}(\fmicrostrain_{2^K}, \macrotemp, \fstatev_{2^K}^n) \right]
\end{equation}
which arise by inserting $\partial \fmicrostress_i / \partial \fmicrostrain \, (\fmicrostrain_i, \macrotemp, \statev^n_i)$ and $\partial \fmicrostress_i / \partial \microtemp \, (\fmicrostrain_i, \macrotemp, \statev^n_i)$, $i=1,\dots,2^K$, into column vectors. To evaluate Expression~\eqref{eq:algorithmic_tangent_ddsdde} and \eqref{eq:algorithmic_tangent_ddsddt}, the partial derivatives of the strain jumps with respect to the macrostrain $\partial \vec{\fa}/\partial \fmacrostrain$ and the absolute temperature $\partial \vec{\fa}/\partial \macrotemp$ need to be computed first. To this end, differentiating the balance of linear momentum~\eqref{eq:euler_lagrange} with respect to the macrostrain $\fmacrostrain$ and the absolute temperature $\macrotemp$ yields the linear systems
\begin{equation}\label{eq:lin_system_da_dE}
	\left[\T\fA \fW \frac{\partial \vec{\fmicrostress}}{\partial \vec{\fmicrostrain}}(\vec{\fmacrostrain} + \fA\vec{\fa}, \macrotemp, \vec{\fstatev}^{\, n}) \fA\right]\frac{\partial \vec{\fa}}{\partial \fmacrostrain} = -\T\fA \fW \frac{\partial \vec{\fmicrostress}}{\partial \fmacrostrain}(\vec{\fmacrostrain} + \fA \vec{\fa}, \macrotemp, \vec{\fstatev}^{\, n})
\end{equation}
and
\begin{equation}\label{eq:lin_system_da_dT}
	\left[\T\fA \fW \frac{\partial \vec{\fmicrostress}}{\partial \vec{\fmicrostrain}}(\vec{\fmacrostrain} + \fA\vec{\fa}, \macrotemp, \vec{\fstatev}^{\, n}) \fA\right]\frac{\partial \vec{\fa}}{\partial \macrotemp} = -\T\fA \fW \frac{\partial \vec{\fmicrostress}}{\partial \macrotemp}(\vec{\fmacrostrain} + \fA \vec{\fa}, \macrotemp, \vec{\fstatev}^{\, n})
\end{equation}
which need to be solved for $\partial \vec{\fa}/\partial \fmacrostrain$ and $\partial \vec{\fa}/\partial \macrotemp$. By comparing Equations~\eqref{eq:lin_system_da_dE} and \eqref{eq:lin_system_da_dT} to \eqref{eq:newton_complex}, we observe that all three problems share the same linear operator, \ie only the right hand sides differ. Using a direct solver, \eg a Cholesky decomposition, the matrix decomposition can be reused to minimize the computational overhead.\\
Derivation of the effective coupling term $\macrorpl$ \wrt the macrostrain $\fmacrostrain$ and absolute temperature $\macrotemp$ gives rise to the DMN's (energy-related) algorithmic tangents
\begin{equation}\label{eq:algorithmic_tangent_drplde}
	\ffD^\textrm{algo}_{\macrostrain} := \frac{\partial \macrorpl}{\partial \fmacrostrain} = \left[1,  \dots, 1\right]^T \fW \left[ \frac{\partial \vec{\microrpl}}{\partial \fmacrostrain}(\vec{\fmacrostrain} + \fA \vec{\fa}, \macrotemp, \vec{\fstatev}^{\, n}) + \frac{\partial \vec{\microrpl}}{\partial \vec{\fmicrostrain}}(\vec{\fmacrostrain} + \fA \vec{\fa}, \macrotemp, \vec{\fstatev}^{\, n}) \fA \frac{\partial \vec{\fa}}{\partial \fmacrostrain} \right]
\end{equation}
and
\begin{equation}\label{eq:algorithmic_tangent_drpldt}
	\ffD^\textrm{algo}_{\macrotemp} := \frac{\partial \macrorpl}{\partial \macrotemp} = \left[1, \dots, 1\right]^T \fW \left[\frac{\partial \vec{\microrpl}}{\partial \macrotemp}(\vec{\fmacrostrain} + \fA \vec{\fa}, \macrotemp, \vec{\fstatev}^{\, n}) + \frac{\partial \vec{\microrpl}}{\partial \vec{\fmicrostrain}}(\vec{\fmacrostrain} + \fA \vec{\fa}, \macrotemp, \vec{\fstatev}^{\, n}) \fA \frac{\partial \vec{\fa}}{\partial \macrotemp} \right].
\end{equation}
As before, $\partial \vec{\microrpl} / \partial \vec{\fmicrostrain} \, (\vec{\fmicrostrain}, \macrotemp, \vec{\fstatev}^{\, n})$ denotes the block-diagonal matrix of phase-wise algorithmic tangents 
\begin{equation}
	\frac{\partial \vec{\microrpl}}{\partial \vec{\fmicrostrain}}(\vec{\fmicrostrain}, \macrotemp, \vec{\fstatev}^{\, n}) = \textrm{block-diag}\left(\frac{\partial \microrpl_1}{\partial \fmicrostrain}(\fmicrostrain_1,  \macrotemp, \fstatev_1^n), \dots, \frac{\partial \microrpl_{2^K}}{\partial \fmicrostrain}(\fmicrostrain_{2^K},  \macrotemp, \fstatev_{2^K}^n) \right).
\end{equation}
Furthermore, for brevity, the vectors of the (energy-related) algorithmic tangents
\begin{equation}
	\frac{\partial \vec{\microrpl}}{\partial \fmacrostrain}(\vec{\fmicrostrain}, \macrotemp, \vec{\fstatev}^{\, n}) = \left[\frac{\partial \microrpl_1}{\partial \fmicrostrain}(\fmicrostrain_1,  \macrotemp, \fstatev_1^n), \dots, \frac{\partial \microrpl_{2^K}}{\partial \fmicrostrain}(\fmicrostrain_{2^K},  \macrotemp, \fstatev_{2^K}^n) \right]
\end{equation}
and
\begin{equation}
	\frac{\partial \vec{\microrpl}}{\partial \macrotemp}(\vec{\fmicrostrain},  \macrotemp, \vec{\fstatev}^{\, n}) = \left[\frac{\partial \microrpl_1}{\partial \microtemp}(\fmicrostrain_1,  \macrotemp, \fstatev_1^n), \dots, \frac{\partial \microrpl_{2^K}}{\partial \microtemp}(\fmicrostrain_{2^K}, \macrotemp, \fstatev_{2^K}^n) \right]
\end{equation}
were introduced. Indeed, to efficiently compute Relations~\eqref{eq:algorithmic_tangent_drplde} and \eqref{eq:algorithmic_tangent_drpldt}, the already computed partial derivatives $\partial \vec{\fa}/\partial \fmacrostrain$ and $\partial \vec{\fa}/\partial \macrotemp$ are reused.\\
Later on in Section~\ref{sec:results_online_validation}, we take a closer look at the effective dissipation $\macrodissipation$ to assess the self-heating of the DMN under cyclic and non-cyclic loading. For this reason, we compute the phase-wise dissipation by
\begin{equation}
	\microdissipation_i(\fmicrostrain_i, \macrotemp, \fstatev_i^n) = - \pd{\condensedpotential_i}{\fstatev}(\fmicrostrain_i, \macrotemp, \fstatev_i^n) \cdot \frac{\fstatev_i - \fstatev^n_i}{\triangle t} \quad \textrm{with} \quad \vec{\microdissipation} = \left[\microdissipation_1, \dots, \microdissipation_{2^K}\right] \in \ffR^{2^K}.
\end{equation}
Then, the effective dissipation is computed by averaging
\begin{equation}\label{eq:effective_dissipation}
	\macrodissipation  = \T{\left[1, \dots, 1\right]} \fW \vec{\microdissipation}(\vec{\fmacrostrain} + \fA \vec{\fa}, \macrotemp, \vec{\fstatev}^{\, n}).
\end{equation}
The pseudo-code summarizing the relevant steps of the algorithm can be found in Algorithm~\ref{alg:online_evaluation}. Please note that the effective stress $\fmacrostress$, the effective thermomechanical coupling term $\macrorpl$, the effective dissipation $\macrodissipation$ and the algorithmic tangents $\ffC^\textrm{algo}_{\macrostrain}$, $\ffC^\textrm{algo}_{\macrotemp}$, $\ffD^\textrm{algo}_{\macrostrain}$ and $\ffD^\textrm{algo}_{\macrotemp}$ are computed after the convergence of Newton's method for reasons of numerical efficiency.

\begin{algorithm}
	
	\caption{Pseudo-code for the offline phase: Fixed parameters: $\textrm{tol}$, $\textrm{maxit}$, $\textrm{maxbacktrack}$, $\gamma$\\
		 Input: $\fmacrostrain^{n+1}$, $\macrotemp^{n+1}$, $\vec{\fstatev}^{\, n}$,  $\vec{\fa}^{\, n}$. Output: $\fmacrostress^{n+1}$, $\macrorpl^{n+1}$, $\ffC^\textrm{algo}_{\macrostrain}$, $\ffC^\textrm{algo}_{\macrotemp}$, $\ffD^\textrm{algo}_{\macrostrain}$, $\ffD^\textrm{algo}_{\macrotemp}$} \label{alg:online_evaluation}
	
	\begin{algorithmic}[1]
		\State $\vec{\fa}^{\, n+1} \gets \vec{\fa}^{\, n}$ 
		\Comment{Reuse old displacement jumps with $\vec{\fa}^{\, 0} \gets \fzero$.}
		\State $\textrm{res} \gets \textsc{Residual}(\fmacrostrain^{n+1}, \macrotemp^{n+1}, \vec{\fa}^{\, n+1}, \vec{\fstatev}^{\, n})$
		\Comment{Compute residual}
		\For{$i = 1$ \textbf{to} $\textrm{maxit}$}
		\State $\left[\T{\fA} \fW \frac{\partial \vec{\fmicrostress}}{\partial \vec{\fmicrostrain}} \fA\right]\triangle \vec{\fa}^{\, n+1} = -\T{\fA} \fW \vec{\fmicrostress}$
		\Comment{Solve linear system}
		\State $\textrm{res} \gets \textsc{Backtracking}(\fmacrostrain^{n+1}, \macrotemp^{n+1}, \vec{\fa}^{\, n+1}, \triangle \vec{\fa}^{\, n+1}, \vec{\fstatev}^{\, n}, \textrm{res})$
		\Comment{Update displacement jumps}
		\If{$\textrm{res} < \textrm{tol}$}
		\State break
		\Comment{Break if residual is smaller than tolerance}
		\EndIf
		\EndFor
		\State Update state variables $\vec{\fstatev}^{\, n}$ to $\vec{\fstatev}^{\, n+1}$ and compute $\fmacrostress^{n+1}$, $\macrorpl^{n+1}$, $\macrodissipation^{n+1}$, $\ffC^\textrm{algo}_{\macrostrain}$, $\ffC^\textrm{algo}_{\macrotemp}$, $\ffD^\textrm{algo}_{\macrostrain}$, $\ffD^\textrm{algo}_{\macrotemp}$
		\State
		\Function{Residual}{$\fmacrostrain^{n+1}$, $\macrotemp^{n+1}$, $\vec{\fa}^{\, n+1}$, $\vec{\fstatev}^{\, n}$}
		\State $\vec{\fmicrostrain}^{\, n+1} \gets \vec{\fmacrostrain}^{\, n+1} + \fA \vec{\fa}^{\, n+1}$
		\Comment{Compute phase strains}
		\For{$i = 1$ \textbf{to} $2^K$}
		\State $\fmicrostress^{n+1}_i \gets \frac{\partial \condensedpotential_i}{\partial \fmicrostrain} \left(\fmicrostrain^{n+1}_i, \macrotemp^{n+1}, \fstatev^{n}_i\right)$
		\Comment{Evaluate material laws}
		\EndFor
		\State $\textrm{res} \gets \frac{\normg{\T\fA \fW\vec{\fmicrostress}^{n+1}}_\textrm{F}}{(2^K - 1) \normg{\fmacrostress^{\, n+1}}_\textrm{F}} \quad$ with $\quad \fmacrostress^{n+1} \gets \sum_{i=1}^{2^K} w^i_{K+1} \fmicrostress^{n+1}_i$
		\Comment{Compute residual}
		\State \Return $\textrm{res}$
		\EndFunction
		\State
		\Function{Backtracking}{$\fmacrostrain^{n+1}$, $\macrotemp^{n+1}$, $\vec{\fa}^{\, n+1}$, $\triangle \vec{\fa}^{\, n+1}$, $\vec{\fstatev}^{\, n}$, $\textrm{res}_\textrm{old}$}
		\State $\vec{\fa}^{\, n+1} \gets \vec{\fa}^{\, n+1} + \triangle \vec{\fa}^{\, n+1}$
		\Comment{Update displacement jumps}
		\State $\textrm{res} \gets \textsc{Residual}(\fmacrostrain^{n+1}, \macrotemp^{n+1}, \vec{\fa}^{\, n+1}, \vec{\fstatev}^{\, n})$
		\Comment{Update residual}
		\For{$i \gets 0$ \textbf{to} $\textrm{maxbacktrack} - 1$}				
		\If{$\textrm{res} < \textrm{res}_\textrm{old}$}
		\State break
		\Comment{Break if residual decreases}
		\EndIf			
		\State $\vec{\fa}^{\, n+1} \gets \vec{\fa}^{\, n+1} - \gamma^i (1 - \gamma) \triangle \vec{\fa}^{\, n+1}$
		\Comment{Backtracking: update displacement jumps}
		\State $\textrm{res} \gets \textsc{Residual}(\fmacrostrain^{n+1}, \macrotemp^{n+1}, \vec{\fa}^{\, n+1}, \vec{\fstatev}^{\, n})$
		\Comment{Backtraking: update residual}
		\EndFor
		\State \Return $\textrm{res}$
		\EndFunction
	\end{algorithmic}
\end{algorithm}

\section{Short fiber reinforced polyamide}\label{sec:material_parameters}

\begin{figure}[h!]
	\centering
	\includegraphics[width=0.4\textwidth]{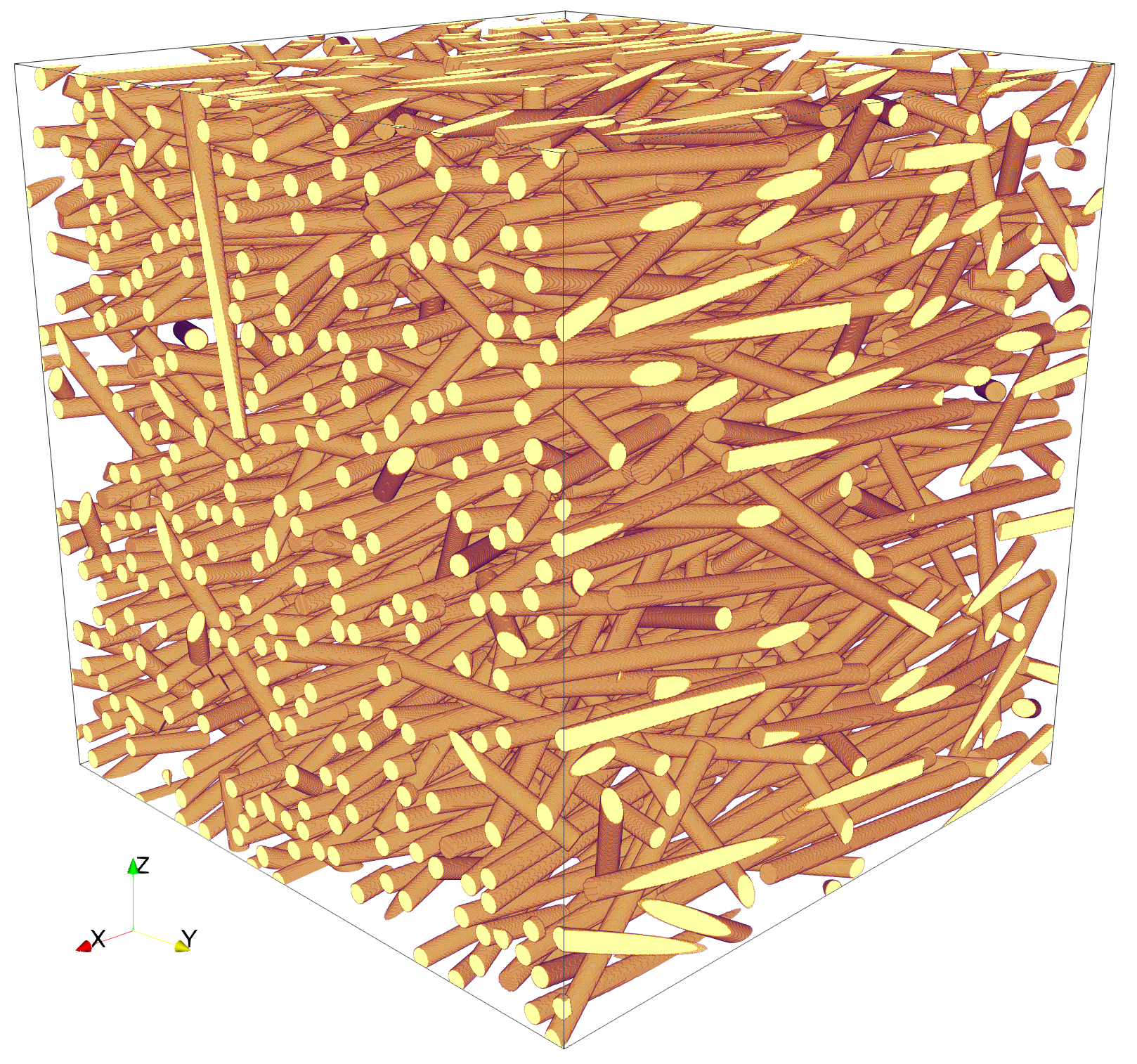}
	\caption{Generated microstructure realization comprising $577$ fibers.}
	\label{fig:microstructure}
\end{figure}

In general, thermoplastic polymers feature a pronounced thermomechanical coupling. For this reason, we study a short fiber reinforced polyamide $6.6$ (PA66) as our benchmark composite. As reinforcement, we consider E-glass fibers with a (uniform) fiber length of $L_\textrm{f} = 200 \unit{\mu m}$ and a fiber diameter of $D_\textrm{f} = 10 \unit{\mu m}$. We choose a fiber volume fraction of $c_\textrm{f} = 16 \%$, which correspond to a fiber mass fraction of approximately $30 \%$. The fiber orientation is described by a transversely isotropic fiber orientation tensor of second order~\cite{AdvaniTucker1987} which reads 
\begin{equation}
	\fA \equalhat \diag{0.8, 0.1, 0.1}
\end{equation}
in Cartesian coordinates, \ie $80\%$ of the fibers point in the $\fe_1$ direction, whereas $20\%$ of the fibers are uniformly distributed in the $\fe_2$-$\fe_3$ plane. Fig.~\ref{fig:microstructure} illustrates an example of such a microstructure comprising 577 straight, cylindrical fibers.\\

\paragraph{E-glass fibers} 

We model the E-glass fibers as isotropic, linear thermoelastic. We rely upon the commonly used additive splitting of the volume-specific Helmholtz free energy density into two parts
\begin{equation}
	\freeenergy(\fmicrostrain, \microtemp) = 	\freeenergy_\textrm{mech}(\fmicrostrain, \microtemp) + \freeenergy_\textrm{heat}(\microtemp).
\end{equation}
The first part \review{$\freeenergy_\textrm{mech}(\fmicrostrain, \microtemp)$} represents the storage of mechanical energy whereas the second part $\freeenergy_\textrm{heat}(\microtemp)$ represent the heat-storage alone. We assume the heat capacity at constant strain to be independent of the deformation $\fmicrostrain$. Thus, the mechanical part of the Helmholtz free energy $\freeenergy_\textrm{mech}(\fmicrostrain, \microtemp)$ may at most be linear in the temperature and 
\begin{equation}
	c_{\microstrain}(\microtemp) = - \microtemp \frac{\partial^2 \freeenergy_\textrm{heat}}{\partial \microtemp^2}(\microtemp)
\end{equation}
holds. For a constant heat capacity at constant strain $c_{\microstrain}(\microtemp) = c_0$, the heat storage part of the Helmholtz free energy reads
\begin{equation}
	\freeenergy_\textrm{heat}(\microtemp) = c_0 \left[ (\microtemp - \microtemp_0) - \microtemp \ln\left(\frac{\microtemp}{\microtemp_0}\right) \right],
\end{equation}
where $\microtemp_0$ stands for the reference temperature. The mechanical part of the Helmholtz free energy is given by the following quadratic form
\begin{equation}
	\freeenergy_\textrm{mech}(\fmicrostrain, \microtemp) = \frac{1}{2} \fmicrostrain : \ffC \left[\fmicrostrain\right] - \fmicrostrain : \ffC[\falpha (\microtemp - \microtemp_0)],
\end{equation}
such that the  stress response of the material computes to
\begin{equation}
	\fsigma = \ffC \left[\fmicrostrain - \falpha (\microtemp - \microtemp_0)\right].
\end{equation}
Both the stiffness $\ffC$ and the coefficient of thermal expansion $\falpha$ are assumed to be isotropic, \ie the following relations
\begin{equation}
	\ffC = 3 K \ffP_1 + 2 G \ffP_2 \quad \textrm{and} \quad \falpha = \alpha_0 \IDTWO
\end{equation}
hold, with the \review{projection operators} $\ffP_1: \Sym{d} \rightarrow \Sph{d}$ and $\ffP_2: \Sym{d} \rightarrow \Dev{d}$ on the spherical and deviatoric subspaces of $\Sym{d}$ \review{which read
\begin{equation}
	\left(\ffP_1\right)_{mnop} =  \frac{1}{3}\delta_{mn} \delta_{op} \quad \textrm{and} \quad \left(\ffP_2\right)_{mnop} =  \frac{1}{2}(\delta_{mo} \delta_{pn} + \delta_{mp} \delta_{on}) - \frac{1}{3} \delta_{mn} \delta_{op}
\end{equation}
in Cartesian coordinates} and $\IDTWO: \ffR^d \rightarrow \ffR^d$ denotes the identity on $\ffR^d$. The bulk modulus $K$ and the shear modulus $G$ may be expressed in terms of the Young's modulus $E$ and the Poisson's ratio $\nu$, \ie
\begin{equation}
	\quad K = \frac{E}{3(1 - 2 \nu)} \quad \textrm{and} \quad \quad G = \frac{E}{2(1 + \nu)}.
\end{equation}
As the material is purely elastic, the dissipation potential vanishes identically $\dissipation(\microtemp) \equiv 0$. Thus, the thermomechanical coupling term
\begin{align}
		\microrpl(\fmicrostrain, \microtemp) &= \microtemp  \pdmixed{\freeenergy}{\microtemp}{\fmicrostrain}(\fmicrostrain, \microtemp) : \dot{\fmicrostrain} = - \microtemp \, \dot{\fmicrostrain} : \ffC \left[ \falpha \right]
\end{align}
is solely dependent on the strain rate $\dot{\fmicrostrain}$ due to the vanishing dissipation, \ie $\microdissipation \equiv 0$ holds. In fact, a non-vanishing strain rate causes self-cooling under hydrostatic extension and self-heating under hydrostatic compression. This effect is commonly referred to as Gough-Joule effect, see, \eg Section $96$ in Truesdell-Noll~\cite{TruesdellNoll}. The material parameters for the E-glass fibers are taken from Tikarrouchine et al.~\cite{Tikarrouchine2019} and summarized in Tab.~\ref{tab:mat_param_glass_fibers}.

\begin{table}[H]
	\centering
	\begin{tabular}{l l}
		\hline
		Young's modulus & $E = 72.0 \unit{GPa}$  \\
		Poisson's ratio       & $\nu = 0.26$ \\
		\hline
		Heat capacity    & $c_0 = 2.1 \times 10^6 \unit{J\; {m}^{-3}\; K^{-1}}$\\
		Thermal expansion &  $\alpha_0 = 9 \times 10^{-6}\unit{K^{-1}}$\\
		Thermal conductivity &  $\kappa_0 = 0.93 \unit{W\; m^{-1}\; K^{-1}}$\\
		\hline
	\end{tabular}
	\caption{Material parameters of the E-glass fibers~\cite{Tikarrouchine2019}}
	\label{tab:mat_param_glass_fibers}
\end{table}


\paragraph{Polyamide $6.6$ matrix} 

For modeling the material behavior of the PA66 matrix, we adapt the model proposed by Krairi and co-workers~\cite{Krairi2019}, which was specifically derived for thermoplastic polymers under non-isothermal conditions. The model couples linear viscoelasticity, viscoplasticty and thermal effects such as thermal softening and dissipative self-heating. More precisely, the linear viscoelastic part of the model is given by a generalized Maxwell model comprising $N$ Maxwell elements, and the viscoplastic part is governed by $J_2$-viscoplasticity. We refer to Krairi et al.~\cite{Krairi2019} for all underlying modeling assumptions and the experimental calibration of the model.\\ 
For the PA66 matrix, we prescribe the following heat-storage related free energy
\begin{equation}
	\freeenergy_\textrm{heat}(\microtemp) = c_0 \left[ (\microtemp - \microtemp_0) - \microtemp \ln\left(\frac{\microtemp}{\microtemp_0}\right) \right].
\end{equation}
Furthermore, the mechanical part of the Helmholtz free energy reads
\begin{align}\label{eq:helmholtz_PA66}
	\begin{aligned}
		\freeenergy_\textrm{mech}(\fmicrostrain, \microtemp, \fstatev) & = \frac{1}{2} (\fmicrostrain - \fmicrostrain_\textrm{vp}) : \ffC_{\infty} [\fmicrostrain - \fmicrostrain_\textrm{vp}] - (\fmicrostrain - \fmicrostrain_\textrm{vp}) : \ffC_{\infty} [\falpha  (\microtemp - \microtemp_0)] + \int_0^{\microstrain_\textrm{p}} H(\microtemp, \bar{\microstrain}_\textrm{p}) \dif\bar{\microstrain}_\textrm{p}\\ 
		& +  \frac{1}{2} \sum_{i=1}^{N} (\fmicrostrain - \fmicrostrain_\textrm{vp} - \fmicrostrain_{\textrm{v},i}) : \ffC_i [\fmicrostrain - \fmicrostrain_\textrm{vp} - \fmicrostrain_{\textrm{v},i}] - \sum_{i=1}^{N} (\fmicrostrain - \fmicrostrain_\textrm{vp} - \fmicrostrain_{\textrm{v},i}) : \ffC_i [\falpha (\microtemp - \microtemp_0)].\\
	\end{aligned}
\end{align}
For readability, we collect the state variables, \ie the accumulated plastic strain $\microstrain_\textrm{p}$, the viscoplastic strain $\fmicrostrain_\textrm{vp}$ and the viscoelastic strains $\set{\fmicrostrain_{\textrm{v},i}}_{i=1}^N$ into the state vector $\fstatev = \left[ \microstrain_\textrm{p}, \fmicrostrain_\textrm{vp}, \fmicrostrain_{\textrm{v}, 1}, \dots, \fmicrostrain_{\textrm{v}, N} \right] \in Z := \ffR_{\ge 0} \times \Dev{d} \times \Sym{d}^{\times^N}$.\\
With the Helmholtz free energy~\eqref{eq:helmholtz_PA66} at hand, the material's stress response computes to 
\begin{equation}
	\fmicrostress =  \ffC_{\infty} [\fmicrostrain - \fmicrostrain_\textrm{vp} - \falpha (\microtemp - \microtemp_0)] +  \sum_{i=1}^{N} \ffC_i [\fmicrostrain - \fmicrostrain_\textrm{vp} - \fmicrostrain_{\textrm{v},i} - \falpha (\microtemp - \microtemp_0)].
\end{equation}
For a fixed viscoplastic strain $\fmicrostrain_{\textrm{vp}}$, we assume the material to be linear and isotropic, both in its long-term elastic and its purely viscoelastic response, and to feature an isotropic thermal expansion. More precisely, the stiffness governing infinitely slow processes $\ffC_{\infty}$, the stiffness $\ffC_i$ associated to the $i$-th dashpot and the coefficient of thermal expansion $\falpha$ admit the representations
\begin{equation}
	\ffC_{\infty} = 3 K_{\infty} \ffP_1 + 2 G_{\infty} \ffP_2, \quad \ffC_i = 3 K_i \ffP_1 + 2 G_i \ffP_2 \quad \textrm{and} \quad \falpha = \alpha_0 \IDTWO.
\end{equation}
The bulk  $K_{\infty}$, $K_{i}$ and shear moduli $G_{\infty}$, $G_{i}$ are expressed in terms of the Young's moduli $E_{\infty}$, $E_i$ and the Poisson's ratio $\nu$, \ie the following relations
\begin{equation}
	K_{\infty} = \frac{E_{\infty}}{3(1 - 2 \nu)}, \quad  K_i = \frac{E_i}{3(1 - 2 \nu)}, \quad G_{\infty} = \frac{E_{\infty}}{2(1 + \nu)} \quad \textrm{and} \quad G_i = \frac{E_i}{2(1 + \nu)}
\end{equation}
hold. Indeed, for the model at hand, the bulk and shear moduli $K_i$ and $G_i$ are coupled due to an assumed constant Poisson's ratio $\nu$, see Krairi et al.~\cite{Krairi2019}. Such an assumption is not unusual if only experimental data from uniaxial experiments are available.\\
Concerning the thermo-viscoelastic behavior, we assume the PA66 to be thermorheologically simple, \ie the viscosity tensor $\ffV_i$ associated to the $i$-th dashpot of the generalized Maxwell model should have the form
\begin{equation}
	\ffV_i = a_\microtemp(\microtemp) \left(3 K_i \, \tau_{\textrm{K},i} \, \ffP_1 + 2 G_i \tau_{\textrm{G},i} \, \ffP_2\right),
\end{equation}
where $a_{\microtemp}: \ffR_{> 0} \rightarrow \ffR_{> 0}$ denotes a temperature-dependent shift function. The volumetric and deviatoric relaxation times
\begin{equation}
	\tau_{\textrm{K},i} = \frac{\tau_i E_i}{K_i} \quad \textrm{and} \quad \tau_{\textrm{G},i} = \frac{\tau_i E_i}{G_i}  
\end{equation}
are expressed in terms of the Young's modulus $E_i$, the bulk and shear moduli $K_i$ and $G_i$ and the relaxation time $\tau_i$. The fluidity tensor $\ffF_i$ is given by the pseudoinverse of the viscosity tensor $\ffF_i = \ffV_i^{\dagger}$, giving rise to the evolution equation for the viscous strain 
\begin{equation}
	\dot{\fmicrostrain}_{\textrm{v},i} = \ffF_i \left[ \fsigma_{\textrm{v},i} \right],
\end{equation}
where $\fsigma_{\textrm{v},i}$ denotes the (viscous) partial stress
\begin{equation}
	\fmicrostress_{\textrm{v},i} = \ffC_i [\fmicrostrain - \fmicrostrain_\textrm{vp} - \fmicrostrain_{\textrm{v},i} - \falpha (\microtemp - \microtemp_0)]
\end{equation}
of the $i$-th dashpot.  As we consider temperatures above the glass transition, the temperature-dependent shift function is assumed to obey the Williams-Landel-Ferry (WLF)~\cite{WLF1955} equation
\begin{equation}
	\log_{10}(a_\microtemp(\microtemp)) = -\frac{C_1 (\microtemp - \microtemp_{\textrm{ref}})}{C_2 + (\microtemp - \microtemp_{\textrm{ref}})}.
\end{equation}
To capture thermal softening of the material, the yield stress 
\begin{equation}
	\microstress_\textrm{Y}: \ffR_{> 0} \rightarrow \ffR_{> 0}, \quad \microtemp \mapsto \Gamma(\microtemp, \beta_1) \, \sigma_\textrm{Y0},
\end{equation}
and the power-law hardening
\begin{equation}
	H: \ffR_{> 0} \times \ffR_{\ge 0} \rightarrow \ffR_{\ge 0}, \quad (\microtemp, \microstrain_\textrm{p}) \mapsto \Gamma(\microtemp, \beta_1) \, k \, \microstrain_\textrm{p}^n,
\end{equation}
feature an explicit temperature-dependence. The temperature-degradation function 
\begin{equation}
	\Gamma: \ffR_{> 0} \times \ffR_{\ge 0} \rightarrow \ffR_{> 0}, \quad (\microtemp, \beta) \mapsto \exp(- \beta (\microtemp - \microtemp_{\textrm{ref}})),
\end{equation}
takes the temperature and the material parameter $\beta_1 \in \ffR_{\ge 0}$ as input and degrades both the yield stress and the isotropic hardening \wrt the temperature. As for classical $J_2$-viscoplasticity, the evolution of the plastic strain
\begin{equation}
	\dot{\fmicrostrain}_{\textrm{vp}} = \sqrt{\frac{3}{2}} \dot{\microstrain}_\textrm{p} \frac{\fmicrostress'}{\norm{\fmicrostress'}}
\end{equation}
is driven by the deviatoric part of the stress tensor $\fmicrostress'$. The accumulated plastic strain rate $\dot{\microstrain}_\textrm{p}$ is given by the following evolution equation
\begin{equation}
	\dot{\microstrain}_\textrm{p} = \frac{\microstress_\textrm{Y}(\microtemp)}{\eta(\microtemp)} \left\langle\frac{ \sqrt{\frac{3}{2}} \norm{\fmicrostress'} - \microstress_\textrm{Y}(\microtemp) - H(\microtemp, \microstrain_\textrm{p}) }{\microstress_\textrm{Y}(\microtemp)}\right\rangle_{+}^m,
\end{equation}
see Krairi et al.~\cite{Krairi2019}, where the reference viscosity
\begin{equation}
	\eta: \ffR_{> 0} \rightarrow \ffR_{> 0}, \quad \microtemp \mapsto \Gamma(\microtemp, \beta_2) \,  \eta_0,
\end{equation}
involves a temperature-dependence as well.\\
In addition to the Helmholtz free energy, the material's (extended-valued) dissipation potential takes the following form
\begin{equation}
	\phi(\microtemp, \dot{\fstatev}) = 
	\left\{ \begin{array}{rl}
		\sigma_\textrm{Y}(\microtemp) \, \dot{\microstrain}_\textrm{p} + \sum_{i=1}^{N} \fsigma_{\textrm{v},i} : \dot{\fmicrostrain}_{\textrm{v},i}, & \dot{\microstrain}_\textrm{p} = \sqrt{\frac{2}{3}} \norm{\dot{\fmicrostrain}_{\textrm{vp}}},\\
		+\infty, & \textrm{otherwise}.\\
	\end{array} \right.
\end{equation}
For the material at hand, the thermomechanical coupling term $D$ computes to
\begin{align}
	\begin{aligned}
	\microrpl(\fmicrostrain, \microtemp, \fstatev) =& \, \microtemp  \pdmixed{\freeenergy}{\microtemp}{\fmicrostrain}(\fmicrostrain, \microtemp, \fstatev) : \dot{\fmicrostrain} +  \microtemp \pdmixed{\freeenergy}{\microtemp}{\fstatev}(\fmicrostrain, \microtemp, \fstatev) \cdot \dot{\fstatev}- \pd{\freeenergy}{\fstatev}(\fmicrostrain, \microtemp, \fstatev) \cdot \dot{\fstatev}\\
	=&  - \microtemp \, (\dot{\fmicrostrain} - \dot{\fmicrostrain}_\textrm{vp}) : \ffC_{\infty} \left[ \falpha \right] - \microtemp \sum_{i=1}^{N} (\dot{\fmicrostrain} - \dot{\fmicrostrain}_\textrm{vp} - \dot{\fmicrostrain}_{\textrm{v},i}) : \ffC_i \left[ \falpha \right] + \microtemp \frac{\partial H}{\partial \microtemp}(\microtemp, \microstrain_\textrm{p}) \, \dot{\microstrain}_\textrm{p}\\ 
	&+ \sigma_\textrm{Y}(\microtemp) \, \dot{\microstrain}_\textrm{p} + \sum_{i=1}^{N} \fmicrostress_{\textrm{v}, i} : \dot{\fmicrostrain}_{\textrm{v}, i},
	\end{aligned}
\end{align}
which is composed of three independent parts. The first two terms are responsible for the Joule-Gough effect. The third term is related to the thermal softening and the last two terms, \ie the dissipation
\begin{equation}
	\microdissipation(\fmicrostrain, \microtemp, \fstatev) = \sigma_\textrm{Y}(\microtemp) \, \dot{\microstrain}_\textrm{p} + \sum_{i=1}^{N} \fmicrostress_{\textrm{v}, i} : \dot{\fmicrostrain}_{\textrm{v}, i},
\end{equation}
comprises the dissipated energy due to viscoplastic and viscoelastic flow. The latter is responsible for the self-heating of the material due to viscoelastic or viscoplastic deformations. The full set of material parameters for the PA66, involving $N=12$ Maxwell elements, are summarized in Tab.~\ref{tab:mat_param_PA66}.
\begin{table}[H]
	\centering
	\begin{tabular}{l l}
		\hline
		Young's modulus & $E_{\infty} = 1.5 \unit{GPa}$  \\
		Poisson's ratio & $\nu = 0.42$ \\
		\hline
		Viscoelastic parameters &
		\begin{tabular}{@{}llll}
			$E_1 = 265 \unit{MPa}$ & $\log_{10}(\tau_1 / \SI{}{\second}) = -4.22$ & $E_{7} = 170 \unit{MPa}$ & $\log_{10}(\tau_{7}  / \SI{}{\second}) = 0.53$\\
			$E_2 = 262 \unit{MPa}$ & $\log_{10}(\tau_2 / \SI{}{\second}) = -3.42$ & $E_{8} = 92 \unit{MPa}$ & $\log_{10}(\tau_{8} / \SI{}{\second}) = 1.32$\\
			$E_3 = 248 \unit{MPa}$ & $\log_{10}(\tau_3 / \SI{}{\second}) = -2.63$ & $E_{9} = 78 \unit{MPa}$ & $\log_{10}(\tau_{9} / \SI{}{\second}) = 2.12$\\
			$E_4 = 231 \unit{MPa}$ & $\log_{10}(\tau_4 / \SI{}{\second}) = -1.84$ & $E_{10} = 65 \unit{MPa}$ & $\log_{10}(\tau_{10} / \SI{}{\second}) = 2.91$\\
			$E_5 = 211 \unit{MPa}$ & $\log_{10}(\tau_5 / \SI{}{\second}) = -1.05$ & $E_{11} = 54 \unit{MPa}$ & $\log_{10}(\tau_{11} / \SI{}{\second}) = 3.70$\\
			$E_6 = 190 \unit{MPa}$ & $\log_{10}(\tau_6 / \SI{}{\second}) = -0.26$ & $E_{12} = 48 \unit{MPa}$ & $\log_{10}(\tau_{12} / \SI{}{\second}) = 4.49$\\
		\end{tabular}\\
		\hline
		WLF parameters &  $C_1 = 26.21$ \quad $C_2 = 446.31 \unit{K}$\\
		\hline
		Isotropic hardening & $\sigma_\textrm{Y0} = 15.5 \unit{MPa}$ \quad $k = 103 \unit{MPa}$ \quad $n = 0.32$\\
		Viscoplastic parameters & $\eta_\textrm{0} = 74 \unit{MPa\; s}$ \quad $m = 2$\\
		Thermal softening &	$\beta_1 = \SI{0.011}{\kelvin^{-1}}$ \quad $\beta_2 = \SI{0.07}{\kelvin^{-1}}$ \quad $\microtemp_{\textrm{ref}} = \SI{298.15}{\kelvin}$\\
		\hline
		Heat capacity    & $c_0 = 1.9 \times 10^6 \unit{J\; {m}^{-3}\; K^{-1}}$\\
		Thermal expansion &  $\alpha_0 = 70 \times 10^{-6}\unit{K^{-1}}$\\
		Thermal conductivity &  $\kappa_0 = 0.27 \unit{W\; m^{-1}\; K^{-1}}$\\
		\hline
	\end{tabular}
	\caption{Material parameters of the PA66~\cite{Krairi2019}}
	\label{tab:mat_param_PA66}
\end{table}

\section{Identifying a DMN surrogate model using FFT-based computational homogenization} \label{sec:surrogate_model_validation}

This section is dedicated to the identification of the DMN surrogate model. First, we consider the sampling of the linear elastic training data. To this end, we start by identifying both the necessary resolution and the size of the representative volume element (RVE). Secondly, we present the offline training of the DMN and the validation of the surrogate model for thermomechanically coupled inelastic computations on the microscopic scale. For all numerical computations, we rely on a workstation equipped with two AMD EPYC 7642 with 48 physical cores each and $\numprint{1024}$ GB of DRAM.

\subsection{Material sampling} \label{sec:sampling}

\begin{figure}[h!]
	\centering
	\begin{subfigure}[t]{0.45\textwidth}
		\centering
		\includegraphics[height=5cm]{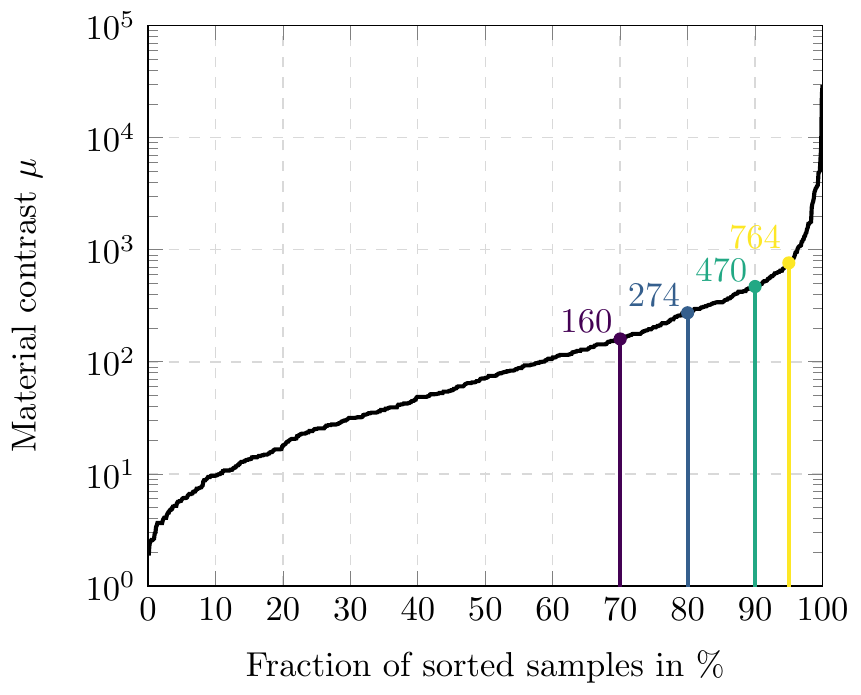}
		\caption{Sorted material contrast vs. samples}
		\label{fig:contrast_sorted}
	\end{subfigure}
	\begin{subfigure}[t]{0.45\textwidth}
		\centering
		\includegraphics[height=5cm]{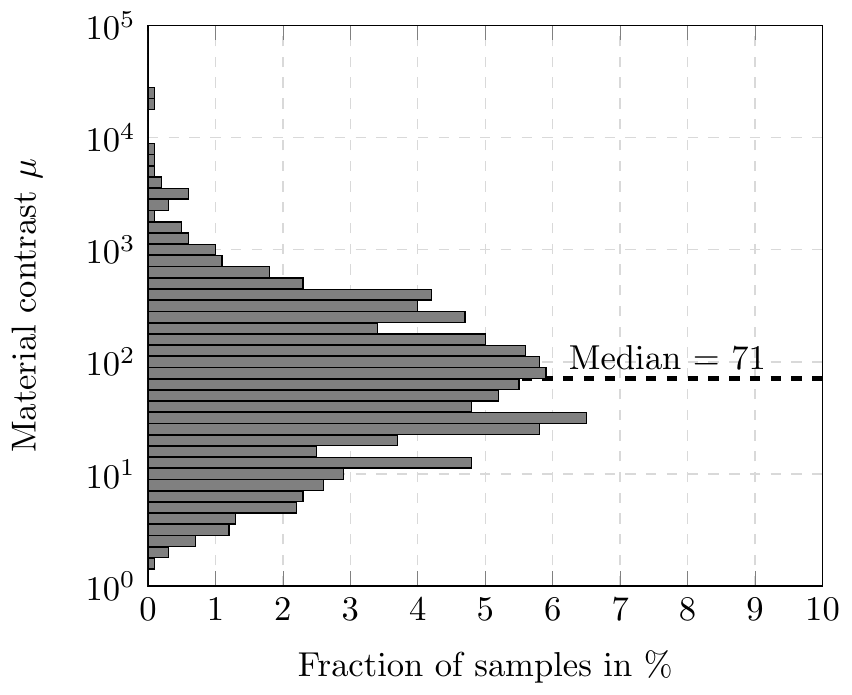}
		\caption{Distribution of material contrast vs. samples}
		\label{fig:contrast_histrogram}
	\end{subfigure}
	\caption{Distribution of the material contrast in the sample set}
	\label{fig:contrast_distribution}
\end{figure}

We start with the sampling of tuples of linear elastic input stiffnesses $\left(\ffC^s_1, \ffC^s_2 \right)$. Indeed, there is some freedom in selecting appropriate sampling strategies. For example, Liu and coworkers~\cite{Liu2018, Liu2019} and Gajek et al.~\cite{Gajek2020} sampled orthotropic stiffnesses. In the work at hand, we follow Gajek et al.~\cite{Gajek2021} who proposed to draw the samples from the space of possible algorithmic tangents occurring during the online evaluation. More precisely, the input stiffnesses $\ffC^s_1$, corresponding to the isotropic, purely thermoelastic glass fibers, are sampled from the set of isotropic stiffnesses, \ie we use a parameterization
\begin{equation}
	\ffC^s_1 = 3 K^s_1 \, \ffP_1 + 2 G^s_1 \, \ffP_2.
\end{equation}
As the polyamide matrix features a thermo-viscoelastic, viscoplastic material behavior, the samples $\ffC^s_2$ are assumed to be isotropic minus a rank-one perturbation, see Chapter 3 in Simo-Hughes~\cite{SimoHughes1998}. Thus, the stiffness $\ffC^s_2$ is assumed to have the form
\begin{equation}
	\ffC^s_2 = 3 K^s_2 \, \ffP_1 + 2 G^s_2 \left(\ffP_2 - a_s \, \fN'_s \otimes \fN'_s\right),
\end{equation}
where $\fN'_s \in \mathcal{N} := \set{\fN \in \Sym{d} \ | \ \tr{\fN} = 0, \ \normg{\fN}_\textrm{F}=1}$ is normalized and deviatoric. In other words, the set of all considered positive definite stiffness tuples $\left(\ffC^s_1, \ffC^s_2\right)$ may be parameterized via
\begin{equation}\label{eq:sample_set}
	\left(K^s_1, G^s_1, K^s_2, G^s_2, a_s, \fN'_s \right) \in \ffR_{>0} \times \ffR_{>0} \times \ffR_{>0} \times \ffR_{>0} \times \left[0, 1\right) \times \mathcal{N}.
\end{equation}
The former set is given in terms of an eight-dimensional continuum. For more details on sampling $\left(K^s_1, G^s_1, K^s_2, G^s_2, a_s, \fN'_s \right)$ from this eight-dimensional space and assembling the stiffnesses $\left(\ffC^s_1, \ffC^s_2 \right)$, we refer to Section 4.4 in Gajek et al.~\cite{Gajek2021}.\\ 
In the following, we assume that $N_\textrm{s} = 1000$ tuples of input stiffnesses $\set{\left( \ffC^s_1, \ffC^s_2 \right)}^{N_s}_{s=1}$ were generated. With these samples at hand, we turn our attention to the computation of the associated effective stiffnesses. For this purpose, a representative volume element (RVE) with a suitable resolution and size needs to be generated first. To this end, we take a closer look at the sampled input stiffnesses. More precisely, we consider the distribution of the material contrast $\mu$ which is defined, for the sample $s$, as
\begin{equation}
	\mu^s = \max\left( \frac{\lambda^s_{1, \textrm{max}}}{\lambda^s_{2, \textrm{min}}}, \frac{\lambda^s_{2, \textrm{max}}}{\lambda^s_{1, \textrm{min}}} \right).
\end{equation}
Here, $\lambda^s_{1/2, \textrm{max}}$ and $\lambda^s_{1/2, \textrm{min}}$ denote the largest and smallest eigenvalues of stiffnesses $\ffC^s_1$ and $\ffC^s_2$, respectively. Fig.~\ref{fig:contrast_sorted} illustrates the sorted material contrast vs. the $\numprint{1000}$ samples. 
We observe that the material contrast starts at around two and goes up to around $\numprint{23000}$. To get a better understanding of how the material contrast is distributed on the sample set, Fig.~\ref{fig:contrast_histrogram} shows the respective histogram with $50$ evenly log-spaced bins. We observe that the median of the distribution is well below a material contrast of $100$ and that only $3 \%$ of the samples exceed a material contrast of $\numprint{1000}$. In the following section, we consider finding a suitable resolution and size of the volume element, taking into account the findings of this section.

\subsection{On the necessary resolution and the size of the RVE} \label{sec:res_RVE_study}

\begin{figure}[H]
	\centering
	\begin{subfigure}{\textwidth}
		\centering
		\includegraphics[height=0.5cm]{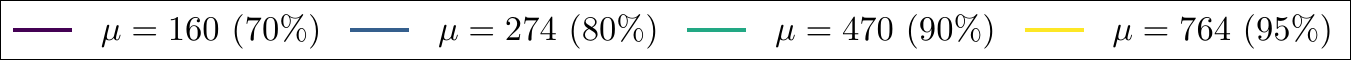}
	\end{subfigure}
	\begin{subfigure}{0.45\textwidth}
		\centering
		\includegraphics[height=5cm]{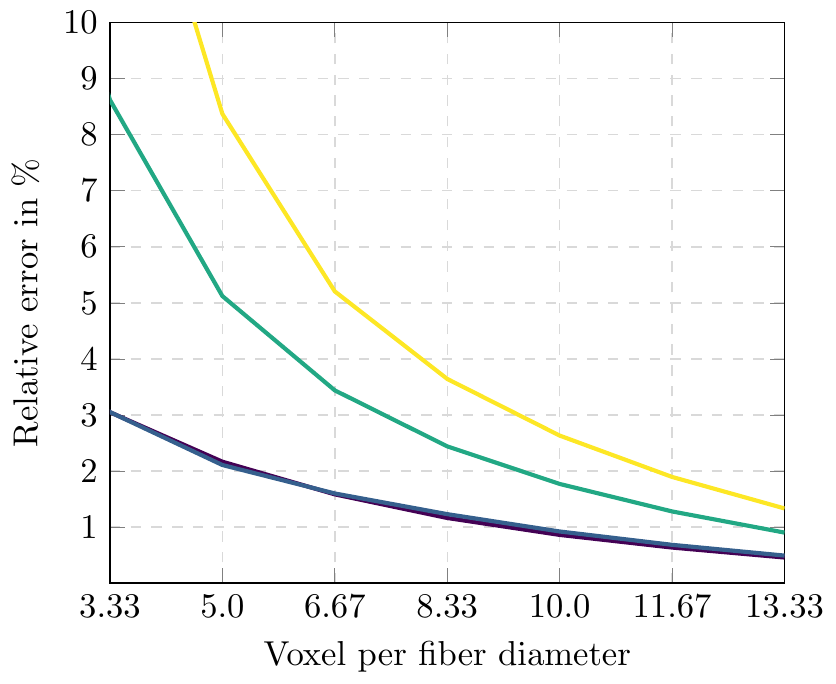}
		\caption{Relative error vs. resolution}
		\label{fig:res_study}
	\end{subfigure}
	\begin{subfigure}{0.45\textwidth}
		\centering
		\includegraphics[height=5cm]{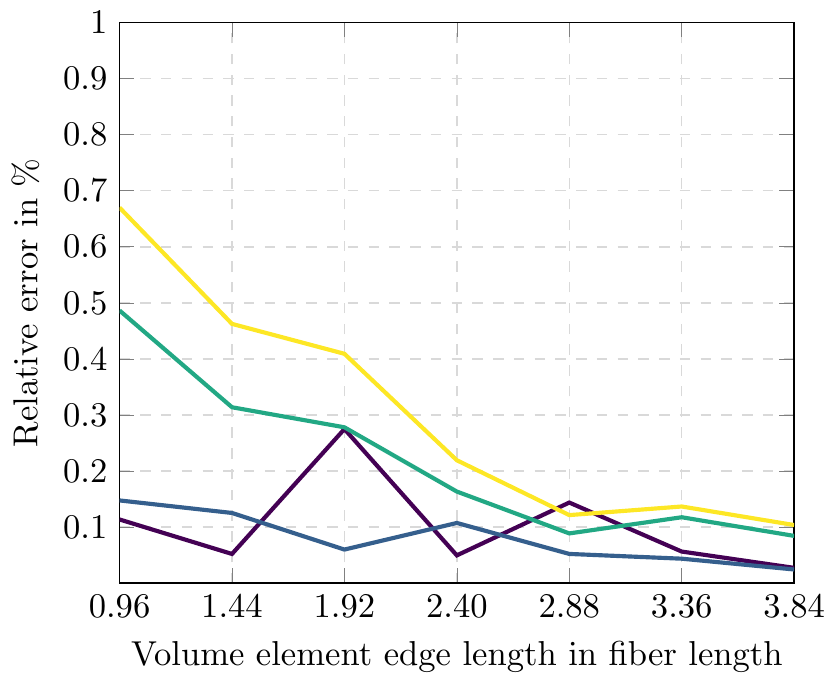}
		\caption{Relative error vs. volume element size}
		\label{fig:RVE_study}
	\end{subfigure}
	\caption{Study to determine necessary resolution and size of the RVE. Shown is the relative error of the computed effective stiffness vs. the resolution and volume element size}
	\label{fig:res_RVE_study}
\end{figure}

Finding a suitable resolution and RVE size is necessary to obtain accurate effective properties. However, performing a resolution and RVE size study for any tuple of input stiffnesses $\left( \ffC^s_1, \ffC^s_2 \right)$ is computationally expensive. The former is especially relevant for samples with a high material contrast, \ie greater than $1000$,  which only occur with a small frequency in the sample set. For this reason, we conduct a resolution and RVE size study for selected samples alone. To be more precise, we choose samples from the sampling set $\set{\left( \ffC^s_1, \ffC^s_2 \right)}^{N_\textrm{s}}_{s=1}$ corresponding to the $70$th, $80$th, $90$th and $95$th percentile \ie samples with a material contrast of $\mu = 160$, $\mu = 274$, $\mu = 470$ and $\mu = 764$, respectively, see Fig.~\ref{fig:contrast_sorted} for an illustration and color coding.\\
For a start, we consider generated cubic microstructures with a variable resolution and with a fixed edge length of $L=384 \unit{\mu m}$, \ie roughly twice the fiber length of $L_\textrm{f} = 200 \unit{\mu m}$. We vary the resolution from $3.3$ to $13.3$ voxels per fiber diameter in equidistant steps. The former corresponds to volume element discretizations with $128^3$ to $512^3$ voxels. We choose a resolution of $20$ voxels per fiber diameter, \ie discretized by $768^3$ voxels, as reference.\\
For generating the volume elements, we rely upon the Sequential Addition and Migration (SAM)~\cite{SAM} method, using the exact closure approximation~\cite{MontgomerySmith2011}. The SAM method takes the fiber length $L_\textrm{f}$, the fiber diameter $D_\textrm{f}$, the fiber volume fraction $c_\textrm{f}$ and the (axis-aligned) fiber orientation tensor $\fA$ as inputs and generates short fiber reinforced microstructure realizations. The effective stiffnesses are computed with the help of an FFT-based computational micromechanics code~\cite{MoulinecSuquet1994,MoulinecSuquet1998} using a conjugate gradient solver \cite{Zeman2010,BrisardDormieux2010} and the staggered grid discretization \cite{willot2014fourier, willot2015fourier}.\\
Fig.~\ref{fig:res_study} shows the relative error of the effective stiffness computed by the Frobenius norm of the corresponding Voigt matrices. For the crudest resolution of $3.33$ voxels per fiber diameter, the relative error exceeds $10\%$. Increasing the resolution decreases the relative error for the four considered  material contrasts. At a resolution of ten voxels per fiber diameter, the relative error of the sample corresponding to the $95$th percentile falls below $3 \%$. For the samples corresponding the $90$th, $80$th/$70$th percentile, the relative error is below $2 \%$ and $1 \%$, respectively. As material contrasts of $\mu = 764$ and above only occur with frequency of less than $5 \%$, we consider the resolution of $10$ voxels per fiber diameter as sufficient. We fix this resolution and focus on finding a suitable size of the RVE.
We investigate volume elements with edge length $L$ ranging from $0.96$ up to $3.84$ fiber lengths. The former corresponds to volume element discretizations with $192^3$ up to $768^3$ voxels. To obtain the reference, we generate a volume element with edge length of $5.76$ fiber lengths and discretized by $1152^3$ voxels. As before, we consider the relative error in the effective stiffness as error measure. For all considered edge lengths, the relative error is well below $1 \%$, see Fig.~\ref{fig:RVE_study}. Indeed, even the smallest volume element, \ie an edge length smaller than the fiber length, the relative error is \review{below} $0.5\%$. Increasing the volume element edge length from $0.96$ to $3.84$ further decreased the error. In the work at hand, we consider a volume elements length of $1.92$ fiber lengths, \ie a edge lengths of $L = 384 \unit{\mu m}$, as sufficient to keep the computational costs for the sampling of the training data reasonable.\\
With the optimal resolution and RVE size at hand, \ie a resolution of $10$ voxels per fiber diameter and a volume element discretization with $384^3$ voxels, we compute the effective stiffnesses of all generated $N_\textrm{s} = 1000$ stiffness samples and turn our attention to the training of the DMN. 

\subsection{Offline training} \label{sec:results_offline_training}

\begin{figure}
	\centering
	\begin{subfigure}{0.49\textwidth}
		\centering
		\includegraphics[height=6cm]{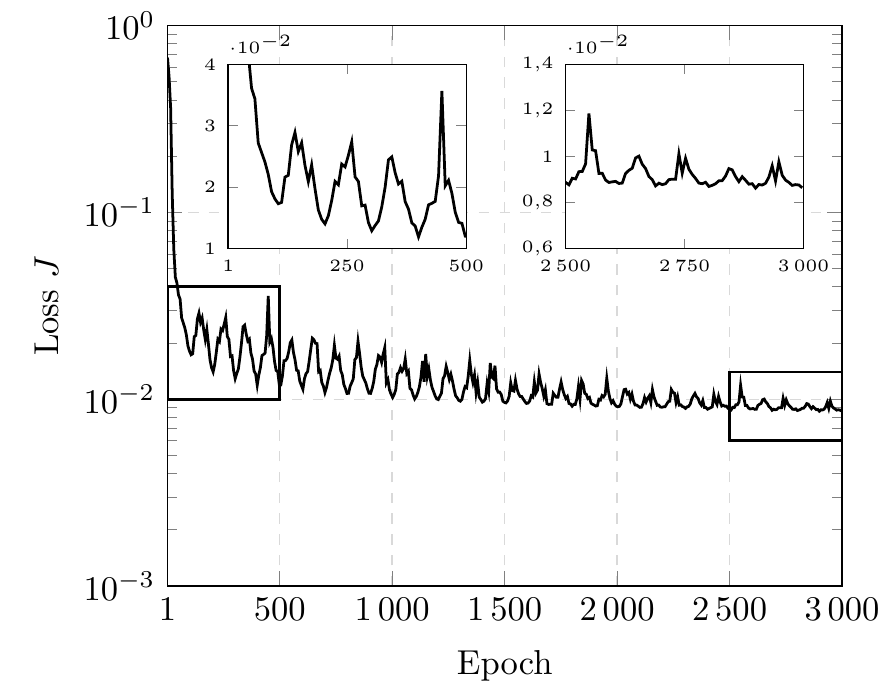}
		\caption{Loss function vs. training epoch}
	\end{subfigure}
	\begin{subfigure}{0.49\textwidth}
		\centering
		\includegraphics[height=6cm]{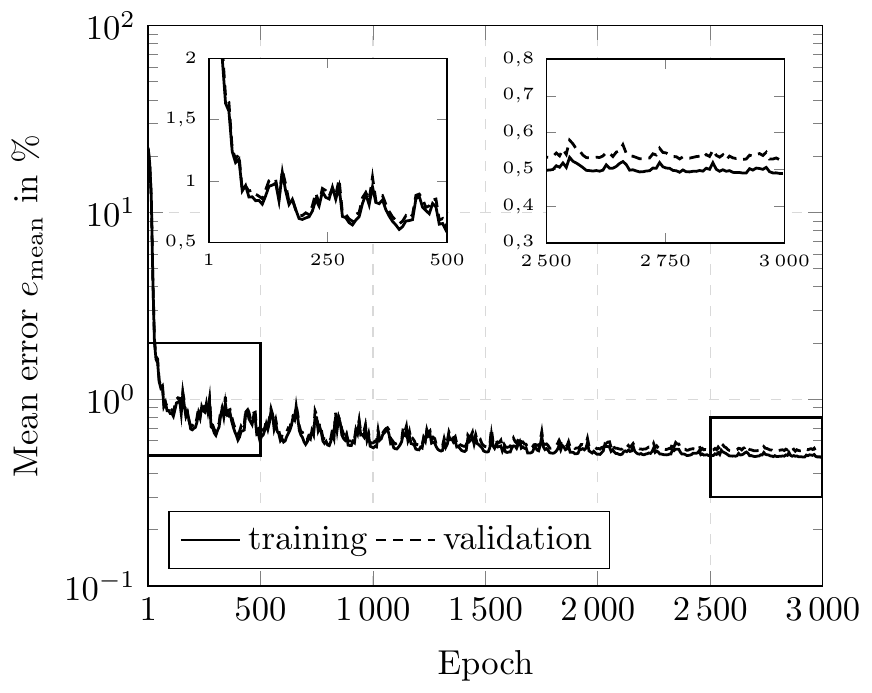}
		\caption{Mean error vs. training epoch}
	\end{subfigure}
	\caption{Loss function (a) and mean training and validation error (b) for the $3000$ training epochs}
	\label{fig:offline_training}
\end{figure}

As explained in Section~\ref{sec:offline_training}, we implement the offline training in PyTorch~\cite{paszke2017automatic} exploiting the frameworks automatic differentiation capabilities, see Gajek et al.~\cite{Gajek2020, Gajek2021} for more details. From previous works~\cite{Liu2018, Liu2019, Gajek2020, Gajek2021}, we know that at least eight layers are necessary to achieve a sufficient approximation quality for inelastic computations, \ie during the online evaluation. For this reason, we restrict to a two-phase DMN with $K = 8$ layers, \ie $255$ individual directions of lamination and $256$ weights as free parameters. \\
We randomly split the training data $\set{\left( \effective{\ffC}^s, \ffC^s_1, \ffC^s_2 \right)}^{N_s}_{s=1}$ into a training and a validation set, comprising $90 \%$ and $10 \%$ samples, respectively. The DMN is trained with mini batches with a batch size of $N_\textrm{b} = 32$ samples, which are drawn randomly from the training set. Batches with less than $32$ samples are discarded. Prior to the offline training, we sample the unknown directions of lamination $\fn^i_k$ from a uniform distribution on the unit sphere and the initial weights $w^i_{K+1}$ are sampled from a uniform distribution on $\left[0, 1\right]$ and subsequently rescaled to sum to unity.\\
For training the DMN, we rely on the AMSGrad method~\cite{Adam, AMSGrad} and determine appropriate learning rates $\alpha_{\vec{n}}$ and $\alpha_{\review{\vec{v}}}$ by a learning rate sweep as suggested by Smith-Topin~\cite{Smith2019}. The learning rate sweep yields almost identical learning rates, \ie $\alpha_{\vec{n}} = \alpha_{\review{\vec{v}}} = 1.5 \cdot 10^{-2}$. To aid finding a suitable minimizer for $J$~\eqref{eq:loss_function}, we employ the warm restart technique as suggested by Loshchilov-Hutter~\cite{Loshchilov2016}. The warm restarts are realized by a harmonic learning rate modulation
\begin{equation}\label{eq:learning_rate_moduluation}
	\alpha: \ffN \rightarrow \ffR, \quad m \mapsto \gamma^m\left(\alpha_\textrm{min} + \frac{1}{2}\left( \alpha_\textrm{max} - \alpha_\textrm{min} \right) \left( 1 + \cos\left(\pi \frac{m}{M}\right) \right)\right)
\end{equation}
of the learning rates $\alpha_{\vec{n}}$ and $\alpha_{\review{\vec{v}}}$ in combination with a geometric decay to enforce convergence. Here, $\alpha_\textrm{min}$ and $\alpha_\textrm{max}$ denote the minimum and maximum learning rate, $2M$ corresponds to the period of the modulation, and $\gamma$ represents the geometric decay rate. The maximum learning rate, both for $\alpha_{\vec{n}}$ and $\alpha_{\review{\vec{v}}}$, is set to $\alpha_\textrm{max} = 1.5 \cdot 10^{-2}$, \ie the result of the learning rate sweep. The minimum learning rate is set to $\alpha_\textrm{min} = 1.5 \cdot 10^{-3}$, \ie one magnitude smaller. In addition, we choose $M = 50$ as well as $\gamma = 0.999$ for the learning rate modulation~\eqref{eq:learning_rate_moduluation} and set $p = 1$, $q = 10$ and $\lambda = 10^3$ for the loss function~\eqref{eq:loss_function}.\\
We measure the accuracy of the fit by the mean error
\begin{equation}
	e_\textrm{mean} = \frac{1}{N_s} \sum_{s=1}^{N_s} \frac{\normg{\DMNLIN{}\left(\ffC^s_1, \ffC^s_2, \vec{\fn}, \vec{w}\right) - \effective{\ffC}^s}_1}{\normg{\effective{\ffC}^s}_1}, 
\end{equation}
 where $\norm{\cdot}_1$ denotes the $\ell^1$-norm \review{of the components in (normalized) Voigt-Mandel} notation, and $N_\textrm{s}$ denotes the number of elements in the training or validation set, depending on the considered scenario.\\
In Fig.~\ref{fig:offline_training}, the training progress in terms of the loss $J$ and the mean error $e_\textrm{mean}$ is illustrated. Overall, the effect of the learning rate modulation becomes apparent. The loss as well as the mean training and validation error fluctuate heavily, especially for the first $500$ epochs. The fluctuation decreases due to the learning rate decay such that in the last $500$ epochs, convergence is ensured. During the training, no significant model over-fitting can be observed as the validation error does not increase noticeably during training.

\subsection{Online validation} \label{sec:results_online_validation}

\begin{figure}[H]
	\begin{subfigure}{\textwidth}
		\centering
		\includegraphics[height=0.48cm]{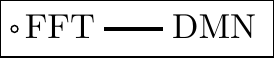}
		\includegraphics[height=0.48cm]{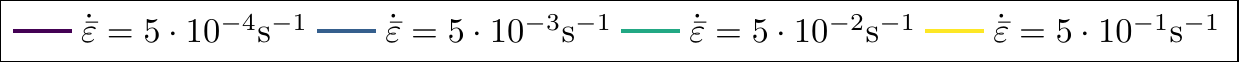}
	\end{subfigure}
	\begin{subfigure}{\textwidth}
		\centering
		\begin{subfigure}[b]{0.32\textwidth}
			\includegraphics[height=3.85cm]{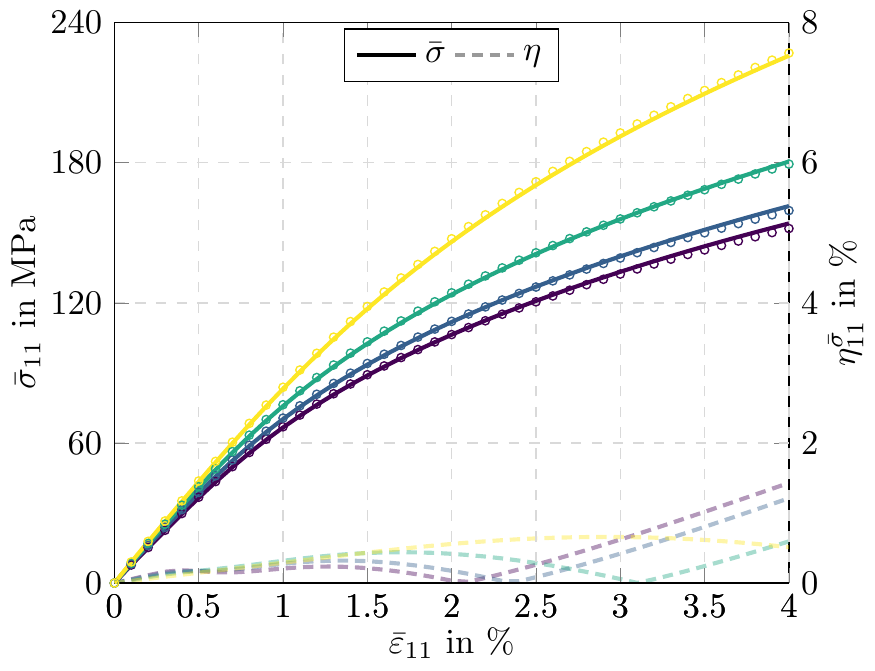}
		\end{subfigure}
		\begin{subfigure}[b]{0.32\textwidth}
			\includegraphics[height=3.85cm]{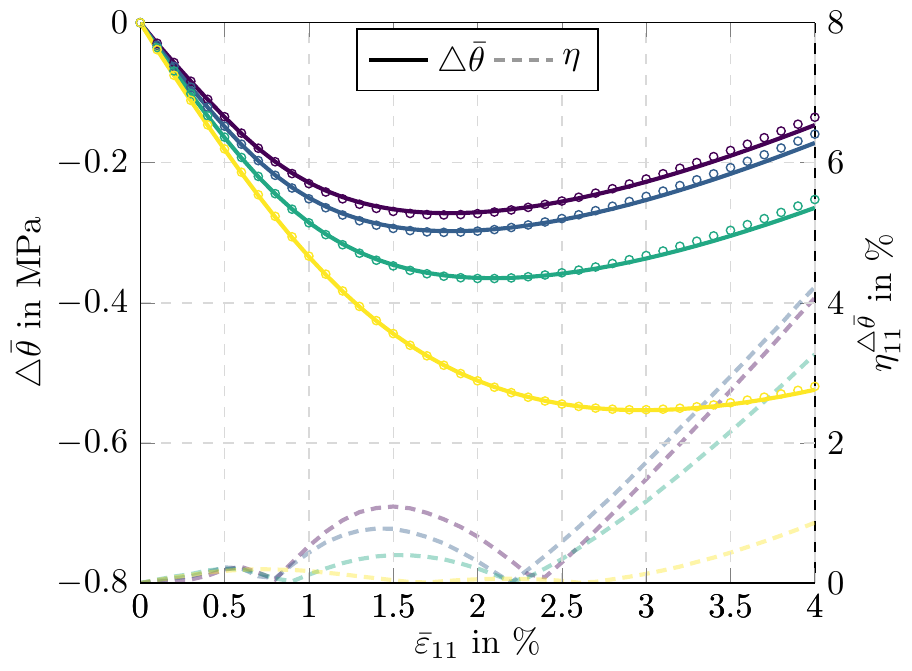}
		\end{subfigure}
		\begin{subfigure}[b]{0.32\textwidth}
			\includegraphics[height=3.85cm]{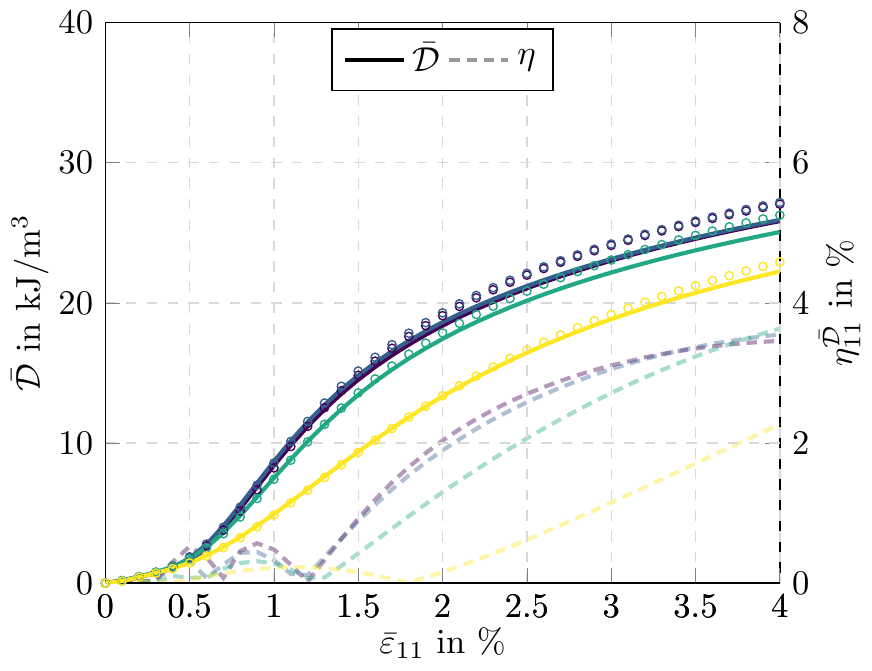}
		\end{subfigure}
	\end{subfigure}
	\caption{Strain-controlled monotonic loading: uniaxial extension in principal fiber direction}
	\label{fig:monotonic_loading}
\end{figure}

\begin{figure}[H]
	\centering
	\begin{subfigure}[t]{0.22\textwidth}
		\includegraphics[width=\textwidth]{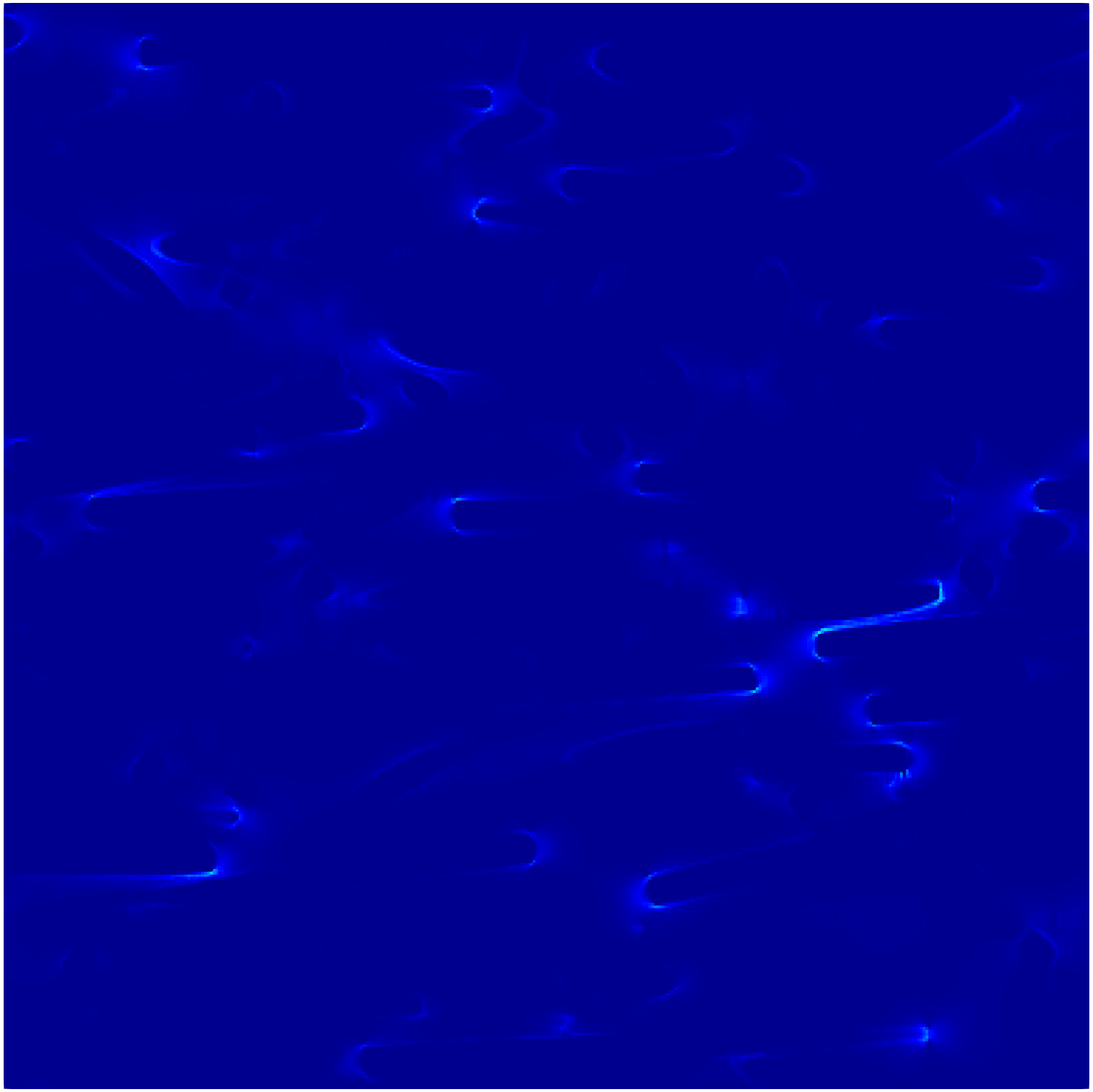}
		\caption{$\macrostrain = 1.0 \%$}
	\end{subfigure}
	\begin{subfigure}[t]{0.22\textwidth}
		\includegraphics[width=\textwidth]{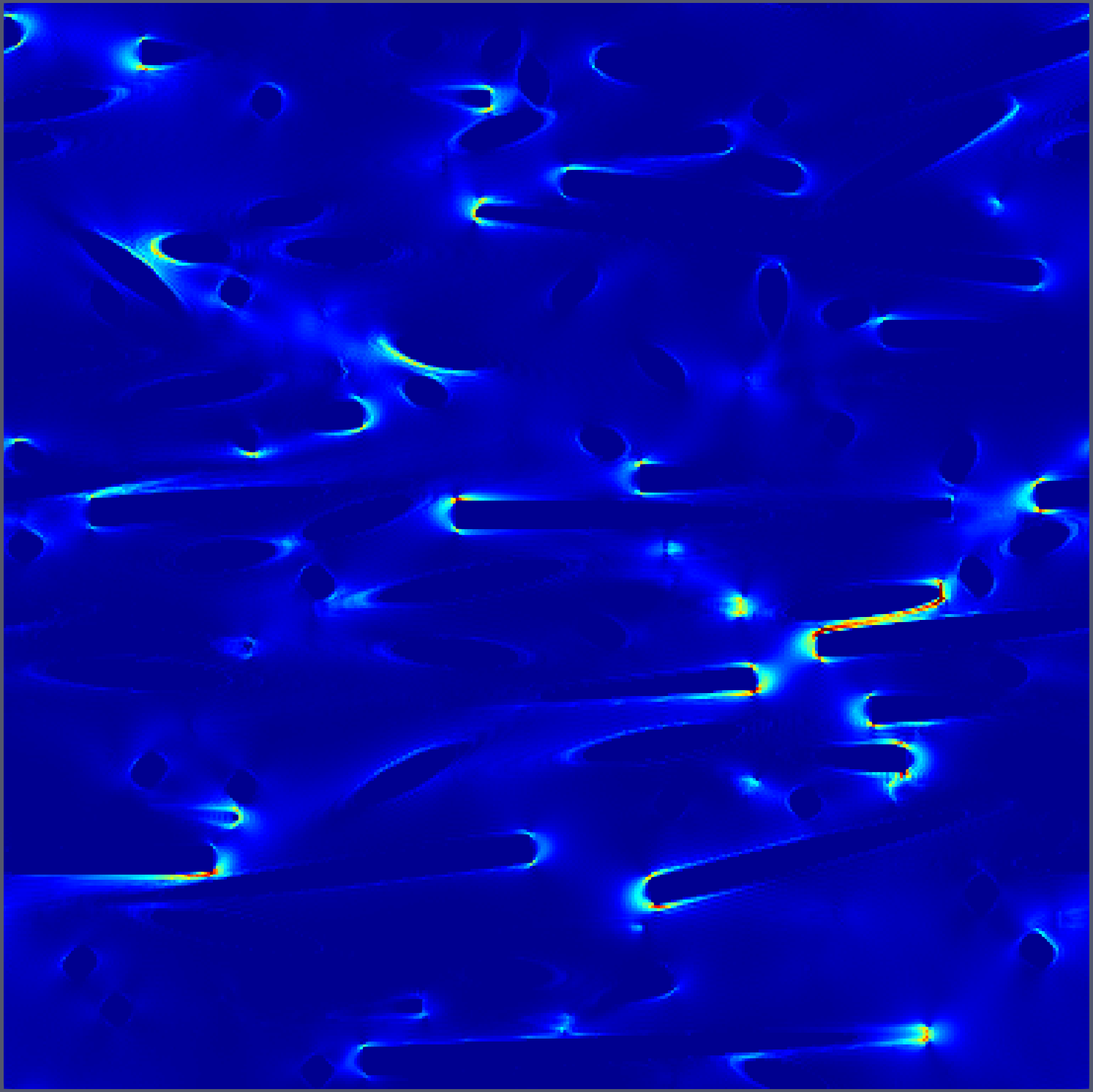}
		\caption{$\macrostrain = 2.0 \%$}
	\end{subfigure}
	\begin{subfigure}[t]{0.22\textwidth}
		\includegraphics[width=\textwidth]{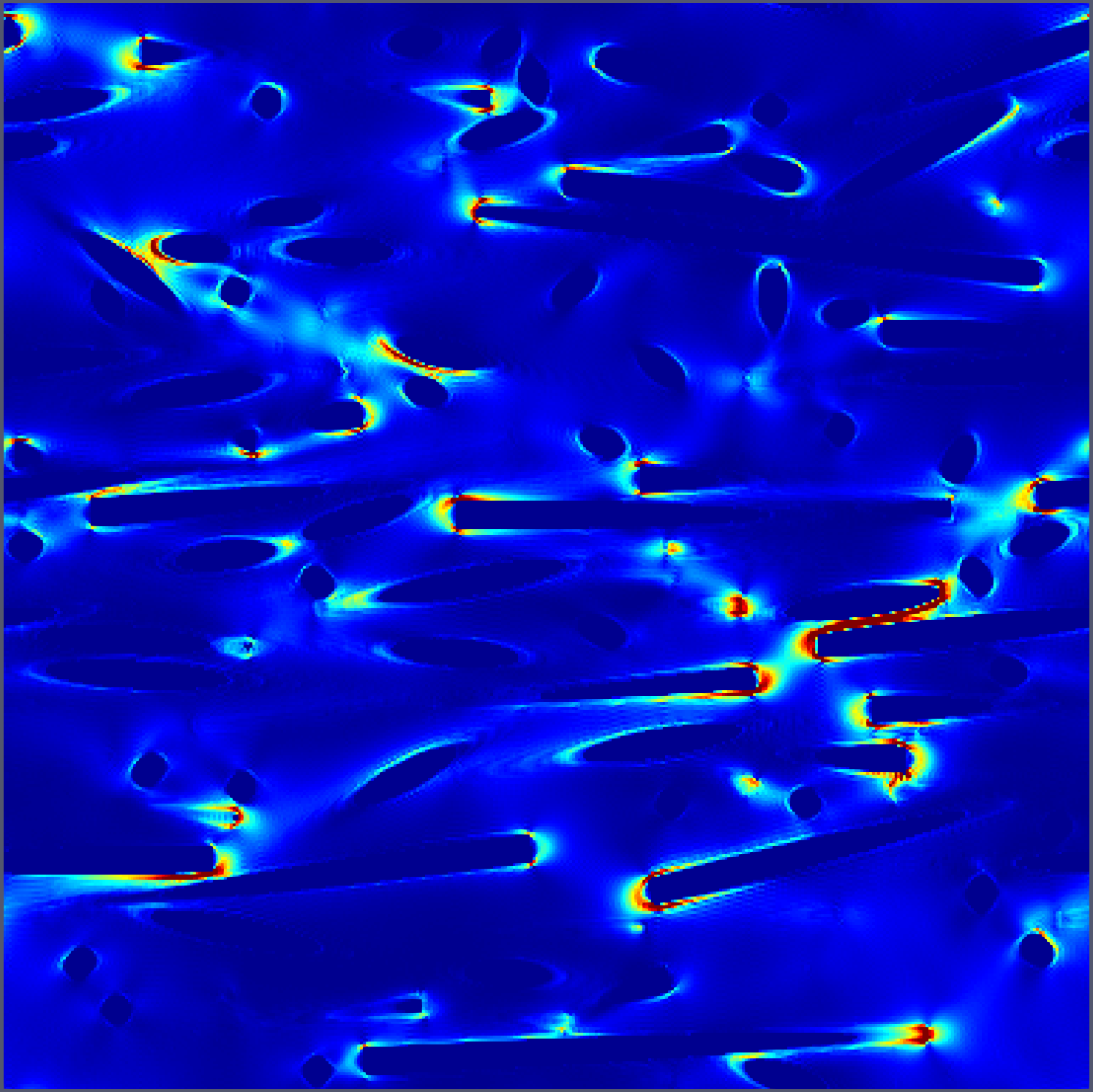}
		\caption{$\macrostrain = 3.0 \%$}
	\end{subfigure}
	\begin{subfigure}[t]{0.22\textwidth}
		\includegraphics[width=\textwidth]{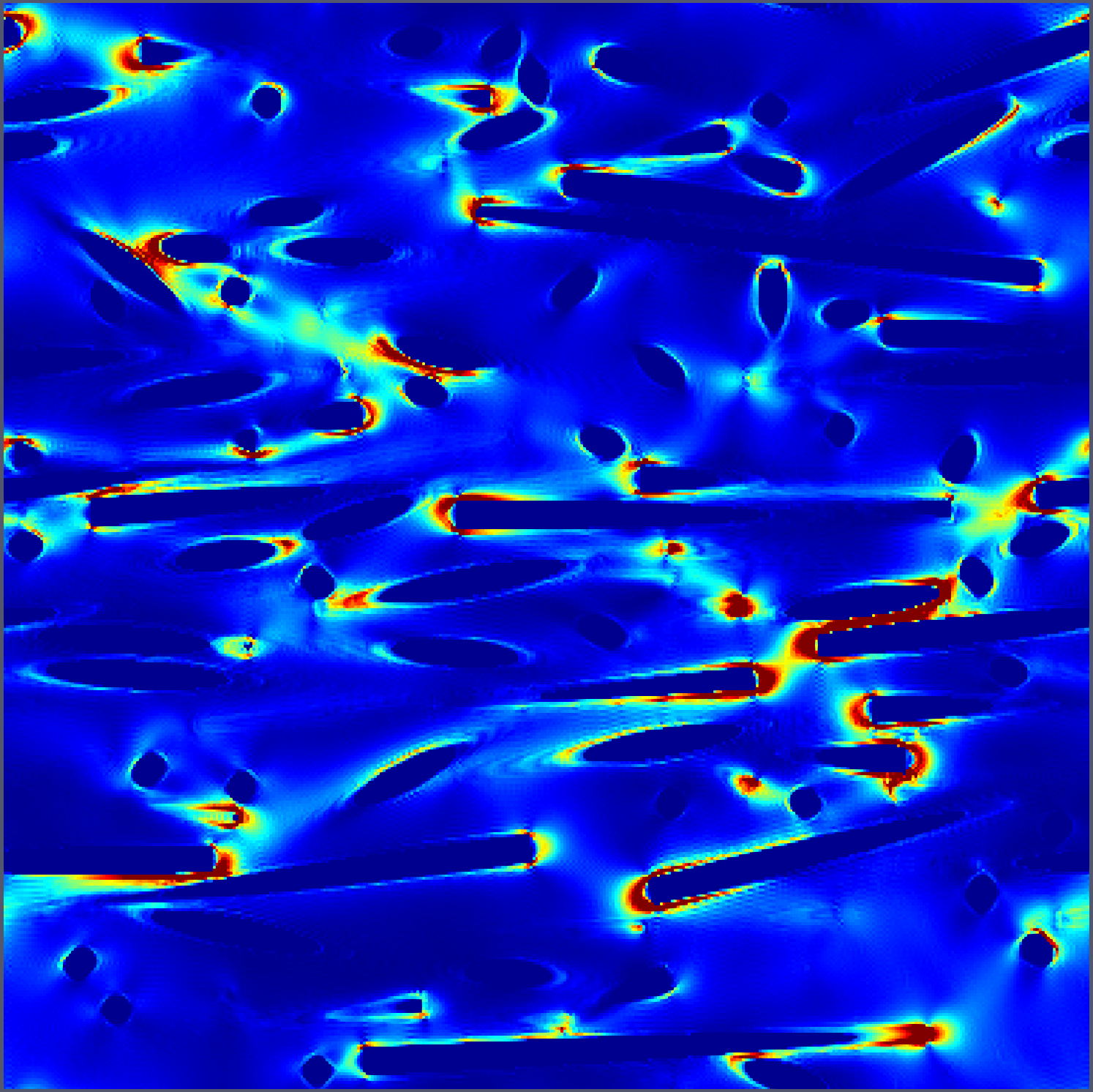}
		\caption{$\macrostrain = 4.0 \%$}
	\end{subfigure}
	\begin{subfigure}[t]{0.067\textwidth}
		\includegraphics[width=\textwidth]{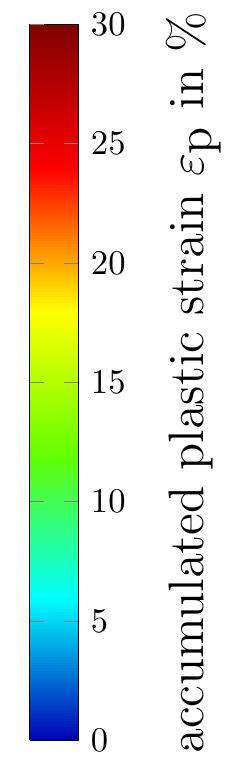}
	\end{subfigure}
	\caption{Accumulated plastic strain $\microstrain_\textrm{p}$ for a $4.0\%$ uniaxial extension in principal fiber direction with a strain rate of $\dot{\macrostrain} = 5 \cdot 10^{-4}  \unit{s^{-1}}$}
	\label{fig:accumulated_plastic_strain_slice}
\end{figure}

This section is concerned with validating the identified DMN surrogate model for the inelastic regime. To this end, we compare the DMN's predicted effective stress $\fmacrostress$, the associated effective dissipation $\macrodissipation$ as well as the change of the absolute temperature $\triangle \macrotemp = \macrotemp - \macrotemp_0$ to a high-fidelity full-field solution on the microscopic scale. To compute the reference solution, we use the implicit staggered solution scheme of Wicht et al.~\cite{WichtThermo2020}, an inexact Newton-CG~\cite{NewtonKrylov} solver and the discretization by trigonometric polynomials as introduced by Moulinec-Suquet~\cite{MoulinecSuquet1994, MoulinecSuquet1998}.\\
First, to obtain accurate inelastic results, a suitable resolution \review{and size} of the RVE needs to be determined \review{first}. \review{In Section~\ref{sec:res_RVE_study}, we learned that the RVE size has a minor influence on the effective elastic response of the composite. For this reason, we fix the volume element's edge length of $L = 384 \unit{\mu m}$ and only} vary the RVE's resolutions from $5$ to $10$ voxels per fiber diameter in equidistant steps. The former corresponds to volume element discretizations with $192^3$ to $384^3$ voxels, respectively. As loading, we consider a uniaxial extension in the principal fiber direction, \ie
\begin{equation}
	\fmacrostrain = \macrostrain \, \fe_1 \! \otimes \fe_1,
\end{equation}
and use mixed boundary conditions~\cite{Kabel2016}, \ie stress free loading perpendicular to the loading direction. The strain loading is applied in $40$ equidistant load steps with a strain rate of $\dot{\macrostrain} = 5 \cdot 10^{-4} \unit{s^{-1}}$. The reference temperature is set to $\macrotemp_0 = \SI{293.15}{\kelvin}$.  For simplicity, we assume adiabatic conditions, as we consider a single macroscopic point without any additional macroscopic heat sources.
\begin{figure}[H]
	\begin{subfigure}{\textwidth}
		\centering
		\includegraphics[height=0.48cm]{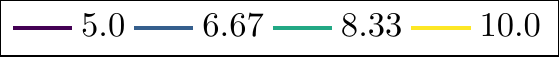}
	\end{subfigure}
	\begin{subfigure}{\textwidth}
		\centering
		\begin{subfigure}[b]{0.32\textwidth}
			\includegraphics[height=4.1cm]{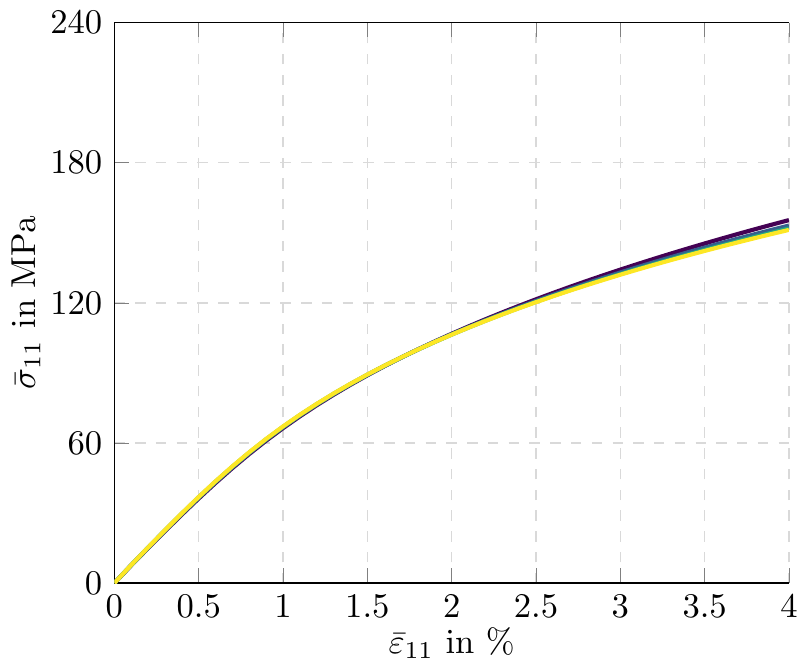}
		\end{subfigure}
		\begin{subfigure}[b]{0.32\textwidth}
			\includegraphics[height=4.1cm]{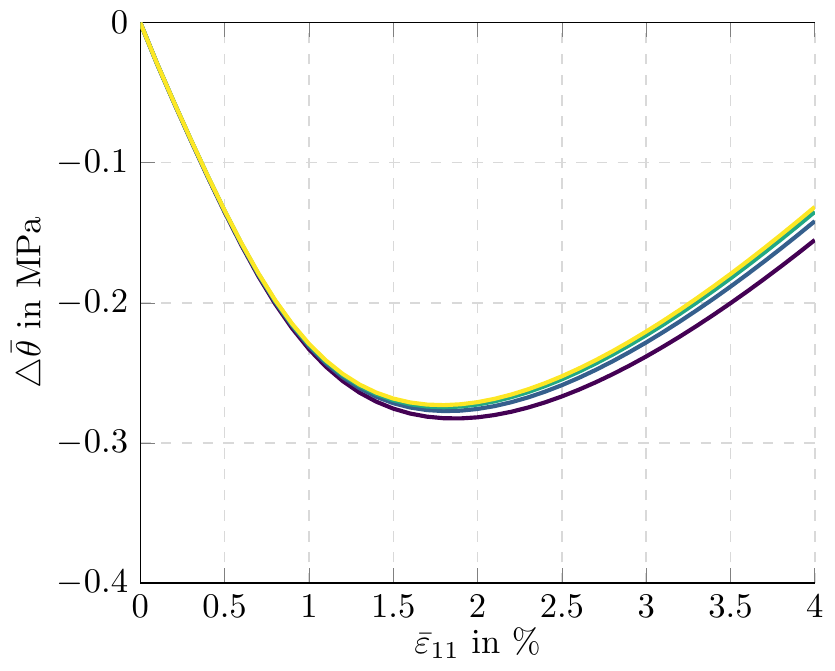}
		\end{subfigure}
		\begin{subfigure}[b]{0.32\textwidth}
			\includegraphics[height=4.1cm]{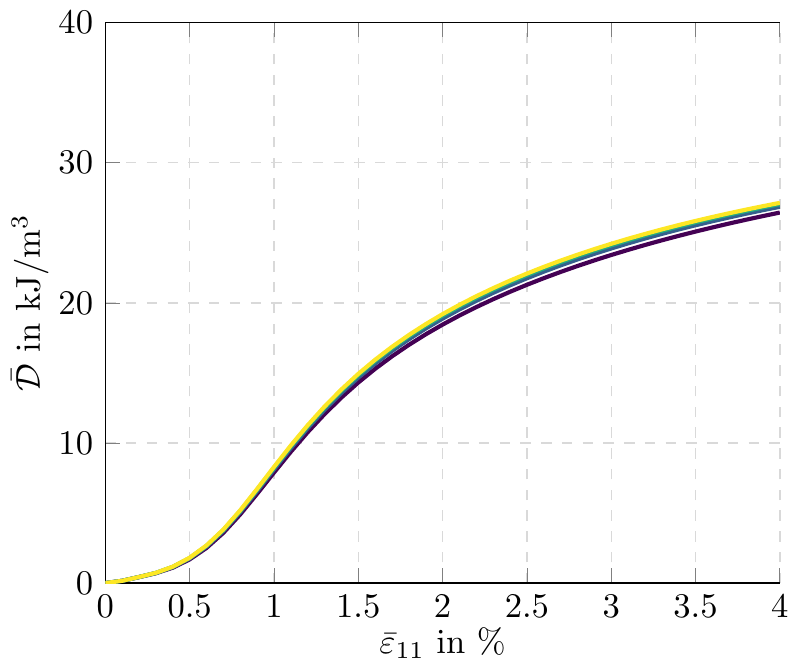}
		\end{subfigure}
	\end{subfigure}
	\caption{Effective stress, temperature change and effective dissipation for the four considered resolutions and a uniaxial extension in the principal fiber direction}
	\label{fig:resoluation_study_inelastic}
\end{figure}
In Fig.~\ref{fig:resoluation_study_inelastic}, the computed effective stress $\fmacrostress$, the change of the absolute temperature $\triangle \macrotemp$ and the effective dissipation $\macrodissipation$ are shown for all four considered resolutions. For a macrostrain of $\macrostrain = 1.0 \%$ and below, the Joule-Gough effect, \ie an almost linear temperature decease due to the hydrostatic extension, becomes apparent. This regime is captured well, even for the coarsest resolution. At around $\macrostrain = 1.0 \%$ macrostrain, the matrix starts to deform plastically. Due to the increasing dissipation, self-heating of the composite occurs and the four solutions start to deviate noticeably. Thus, to accurately capture self-heating effects, a resolution of at least $8.33$ fibers per fiber diameter is necessary. Such a resolution suffices to accurately compute the effective stress and the effective dissipation as well. For this reason, we consider a resolution of $8.33$ voxels per fiber diameter, \ie a volume element discretization with $320^3$ voxels, as sufficient for the inelastic computations. \\
With the identified resolution at hand, we turn back to the validation of the DMN surrogate model. For this purpose, we implemented the procedure introduced in Section~\ref{sec:online_evaluation} as an implicit user-material subroutine. A computationally efficient implementation of the UMAT is critical. For this reason, we use the binary tree compression as explained in Gajek et al.~\cite{Gajek2021} and exploit the sparsity pattern of the gradient operator $\fA$ and the Jacobians $\partial \vec{\fmicrostress} / \partial \vec{\fmicrostrain}$ and $\partial \vec{\microrpl} / \partial \vec{\fmicrostrain}$. For this reason, we rely upon the Eigen3~\cite{eigen3} library for all linear algebra operations. We set the tolerance for the convergence criterion to $\textrm{tol} = 10^{-12}$ and solve the linear system with the help of a sparse Cholesky decomposition. The former allows to reuse the decomposition for computing the algorithmic tangents $\ffC^\textrm{algo}_{\macrostrain}$, $\ffC^\textrm{algo}_{\macrotemp}$, $\ffD^\textrm{algo}_{\macrostrain}$, $\ffD^\textrm{algo}_{\macrotemp}$ with minimal computational overhead, see Section~\ref{sec:online_evaluation}.\\
\paragraph{Strain-controlled monotonic and non-monotonic virtual experiments} We first consider strain-controlled virtual experiments. Using the material parameters of Section~\ref{sec:material_parameters}, we investigate six monotonic uniaxial strain loadings
\begin{equation}
	\fmacrostrain = \frac{\macrostrain}{2} \left(\fe_i \otimes \fe_j + \fe_j \otimes \fe_i\right) \quad \textrm{with} \quad  (i,j) \in L_1 := \set{(1,1),(2,2),(3,3),(1,2),(1,3),(2,3)}.
\end{equation}
For every uniaxial strain loading direction in the index set $L_1$, a monotonic strain loading amplitude of $\macrostrain = 4.0\%$ is applied in $40$ equidistant load steps. To capture the rate dependence of the polyamide matrix, we investigate four individual strain rates which are logarithmically spaced from $\dot{\macrostrain} = 5 \cdot 10^{-4} \unit{s^{-1}}$ to $\dot{\macrostrain} = 5 \cdot 10^{-1} \unit{s^{-1}}$.\\
To evaluate the approximation errors of the DMN in a quantitative way, we introduce the following error measures. For a load in direction $(i,j)$, we define the relative error in the effective stress component $\macrostress_{ij}$, the change in absolute temperature $\triangle \macrotemp$ and the effective dissipation $\macrodissipation$ as
\begin{equation}\label{eq:rel_stress_errors}
	\eta^{\macrostress}_{ij}(t) = \frac{\left|\macrostress^{\, \textrm{DMN}}_{ij}(t) - \macrostress^{\, \textrm{FFT}}_{ij}(t)\right|}{\underset{t \in \mathcal{T}}{\max}\left|\macrostress^{\, \textrm{FFT}}_{ij}(t)\right|}, \quad \eta^{\triangle \macrotemp}_{ij}(t) = \frac{\left|\triangle \macrotemp^{\, \textrm{DMN}}(t) - \triangle \macrotemp^{\, \textrm{FFT}}(t)\right|}{\underset{t \in \mathcal{T}}{\max}\left|\triangle \macrotemp^{\, \textrm{FFT}}(t)\right|}, \quad \eta^{\macrodissipation}_{ij}(t) = \frac{\left|\macrodissipation^{\, \textrm{DMN}}(t) - \macrodissipation^{\, \textrm{FFT}}(t)\right|}{\underset{t \in \mathcal{T}}{\max}\left|\macrodissipation^{\, \textrm{FFT}}(t)\right|},
\end{equation}
where $\mathcal{T}=[0,T]$ denotes the considered time interval of the simulation. Furthermore, the mean and the maximum error are defined by
\begin{equation} \label{eq:mean_and_maximum_error_measures}
	\eta^{(\cdot)}_\textrm{mean} = \underset{i,j \in \{1,2,3\}}{\max} \frac{1}{T} \int_{0}^{T} \eta^{(\cdot)}_{ij}(t) \dif t \quad \textrm{and} \quad \eta^{(\cdot)}_\textrm{max} = \underset{i,j \in \{1,2,3\}}{\max} \underset{t \in \mathcal{T}}{\max} \,\, \eta^{(\cdot)}_{ij}(t).
\end{equation}
In Fig.~\ref{fig:monotonic_loading}, the results for the monotonic loading in the principal fiber direction, \review{\ie $(i,j) \equiv (1,1)$,} are shown. We observe that, up to the maximum load of $\macrostrain = 4.0\%$, the effective stresses predicted by the DMN and the full-field solution are almost indistinguishable. 
The relative stress error for all four considered strain rates is well below $2.0 \%$. The same holds for the temperature change $\triangle \macrotemp$ and the effective dissipation $\macrodissipation$ for a strain loadings up to $2.0 \%$. Only from $2.0 \%$ macroscopic strain and above, deviations in the effective dissipation, and, thus, also the temperature change, become noticeable. The former is a result of the power-law hardening of the polyamide matrix. Indeed, due to the power-law hardening, local clusters of significant plastic deformation form in the microstructure. Fig.~\ref{fig:accumulated_plastic_strain_slice} visualizes this effect by showing the evolution of the accumulated plastic strain $\microstrain_\textrm{p}$, for the strain rate $\dot{\macrostrain} = 5 \cdot 10^{-4} \unit{s^{-1}}$, on a $\fe_1$-$\fe_2$ slice of the 3D microstructure. 
Fig.~\ref{fig:accumulated_plastic_strain_slice} illustrates that for the chosen loading, clusters with more than $30 \%$ accumulated plastic strain form in the vicinity of fiber ends. This strong plastification leads to a pronounced energy dissipation, which is slightly underestimated by the DMN, see Fig.~\ref{fig:monotonic_loading}. For this reason, the DMN underestimates the self-heating of the composite as it does not fully capture such localization phenomena.\\
To account for more complex loading conditions, \ie load reversal or biaxial loadings, we investigate six independent non-monotonic loadings and six independent biaxial loadings in Appendix~\ref{sec:appendix}. The relative errors in the effective stress, the temperature change and the effective dissipation for all four considered strain rates and all considered load cases are summarized in Tab.~\ref{tab:mat_val_error_monotonic}. 
\begin{table}[h!]
	\centering
	\begin{tabular}{l  c  c  c }
		\hline
		& $\eta^{\macrostress}_\textrm{mean}$ / $\eta^{\macrostress}_\textrm{max}$ & $\eta^{\triangle \macrotemp}_\textrm{mean}$ / $\eta^{\triangle \macrotemp}_\textrm{max}$ & $\eta^{\macrodissipation}_\textrm{mean}$ / $\eta^{\macrodissipation}_\textrm{max}$\\
		\hline
		6 monotonic loadings & $1.35$ \% / $ 3.17$ \% & $1.11$ \% / $ 4.23$ \% & $2.33$ \% / $ 4.57$ \% \\
		\hline
		6 non-monotonic loadings & $0.77$ \% / $ 1.42$ \% & $0.56$ \% / $ 1.23$ \% & $0.98$ \% / $ 4.68$ \% \\
		\hline
		6 biaxial loadings & $1.00$ \% / $ 2.02$ \% & $1.10$ \% / $ 1.82$ \% & $1.24$ \% / $ 3.63$ \% \\
		\hline
	\end{tabular}
	\caption{Mean and maximum relative errors for the investigated strain-controlled uniaxial and biaxial loadings}
	\label{tab:mat_val_error_monotonic}
\end{table}

\paragraph{Stress-controlled cyclic loading}

\begin{figure}[H]
	\begin{subfigure}{\textwidth}
		\centering
		\includegraphics[height=0.48cm]{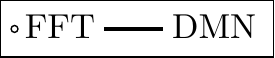}
		\includegraphics[height=0.48cm]{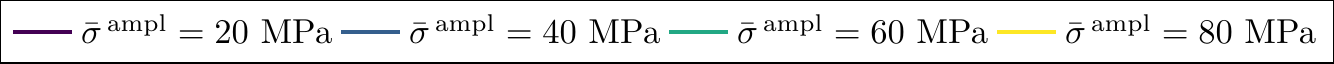}
	\end{subfigure}
	\begin{subfigure}{\textwidth}
		\centering
		\begin{subfigure}[b]{0.32\textwidth}
			\includegraphics[height=3.85cm]{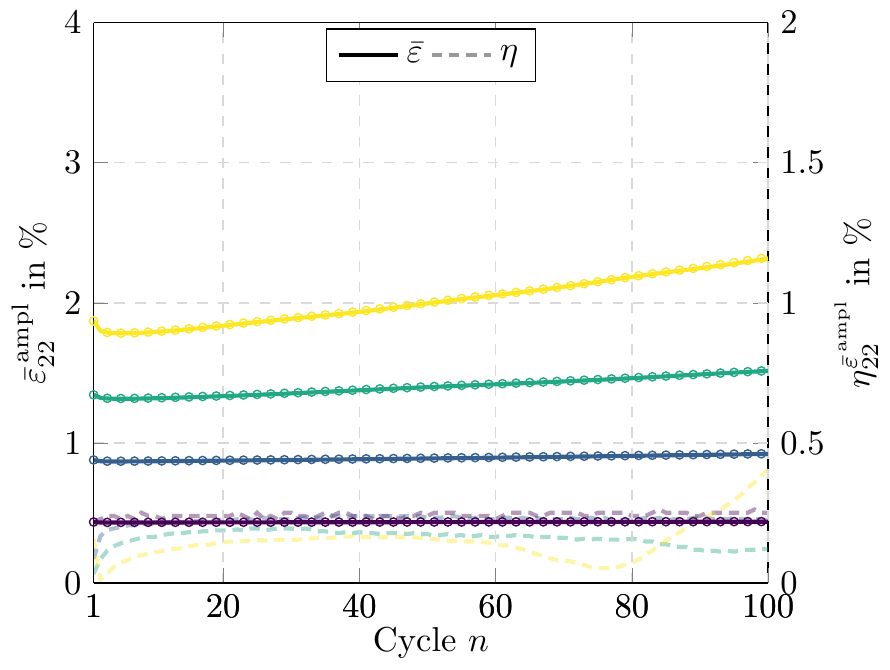}
		\end{subfigure}
		\begin{subfigure}[b]{0.32\textwidth}
			\includegraphics[height=3.85cm]{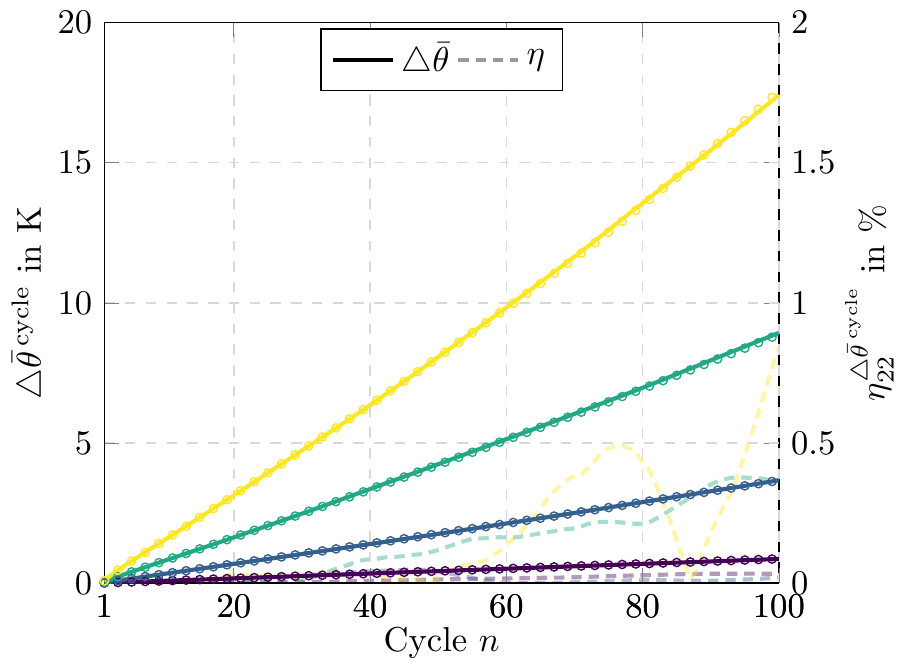}
		\end{subfigure}
		\begin{subfigure}[b]{0.32\textwidth}
			\includegraphics[height=3.85cm]{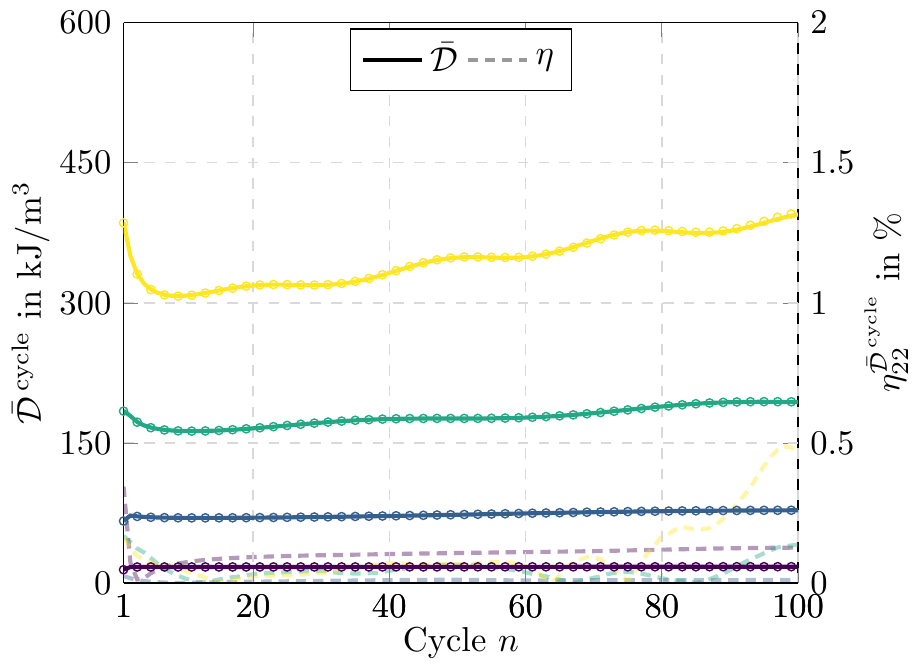}
		\end{subfigure}
	\end{subfigure}
	\caption{Stress-controlled cyclic loading: uniaxial extension perpendicular to the principal fiber direction}
	\label{fig:cyclic_loading_short}
\end{figure}

In the previous section, we investigated the identified DMN surrogate model for monotonic and non-monotonic, uniaxial and biaxial loadings. Indeed, for such loadings, self-heating effects played a minor role. However, polymers, in general, show a significant self-heating under cyclic loading, see, \eg Benaarbia et al~\cite{Benaarbia2015}. For this reason, we conclude this section with the validation of the DMN surrogate model for cyclic loading and conduct stress-controlled virtual experiments
\begin{equation}
	\fmacrostress(t) = \frac{\effective{\microstress}(t)}{2}  \, \left(\fe_i \otimes \fe_j + \fe_j \otimes \fe_i\right) \quad \textrm{with} \quad \effective{\microstress}(t) = \macrostress^{\, \textrm{ampl}} \sin\left(2 \pi \frac{t}{T_\textrm{c}} \right), \quad (i,j) \in L_4 := \set{(1,1), (2,2)}.
\end{equation}
More precisely, for both loading directions in the index set $L_4$, we apply a uniaxial, sinusoidal stress load. Here, $\macrostress^{\, \textrm{ampl}}$ denotes the stress amplitude and $T_\textrm{c}$ represent the period of the harmonic loading. As self-heating effects depend on the loading amplitude, we consider four linearly spaced stress amplitudes, ranging from $\macrostress^{\, \textrm{ampl}} = \SI{20}{\mega\pascal}$ to $\macrostress^{\, \textrm{ampl}} = \SI{80}{\mega\pascal}$. We simulate $100$ cycles, where every cycle is discretized with $20$ equidistant load steps, \ie $\numprint{2000}$ load steps in total. The stress load is applied with a frequency of $f = \SI{10}{\hertz}$, \ie the period is $T_\textrm{c} = \SI{0.1}{\second}$, and adiabatic conditions are assumed due to the short simulation time of $\SI{10}{\second}$. Please note that we consider small stress amplitudes up to $\macrostress^{\, \textrm{ampl}} = \SI{80}{\mega\pascal}$ resulting in strain amplitudes well below $2.5\%$. For this loading, resolutions of the volume element smaller than $8.33$ voxels per fiber diameter are admissible, see Fig.~\ref{fig:resoluation_study_inelastic}. For this purpose, we use a volume element resolved with $256^3$ voxels, corresponding to $6.67$ voxels per diameter, to keep computational costs reasonable.\\
In Fig.~\ref{fig:cyclic_loading_short}, the results for the cyclic loading perpendicular to the principal fiber direction, \review{\ie $(i,j) \equiv (2,2)$,} are shown. The strain amplitude \review{$\macrostrain^{\, \textrm{ampl}}_{22}$}, which is computed by 
\begin{equation}
	\review{\macrostrain^{\, \textrm{ampl}}_{22}(n) = \frac{1}{2} \left(\max_{t \in \mathcal{T}_\textrm{c}(n)}(\fmacrostrain(t) \cdot \fe_2 \otimes \fe_2) - \min_{t \in \mathcal{T}_\textrm{c}(n)}(\fmacrostrain(t) \cdot \fe_2 \otimes \fe_2) \right)} \quad \textrm{with} \quad \mathcal{T}_\textrm{c}(n) := [(n-1) T_\textrm{c}, n T_\textrm{c}]
\end{equation}
\review{for a cycle $n$}, is shown vs.\ the cycles for all four considered amplitudes. We observe that for stress amplitudes of $\macrostress^{\, \textrm{ampl}} = \SI{60}{\mega\pascal}$ and above, the composite exhibits viscoplastic flow, resulting in a decrease of the strain amplitude in the first few cycles due to hardening. Subsequently, for the two largest amplitudes, the strain amplitude increases again due to the self-heating induced thermal softening of the composite. Beside the strain amplitude, the temperature change and the dissipated energy are illustrated as well in Fig.~\ref{fig:cyclic_loading_short}. Here, for cycle $n$, $\triangle \macrotemp^{\, \textrm{cycle}}$ denotes the mean temperature change and $\macrodissipation^{\, \textrm{cycle}}$ expresses the total dissipated energy, \ie
\begin{equation}
	\triangle \macrotemp^{\, \textrm{cycle}}(n) = \frac{1}{T_\textrm{c}} \int_{\mathcal{T}_\textrm{c}(n)} \triangle \macrotemp(t) \dif t \quad \textrm{and} \quad \macrodissipation^{\, \textrm{cycle}}(n) = \int_{\mathcal{T}_\textrm{c}(n)} \macrodissipation(t) \dif t
\end{equation}
hold. We observe an almost linear self-heating of the composite for all considered amplitudes. In the first few cycles, the total dissipated energy is dominated by viscoplastic flow which decreases for an increasing number of cycles due to hardening. Furthermore, we observe a noticeable temperature-dependence of the dissipated energy, \ie a noticeable oscillation of the dissipation starting at around \review{$10$} cycles. Both  is especially visible for the two highest amplitudes. The former can be attributed to the Maxwell elements which are, due to the employed WLF shift function, activated and deactivated depending on the temperature. 
With an increasing number of cycles, the dissipated energy increases again as the material starts to soften resulting in higher strain amplitudes and thus a higher dissipation.\\ 
Comparing the DMN and the full-field solution, we observe an excellent agreement. The strain amplitude, temperature change and dissipated energy of the DMN compared to the full-field solution are almost indistinguishable. To quantify the approximation errors, we evaluate the mean $\eta^{(\cdot)}_{\textrm{mean}}$ and maximum $\eta^{(\cdot)}_{\textrm{max}}$ error~\eqref{eq:mean_and_maximum_error_measures}, for the strain amplitude $\macrostrain^{\, \textrm{ampl}}$, the mean temperature change $\triangle \macrotemp^{\, \textrm{cycle}}$ and the total dissipation $\macrodissipation^{\, \textrm{cycle}}$, respectively. These results are summarized in Tab.~\ref{tab:cyclic_loading_error} for the cyclic loading parallel and perpendicular to the principal fiber direction.

\begin{table}[H]
	\centering
	\begin{tabular}{c  c  c }
		\hline
		$\eta^{\macrostrain^{\, \textrm{ampl}}}_\textrm{mean}$ / $\eta^{\macrostrain^{\, \textrm{ampl}}}_\textrm{max}$ & $\eta^{\triangle \macrotemp^{\, \textrm{cycle}}}_\textrm{mean}$ / $\eta^{\triangle \macrotemp^{\, \textrm{cycle}}}_\textrm{max}$ & $\eta^{\macrodissipation^{\, \textrm{cycle}}}_\textrm{mean}$ / $\eta^{\macrodissipation^{\, \textrm{cycle}}}_\textrm{max}$\\
		\hline
		$0.37$ \% / $ 0.43$ \% & $1.11$ \% / $ 2.32$ \% & $2.17$ \% / $ 2.41$ \% \\
		\hline
	\end{tabular}
	\caption{Mean and maximum relative errors for the investigated stress-controlled uniaxial cyclic loadings}
	\label{tab:cyclic_loading_error}
\end{table}

Summing up, we investigated monotonic, non-monotonic uniaxial, biaxial and cyclic loading \review{scenarios} to validate the identified DMN surrogate model for thermomechanically coupled simulations on the microscopic scale. Indeed, the DMN is able to provide a digital twin for the investigated short fiber reinforced plastic microstructure of thermomechanically coupled constituents. The approximation errors for the effective stress in the inelastic setting were well below $3.5 \%$, for every investigated loading condition. Even the effective dissipation and the predicted temperature change \review{only} range up to $5 \%$, depending on the considered scenario. 

\section{A computational example} \label{sec:computational_example}

\begin{figure}[h!]
	\centering
	\includegraphics[width=0.45\textwidth]{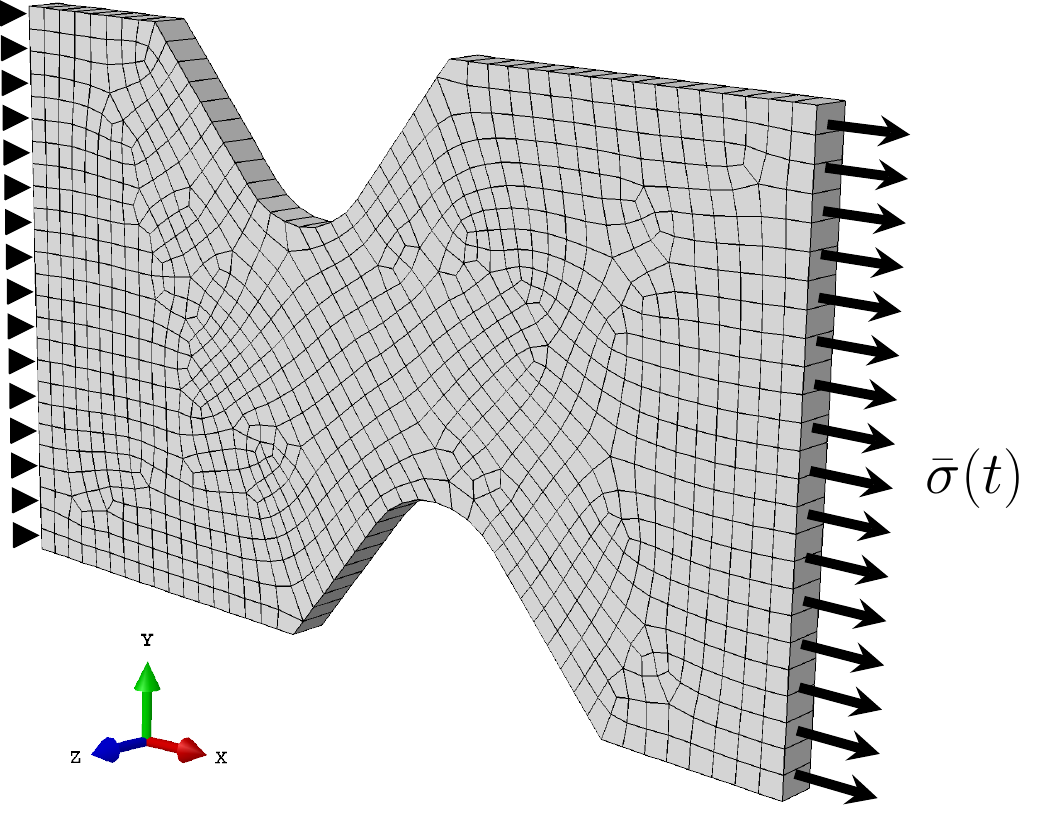}
	\caption{Non-symmetric, notched plate subjected to a cyclic loading~\cite{Tikarrouchine2019}}
	\label{fig:notched_specimen}
\end{figure}

\begin{figure}[H]
	\centering
	\begin{subfigure}[b]{0.85\textwidth}
		\centering
		\begin{subfigure}[b]{0.32\textwidth}
			\includegraphics[width=\textwidth]{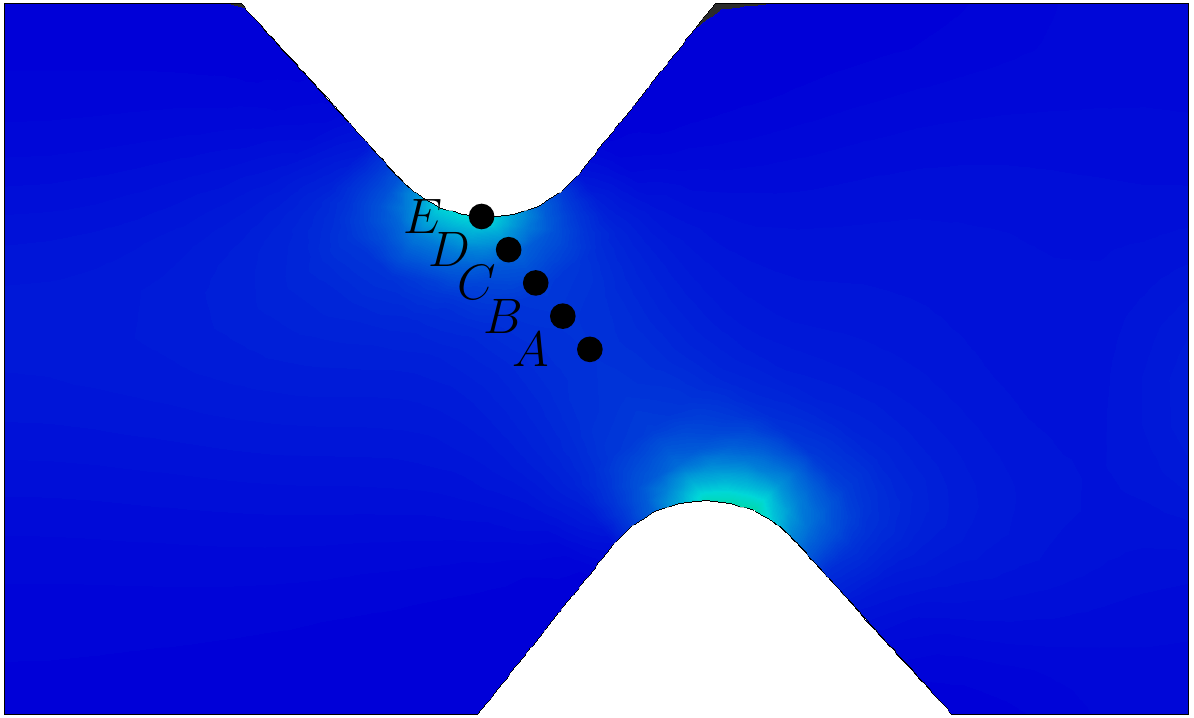}
			\caption{50 cycles}
			\label{fig:points_countour}
		\end{subfigure}
		\begin{subfigure}[b]{0.32\textwidth}
			\includegraphics[width=\textwidth]{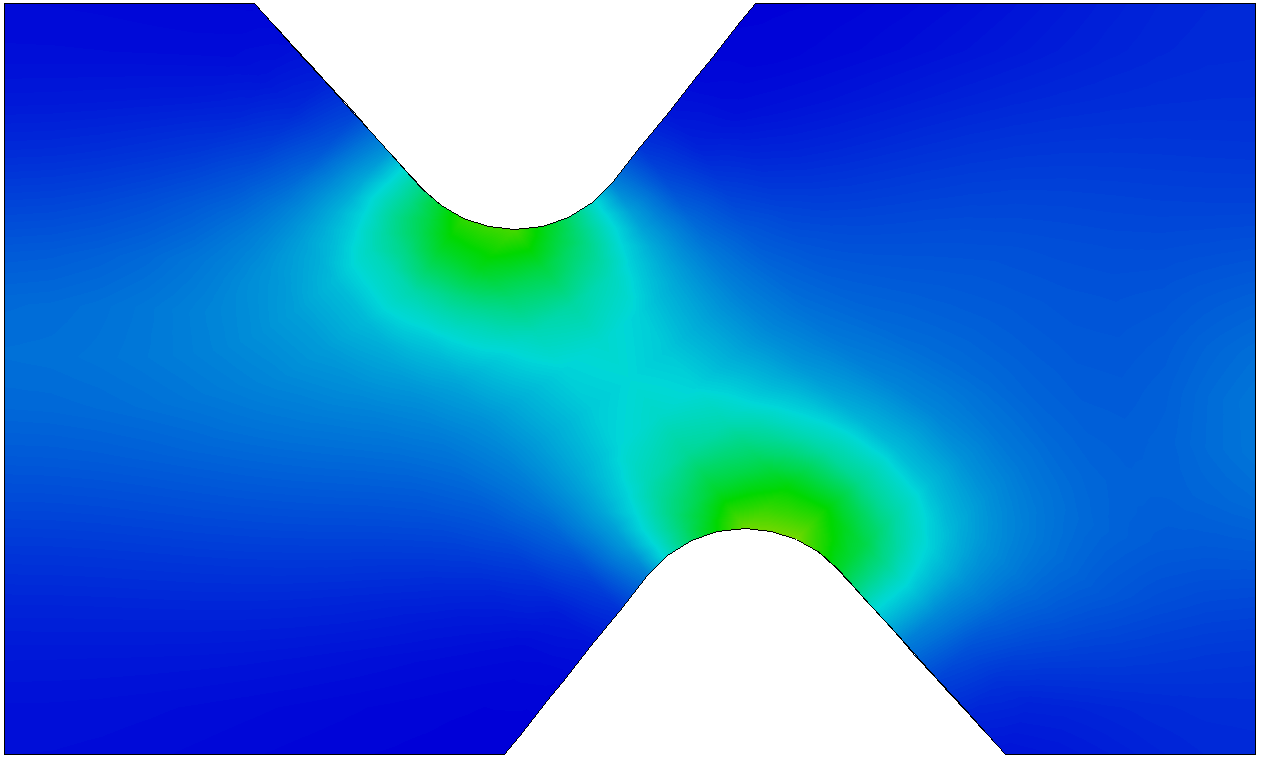}
			\caption{250 cycles}
		\end{subfigure}
		\begin{subfigure}[b]{0.32\textwidth}
			\includegraphics[width=\textwidth]{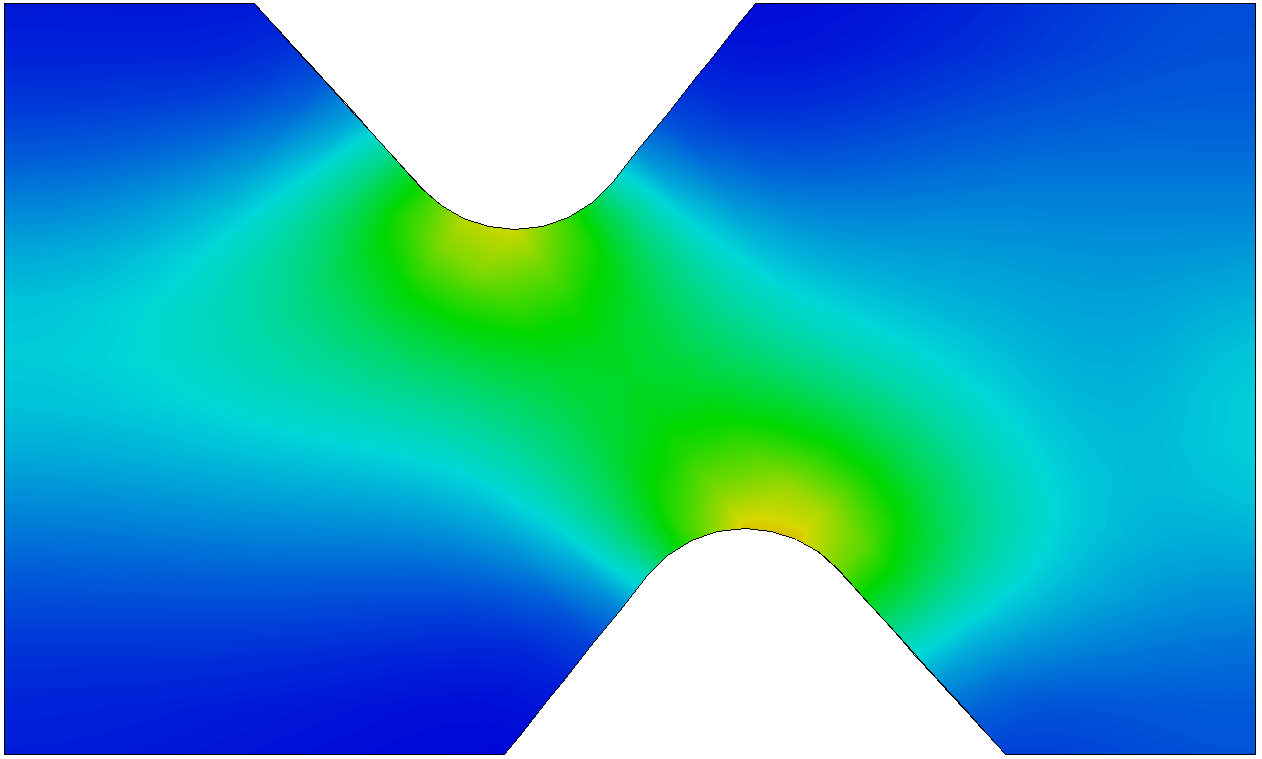}
			\caption{500 cycles}
		\end{subfigure}
		\begin{subfigure}[b]{0.32\textwidth}
			\includegraphics[width=\textwidth]{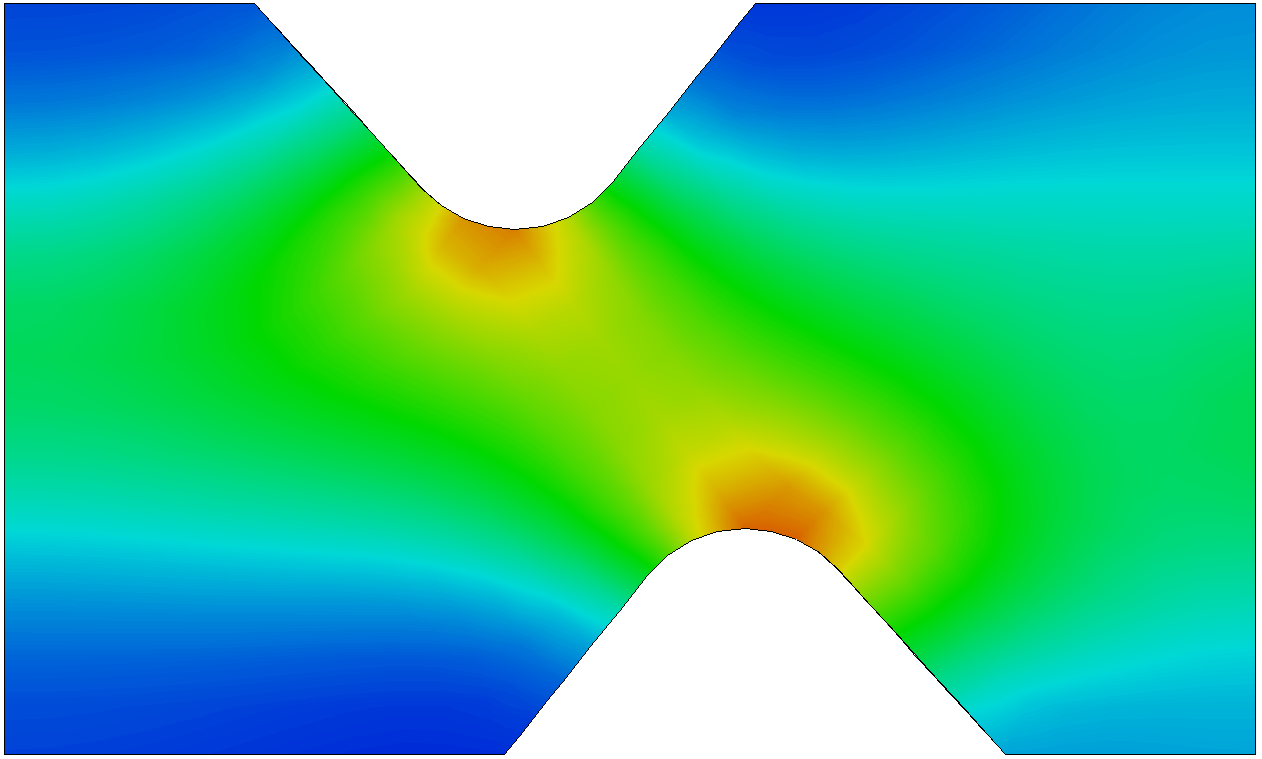}
			\caption{1000 cycles}
		\end{subfigure}
		\begin{subfigure}[b]{0.32\textwidth}
			\includegraphics[width=\textwidth]{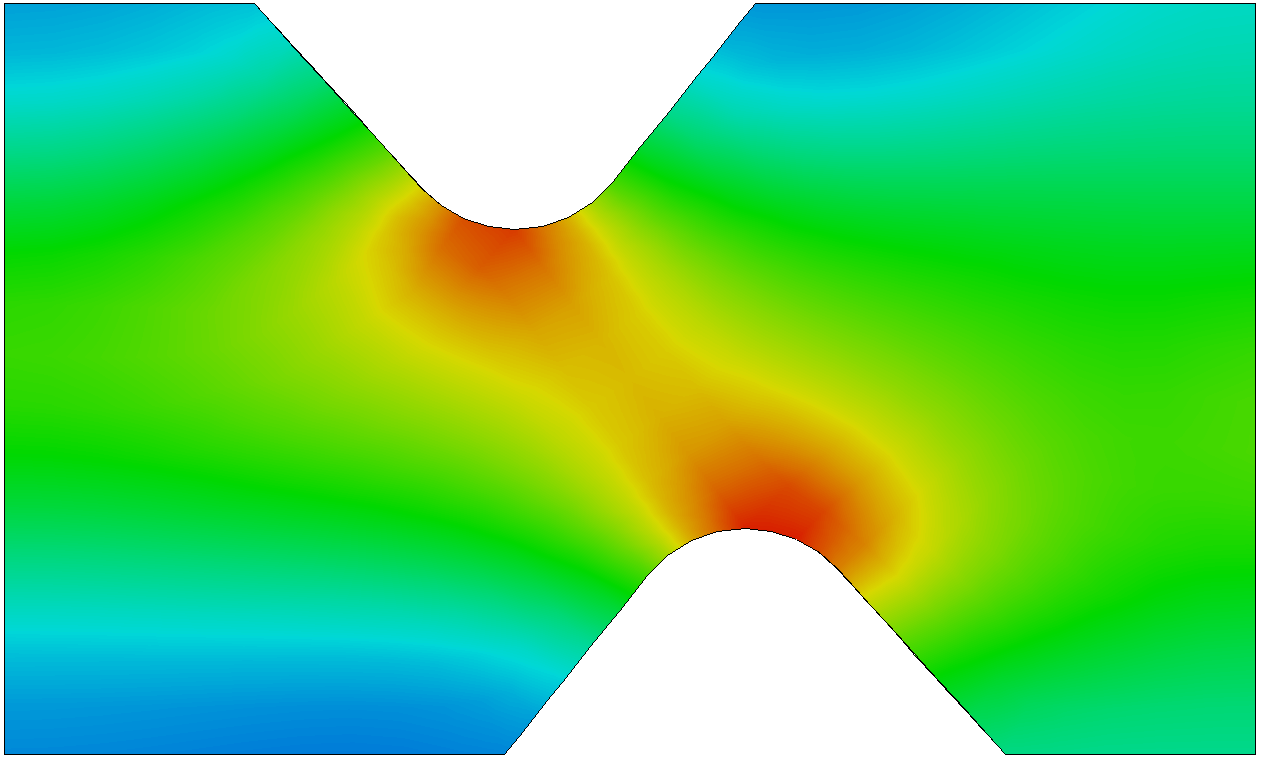}
			\caption{2000 cycles}
		\end{subfigure}
		\begin{subfigure}[b]{0.32\textwidth}
			\includegraphics[width=\textwidth]{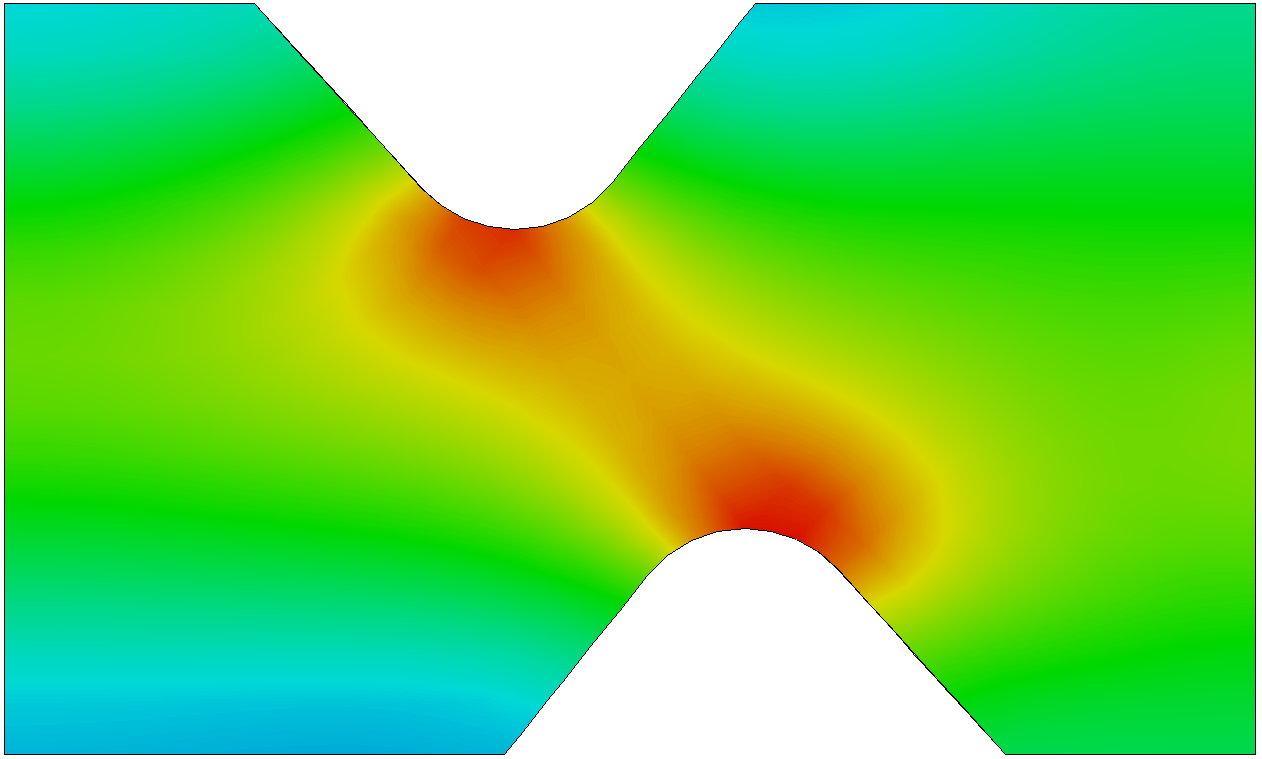}
			\caption{3000 cycles}
		\end{subfigure}
	\end{subfigure}
	\begin{subfigure}[b]{0.09\textwidth}
		\includegraphics[width=\textwidth]{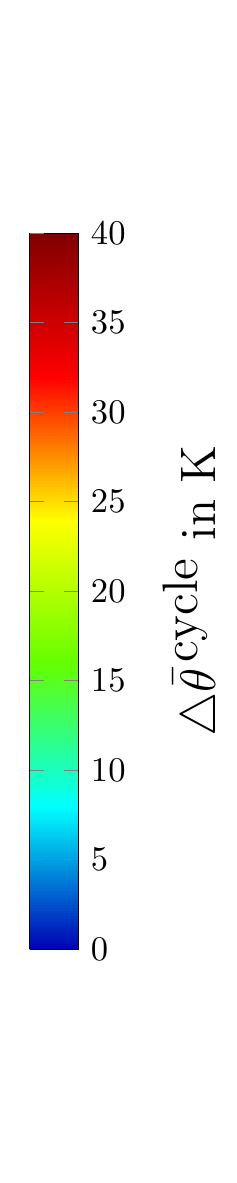}
	\end{subfigure}
	\caption{Evolution of the absolute temperature on the surface of the notched plate subjected to a cyclic loading}
	\label{fig:notched_plate_contour_cyclic_loading}
\end{figure}

\begin{figure}[H]
	\begin{subfigure}{\textwidth}
		\centering
		\includegraphics[height=0.48cm]{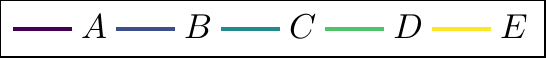}
	\end{subfigure}
	\begin{subfigure}{\textwidth}
		\centering
		\begin{subfigure}[b]{0.32\textwidth}
			\includegraphics[height=3.85cm]{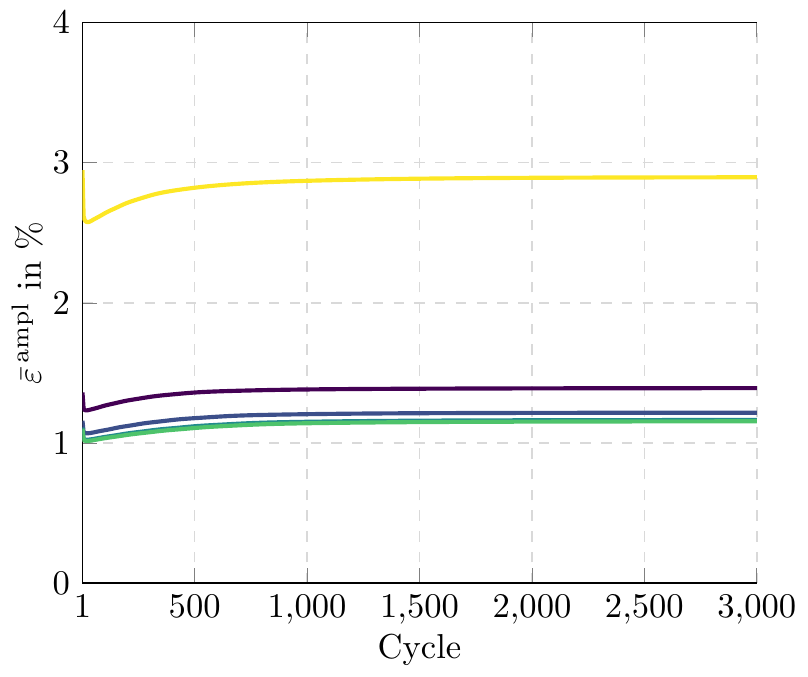}
		\end{subfigure}
		\begin{subfigure}[b]{0.32\textwidth}
			\includegraphics[height=3.85cm]{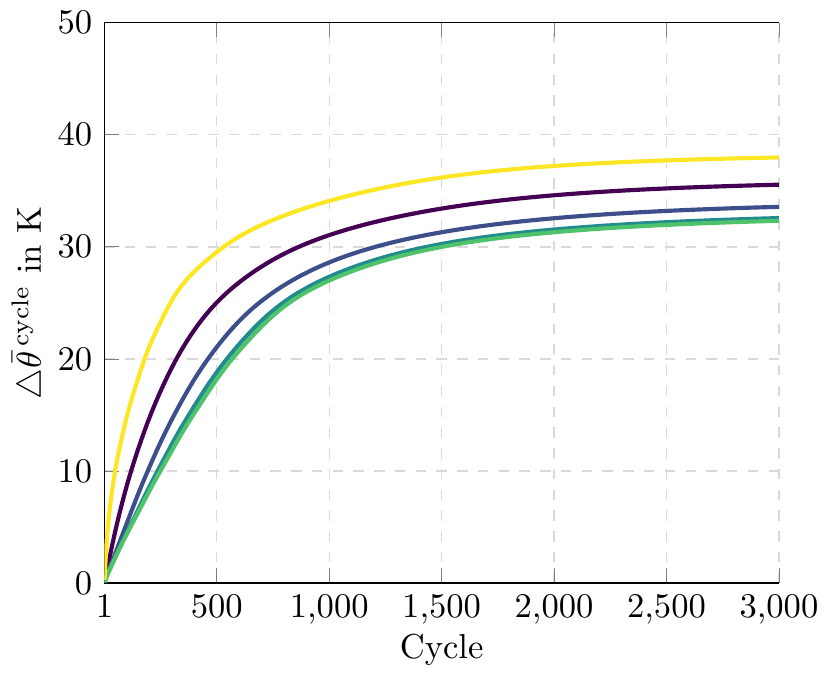}
		\end{subfigure}
		\begin{subfigure}[b]{0.32\textwidth}
			\includegraphics[height=3.85cm]{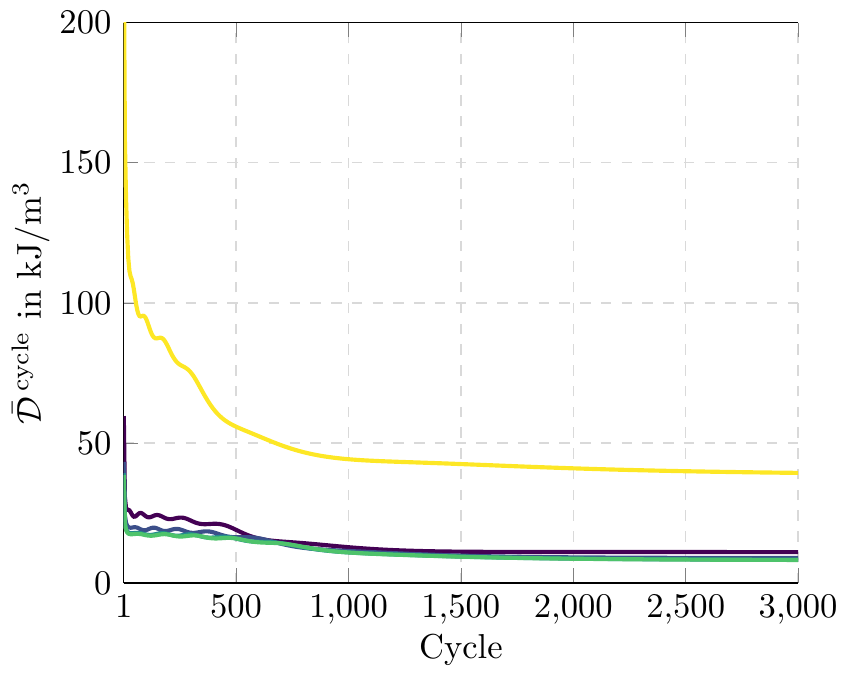}
		\end{subfigure}
	\end{subfigure}
	\caption{The strain amplitude, the absolute temperature and the dissipated energy vs. the number of cycles for the five locations shown in Fig.~\ref{fig:points_countour}}
	\label{fig:notched_plate_plot_cyclic_loading}
\end{figure}

With the identified DMN at hand, we turn our attention to conducting a DMN-accelerated two-scale concurrent thermomechanical simulation. More precisely, we study the macroscopic response of a non-symmetric, notched plate subjected to cyclic loading using the FE software \abq. The effective material response of the short-fiber reinforced polyamide is provided by the identified DMN surrogate model. The local orientation of the material, \ie the principal fiber direction, aligns with the loading direction. The geometry of the structure is similar to Tikarrouchine et al.~\cite{Tikarrouchine2019} and is illustrated in Fig.~\ref{fig:notched_specimen}. 
The structure is clamped on the left hand side and is subjected to a cyclic stress load
\begin{equation}
	\macrostress(t) = \macrostress^{\, \textrm{ampl}} \sin\left(2 \pi \frac{t}{T_\textrm{c}} \right)
\end{equation}
with $\macrostress^{\, \textrm{ampl}} = \SI{50}{\mega\pascal}$ on the right hand side of the plate. We simulate $\numprint{3000}$ cycles, and every cycle is discretized by $20$ equidistant load steps, \ie $\numprint{60000}$ load steps in total. The stress load is applied with a frequency of $f = \SI{10}{\hertz}$, \ie the period is $T_\textrm{c} = \SI{0.1}{\second}$. Due to the long simulation time of $\SI{200}{\second}$, the assumption of adiabatic conditions is not appropriate. For this reason, we prescribe a convective boundary condition on the free surfaces of the plate, \ie the heat flux across the surface of the plate
\begin{equation}
	\effective{q}_{\textrm{s}} = - h (\macrotemp_{\textrm{s}} - \macrotemp_0)
\end{equation}
is a function of the film coefficient $h$ and the difference of the surface temperature $\macrotemp_{\textrm{s}}$ and the ambient temperature $\macrotemp_0 = \SI{293.15}{\mega\pascal}$. We assume a free convection. Thus, the film coefficient for air is set to 
\begin{equation}
	h = \unit{10 \, \frac{W}{m^2 \, K}},
\end{equation}
see Kosky et al.~\cite{Kosky2013}. To account for heat conduction on the macroscopic level, we assume Fourier's law 
\begin{equation}
	\effective{\fq} = - \effective{\fkappa} \, \nabla_{\!\! \effective{x}}\,\macrotemp
\end{equation}
to hold, where $\nabla_{\!\! \effective{x}}$ denotes the gradient operator \wrt the macroscopic point $\effective{\fx} \in \Omega$. We use the thermal conductivities of the glass fibers and the polyamide matrix from Tab.~\ref{tab:mat_param_glass_fibers} and \ref{tab:mat_param_PA66} and compute the effective thermal conductivity tensor $\effective{\fkappa}$ by means of an FFT-based computational homogenization code~\cite{MoulinecSuquet1994, MoulinecSuquet1998, Dorn2019}. Indeed, the effective thermal conductivity is \review{almost} isotropic and reads
\begin{equation}
	\effective{\fkappa} \equalhat \diag{0.361, 0.33, 0.33}  \unit{\frac{W}{m \, K}}
\end{equation}
in Cartesian coordinates. The notched plate is discretized by $\numprint{1099}$ thermally coupled  quadratic hexahedron elements. In every Gauss point, a DMN is integrated implicitly. For solving the global system, we rely on the direct Newton solver in \abq, which solves for the displacements and absolute temperature simultaneously.\\
In Fig.~\ref{fig:notched_plate_contour_cyclic_loading}, the evolution of the mean temperature change $\triangle \macrotemp^{\, \textrm{cycle}}$ is shown. For the first $250$ cycles, a slight self-heating of the plate is observed in the vicinity of the two notches where the viscoelastic and viscoplastic deformation localizes. For an increasing number of cycles, the inner part of the plate starts to heat up as well both due to energy dissipation as well as heat conduction.\\
In addition to the contour plots, Fig.~\ref{fig:notched_plate_plot_cyclic_loading} shows the temporal evolution of the strain amplitude $\macrostrain^{\, \textrm{ampl}}$, mean temperature change $\triangle \macrotemp^{\, \textrm{cycle}}$ and the dissipated energy $\macrodissipation^{\, \textrm{cycle}}$ for five distinct points $A$ to $E$, see Fig.~\ref{fig:points_countour}. Here, in the macroscopic setting, we compute the strain amplitude $\macrostrain^{\, \textrm{ampl}}$ of cycle $n$ by
\begin{equation}
	\macrostrain^{\, \textrm{ampl}}(n) = \frac{1}{2} \left(\max_{t \in \mathcal{T}_\textrm{c}(n)}(\lambda^{\textrm{max}}_{\macrostrain}(t)) - \min_{t \in \mathcal{T}_\textrm{c}(n)}(\lambda^{\textrm{min}}_{\macrostrain}(t)) \right),
\end{equation}
where $\lambda^{\textrm{min}}_{\macrostrain}$ and $\lambda^{\textrm{max}}_{\macrostrain}$ denote the smallest and the largest eigenvalue of the macroscopic strain tensor $\fmacrostrain$. 
A closer look at the evolution of the strain amplitude shows that, as in the microscopic setting, the hardening of the viscoplastic matrix results into a decrease of the strain amplitude in the first few cycles for all five investigated points. Afterwards, the strain amplitude increases again until a steady-state is reached. The reason for the renewed increase of the strain amplitude and subsequent saturation becomes clear by inspecting the evolution of the absolute temperature. In the first few cycles, the temperature increases rapidly in all considered points as a result of the dissipated energy due to the viscoelastic and viscoplastic flow. This increase is more pronounced in the vicinity of the two notches and decreases towards the inside of the plate. The temperature increase results in the thermal softening of the material, and, in turn, the the strain amplitude increases. After about $\numprint{1000}$ cycles, the temperature increase saturates and a steady-state is reached. This steady-state is the result of two effects. One one hand, the dissipated energy in a cycle decreases with increasing temperature due to thermal softening. On the other hand, the heat conduction due to the free convection increases with an increasing surface temperature.\\
These results, \ie the saturating temperature increase and the temperature-dependent mechanical behavior, can only be reproduced in a macroscopic setting, since heat conduction and convection have to be considered. Therefore, only relying on microscopic simulations for characterizing thermomechanically coupled composites by simulative means does not suffice. For this reason, DMNs are a promising technique to enable thermomechanically coupled concurrent two-scale simulations with reasonable computational resources. Last but not least, we consider the computational costs of our approach, accounting for the offline training and the online validation in the following section.\\

\section{Computational costs}

The material sampling was performed in parallel, \ie six independent load steps for computing the effective stiffnesses using $16$ threads each. The training of the DMN was carried out on four threads. The wall-clock times of the sampling and the offline training are summarized in Tab.~\ref{tab:wall_time_offline_training}. Indeed, sampling of the training data took $\SI{74}{\hour}$ whereas the training finished in under $\SI{2}{\hour}$. As we only considered DMNs with a depth of $K=8$, $765$ independent fitting parameters were determined during the offline training.
\begin{table}[h!]
	\begin{center}
		\begin{tabular}{l c c}
			\hline
			& Wall-clock time & \#\textrm{Fitting parameters}\\
			\hline
			Sampling & $\SI{74}{\hour}$ & $-$\\
			Training & $\SI{1.5}{\hour}$ & $765$\\
			\hline
		\end{tabular}
	\end{center}
	\caption{Wall-clock times for sampling, training and number of fitting parameters}
	\label{tab:wall_time_offline_training}
\end{table}
Turning our attention to the online evaluation, we focus on the computational costs of the DMN evaluated at a single Gauss point. Solving the thermomechanical cell problem for a microstructure discretized by $320^3$ voxels for a prescribed macrostrain and absolute temperature takes about $\SI{2737}{\second}$ on average on a single thread. In contrast, integrating a DMN at a single Gauss point takes less than $\SI{6}{\milli\second}$. Thus, we achieve a speed-up of about half a million times compared to solving the cell problem by means of an FFT-based micromechanics solver. For applications which admit using DMNs with less than eight layers, speed-ups in the range of several millions may be possible.\\
\begin{table}[h!]
	\begin{center}
		\begin{tabular}{l c c c c c c }
			\hline
			& FFT ($1$ thread) & DMN ($1$ thread) \\
			\hline
			Wall-clock time		& \review{$45.62 \unit{min}$} & $5.64 \unit{ms}$\\
			Speed-up		& $-$ & $\numprint{485284}$\\
			\hline
		\end{tabular}
	\end{center}
	\caption{Wall-clock times and speed-up (compared to an FFT-base computational micromechanics solver) for a single time step of the inelastic micro simulation}
	\label{tab:wall_time_online_evaluation_gauss_point}
\end{table}\\
Wall-clock time and memory consumption for the component scale simulation are summarized in Tab.~\ref{tab:wall_time_online_evaluation_component_scale}. Indeed, the macroscopic FE model was discretized by $\numprint{1099}$ elements resulting in $\numprint{9706}$ degrees of freedom. Computing all $\numprint{60000}$ time steps involved $\numprint{161240}$ total Newton iterations and took about $\SI{117}{\hour}$ on $96$ threads and required about $\SI{2}{\giga\byte}$ of DRAM. Indeed, \abq only required about $2.7$ Newton iterations (on average) per load increment, indicating a robust quadratic convergence.
\begin{table}[h!]
	\begin{center}
		\begin{tabular}{l c c c c c}
			\hline
			& \abq ($96$ threads) \\
			\hline
			Elements & $\numprint{1099}$\\
			\#DOF	& $\numprint{9706}$\\
			Increments & $\numprint{60000}$\\
			Wall-clock time	& $\SI{117}{\hour}$\\
			Memory consumption & $\SI{1.8}{\giga\byte}$\\
			Total Newton iterations & $\numprint{161240}$\\
			\hline
		\end{tabular}
	\end{center}
	\caption{Wall-clock time, memory consumption and total Newton iterations of the concurrent two-scale simulation}
	\label{tab:wall_time_online_evaluation_component_scale}
\end{table}
These results indicate that DMNs are a promising technique for accelerating thermomechanical two-scale simulations. This holds for structures of moderate complexity resolved by $\numprint{10000}$ time steps (and more), as in the previous examples. Alternatively, DMNs can be used in large-scale concurrent multiscale simulations consisting of millions of elements and a smaller number of time steps, see Gajek et al.~\cite{Gajek2021}. 


\section{Conclusion}

In the work at hand, we extended the framework of direct DMNs to fully coupled thermomechanical two-scale simulations. More precisely, we incorporated the intrinsic two-way thermomechanical coupling between the microscopic and macroscopic scale into the framework. Considering the former is essential to accurately capture the mechanical response of common engineering materials, \eg short-fiber reinforced thermoplastics, in structural simulations.\\
For this purpose, we built upon the first-order homogenization framework of thermomechanical composites established by Chatzigeorgiou et al.~\cite{Chatzigeorgiou2016}, who showed that there is no fluctuation of the absolute temperature on the microscopic scale. For this reason, both the absolute temperature and the macrostrain are regarded as inputs to the DMN's (microscopic) balance of linear momentum. This way, the one-way thermomechanical coupling from the macroscopic onto the microscopic scale is accounted for. Furthermore, we incorporated the back-coupling from the microscopic scale onto the evolution of the macroscopic temperature into the framework. To this end, changes of entropy and dissipated energy are computed and propagated to the macroscopic scale where both combined, act as an additional source term to the macroscopic heat equation. This way, the two-way thermomechanical coupling was incorporated into the framework of direct DMNs. To accelerate a thermomechanically coupled two-scale simulation, we explained how our approach was implemented as an implicit user-material subroutine.\\
Choosing a short-fiber reinforced polyamide $6.6$ with industrial aspect ratio and filler fraction, we demonstrated that the trained DMN was able to predict, for a macroscopic point, the effective stress, the effective dissipation and the ensuing temperature change of the composite with high accuracy for a set of different strain rates and loading conditions. Indeed, DMNs are trained on linear elastic data alone. Predicting the dissipated energy at a macroscopic point, which in turn is intrinsically associated to nonlinear effects on the underlying microstructure, \eg plasticity, is a remarkable result which can be attributed to the DMNs internal structure. As DMNs rely on laminates as building blocks which are combined in a hierarchical manner, it is ensured that DMNs naturally inherit thermodynamic consistency and stress strain monotonicity from their phases. The former constitutes a key feature both in terms of physics as well as numerical implementation and represents one reason for the DMNs approximation capabilities, even for thermomechanically coupled problems.\\ 
To evaluate the performance of our approach in a concurrent two-scale setting, we conducted a thermomechanically coupled simulation of an asymmetric notched plate. The notched plate was subjected to a cyclic stress load also considering heat conduction and convection. Indeed, our results indicate that the \textrm{FE}-\textrm{DMN} method is a powerful piece of technology for accelerating two-scale concurrent simulations. With the possibility of providing speed-ups of five to six magnitudes, DMNs promise to become a standard tool for industrial applications. This way, the \textrm{FE}-\textrm{DMN} method finally realizes the promise of fully coupled thermomechanical two-scale simulations of large-scale industrial problems as envisioned by Chatzigeorgiou et al.~\cite{Chatzigeorgiou2016}.\\ 
In terms of future works, it would of interest to formulate the underlying material models directly in cycle space in the fashion of Köbler et al.~\cite{Fatigue2020}. The former alleviates the need to resolve every load cycle with multiple load steps enabling the stimulative characterization of thermomechanical composites in the regime of high-cycle fatigue. Furthermore, the combination of our approach with the fiber-orientation interpolation scheme~\cite{Kobler2018, Gajek2021} in order to arrive at a DMN surrogate model applicable to short-fiber reinforced polymers with a locally varying fiber orientation would further increase the applicability. Also, extensions to problems involving damage~\cite{Goerthofer2021, Benaarbia2019} and fracture~\cite{PhasefieldFracture2019,HomFrac2019,SandMultiPhysics} are of interest.

\section*{Acknowledgements}
{SG}, {MS} and {TB} acknowledge financial support by the German Research Foundation (DFG) within the International Research Training Group “Integrated engineering of continuous-discontinuous long fiber reinforced polymer structures” (GRK 2078/2). The support by the German Research Foundation (DFG) is gratefully acknowledged. \review{We thank the anonymous reviewers for their helpful comments.}

\section*{Author's contributions}

SG, MS and TB were responsible for the development of the model and methodology presented in this publication. Conceptualization was taken over by MS, and TB. The software implementation and the validation, investigation of the results, formal analysis and the subsequent visualization were performed by SG. Resources were provided by MS and TB. The original manuscript draft was written by SG and extensively reviewed and edited by SG, MS, and TB. The research project was administrated by SG, MS and TB. Funding was acquired by MS and TB. The research was supervised by MS, and TB.

\section*{Conflict of interest}

The authors declare that they have no known competing financial interests or personal relationships that could have appeared to influence the work reported in this publication.

%

\appendix
\section{Additional strain-controlled virtual experiments} \label{sec:appendix}

In this section, we further validate the identified DMN surrogate model for more complex loading conditions, see Section~\ref{sec:results_online_validation}. For this reason, we additionally consider non-monotonic and biaxial loadings.

\begin{figure}[H]
	\begin{subfigure}{\textwidth}
		\centering
		\includegraphics[height=0.48cm]{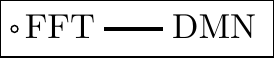}
		\includegraphics[height=0.48cm]{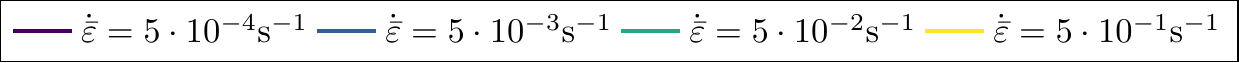}
	\end{subfigure}
	\begin{subfigure}{\textwidth}
		\centering
		\begin{subfigure}[b]{0.32\textwidth}
			\includegraphics[height=3.85cm]{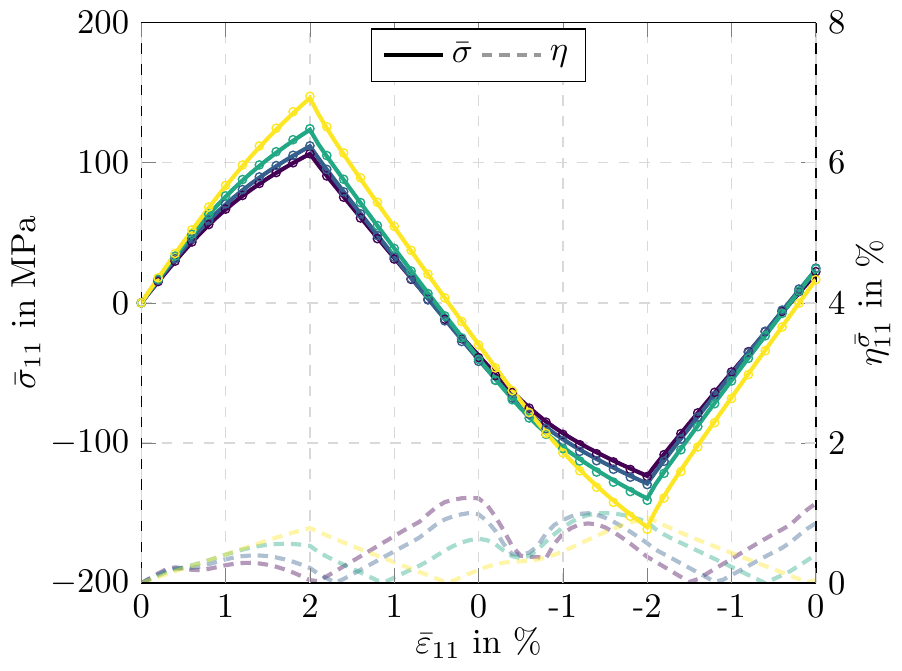}
		\end{subfigure}
		\begin{subfigure}[b]{0.32\textwidth}
			\includegraphics[height=3.85cm]{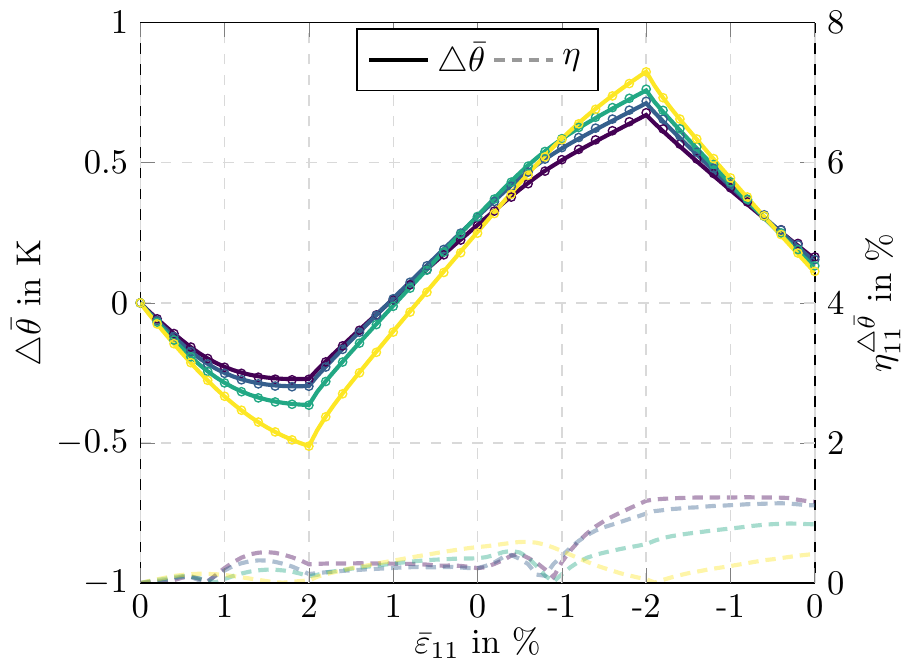}
		\end{subfigure}
		\begin{subfigure}[b]{0.32\textwidth}
			\includegraphics[height=3.85cm]{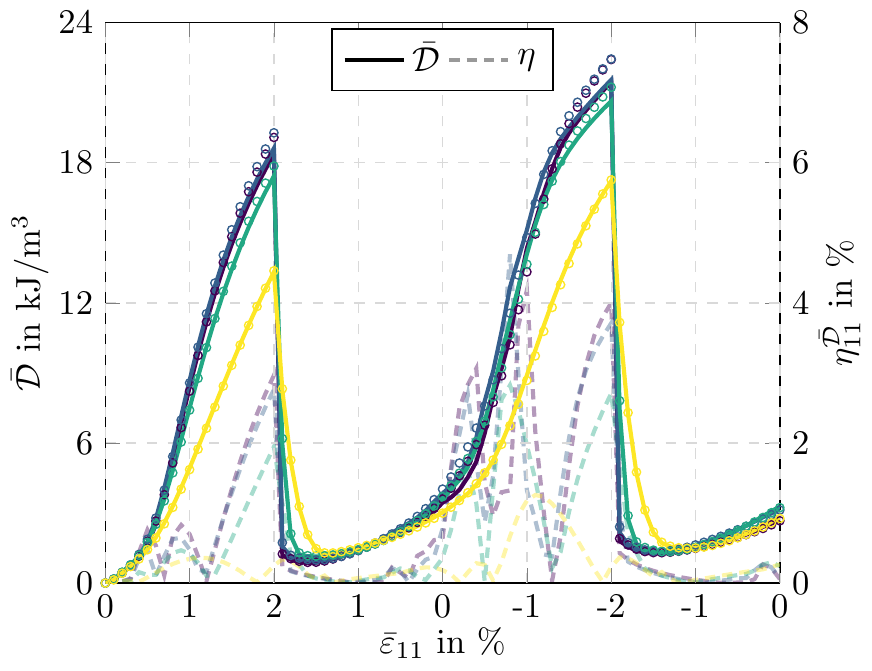}
		\end{subfigure}
	\end{subfigure}
	\caption{Strain-controlled non-monotonic loading: uniaxial extension in principal fiber direction}
	\label{fig:non-monotonic_loading}
\end{figure}

\paragraph{Strain-controlled non-monotonic loading} To account for load reversal, we investigate six non-monotonic loadings
\begin{equation}
	\fmacrostrain = \frac{\macrostrain}{2} \left(\fe_i \otimes \fe_j + \fe_j \otimes \fe_i\right) \quad \textrm{with} \quad (i,j) \in L_2 := \set{(1,1),(2,2),(3,3),(1,2),(1,3),(2,3)}.
\end{equation}
For every direction in the index set $L_2$, a full hysteresis with a strain amplitude of $\macrostrain = 2.0\%$ in $80$ equidistant load steps is computed. Additionally, we use mixed boundary conditions~\cite{Kabel2016} to ensure a stress-free loading perpendicular to the loading direction. As in Section~\ref{sec:results_online_validation}, we investigate four individual strain rates to capture the rate dependence of the composite. In Fig.~\ref{fig:non-monotonic_loading}, the results for the non-monotonic loading in the principal fiber direction, \review{\ie $(i,j) \equiv (1,1)$,} are shown for all four considered strain rates. We observe that for the full hysteresis, the DMN and the full-field solution are almost indistinguishable in terms of the effective stress and temperature change. The corresponding relative errors are around $1 \%$ for all considered strain rates. Only for the effective dissipation we observe slight disagreements between the DMN and the full-field predictions. Still, these deviations are \review{around} $4 \%$, \ie in the range of engineering requirements. The relative errors \wrt to the effective stress, temperature change and effective dissipation are summarized in Tab.~\ref{tab:mat_val_error_monotonic} for all six considered loading directions.

\paragraph{Strain-controlled biaxial loading} In addition to the monotonic and non-monotonic loadings, we investigate six independent biaxial strain loadings
\begin{equation}
	\fmacrostrain = \macrostrain_1 \, \fe_i \otimes \fe_i + \macrostrain_2 \, \fe_j \otimes \fe_j, \quad (i,j) \in L_3 := \set{(1,2),(1,3),(2,1),(2,3),(3,1),(3,2)}.
\end{equation}
For every loading direction in the index set $L_3$, a strain loading of $\macrostrain_1 = 2.0\%$ is applied while the strain in the second direction is held constant $\macrostrain_2 = 0 \%$. Afterwards, a strain load of $\macrostrain_2 = 2.0 \%$ is applied in the second direction, as well. Meanwhile, the strain in the first direction is held constant $\macrostrain_1 = 2.0\%$. The biaxial loadings are computed in $40$ equidistant load steps, and mixed boundary conditions~\cite{Kabel2016} are applied. 

\begin{figure}[H]
	\begin{subfigure}{\textwidth}
		\centering
		\includegraphics[height=0.48cm]{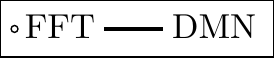}
		\includegraphics[height=0.48cm]{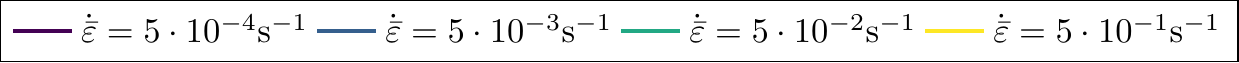}
	\end{subfigure}
	\begin{subfigure}{\textwidth}
		\centering
		\begin{subfigure}[b]{0.32\textwidth}
			\centering
			\includegraphics[height=4.1cm]{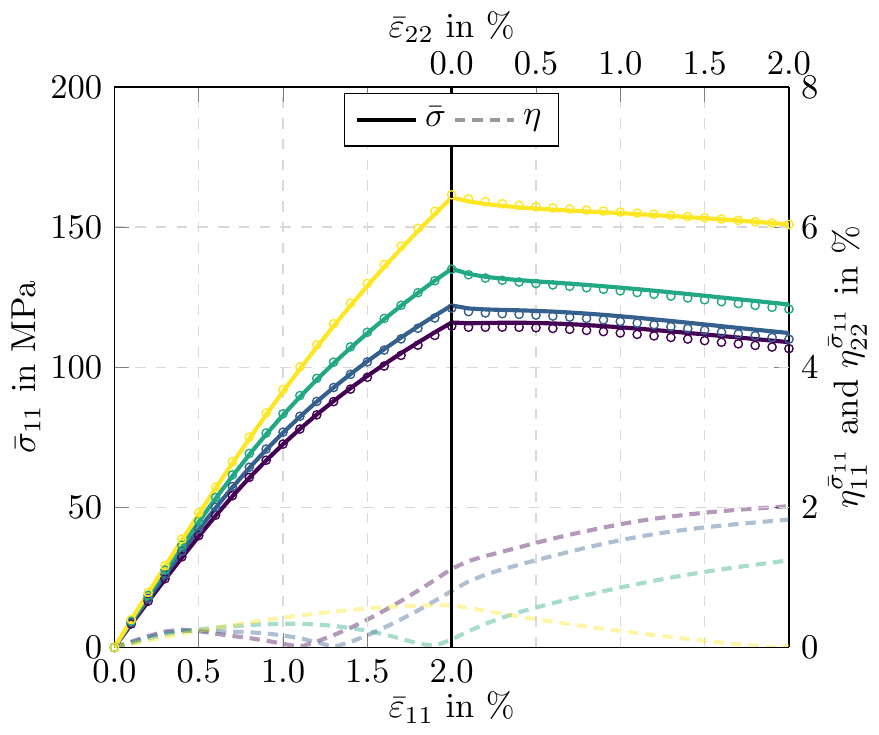}
		\end{subfigure}
		\begin{subfigure}[b]{0.32\textwidth}
			\centering
			\includegraphics[height=4.1cm]{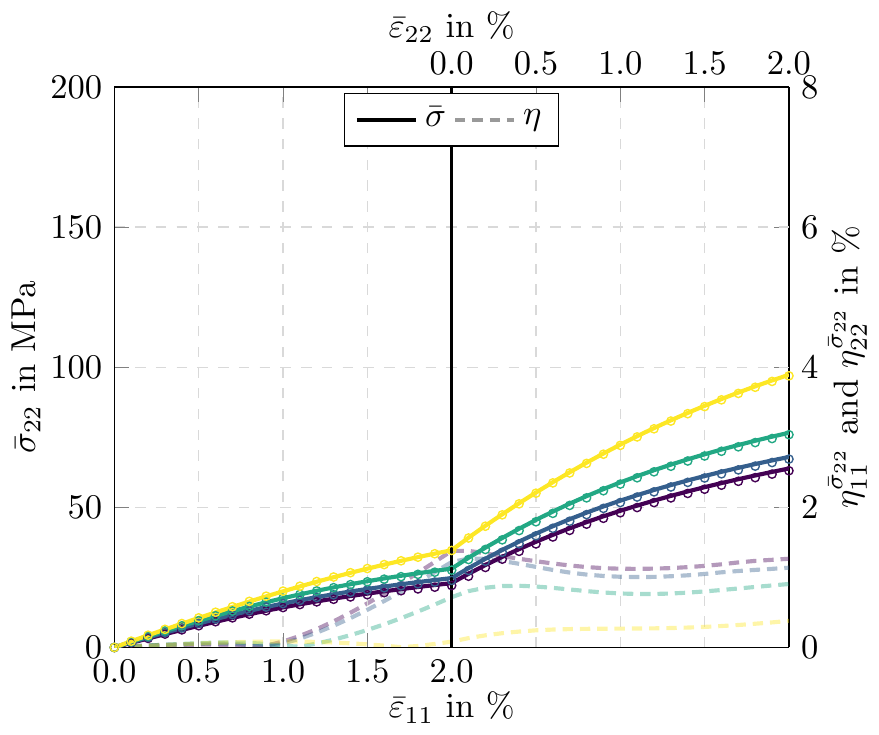}
		\end{subfigure}
	\end{subfigure}
	\begin{subfigure}{\textwidth}
		\centering
		\begin{subfigure}[b]{0.32\textwidth}
			\centering
			\includegraphics[height=4.1cm]{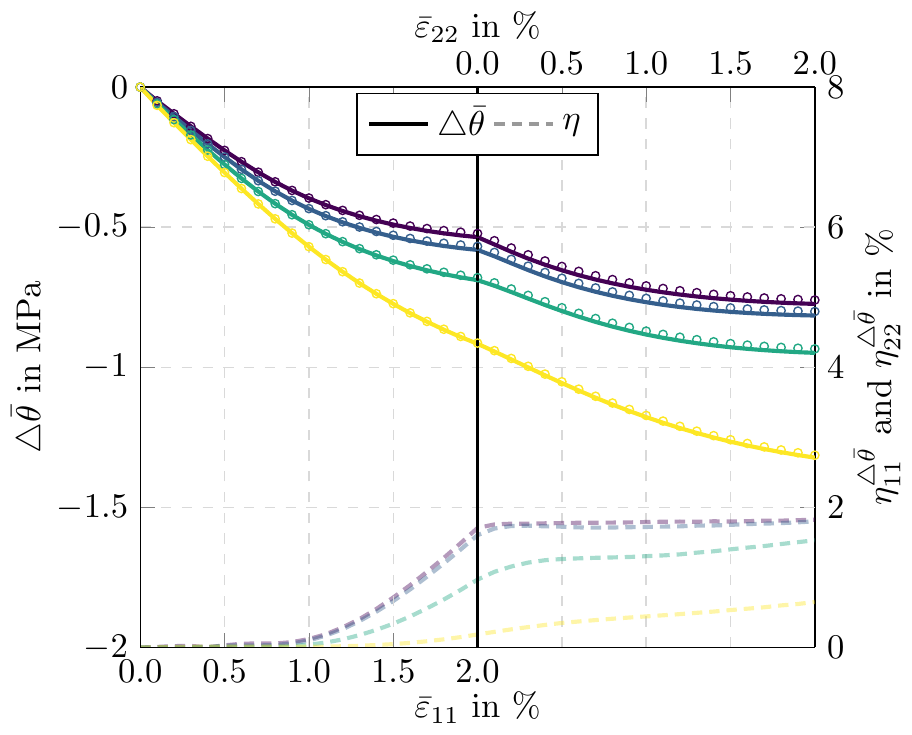}
		\end{subfigure}
		\begin{subfigure}[b]{0.32\textwidth}
			\centering
			\includegraphics[height=4.1cm]{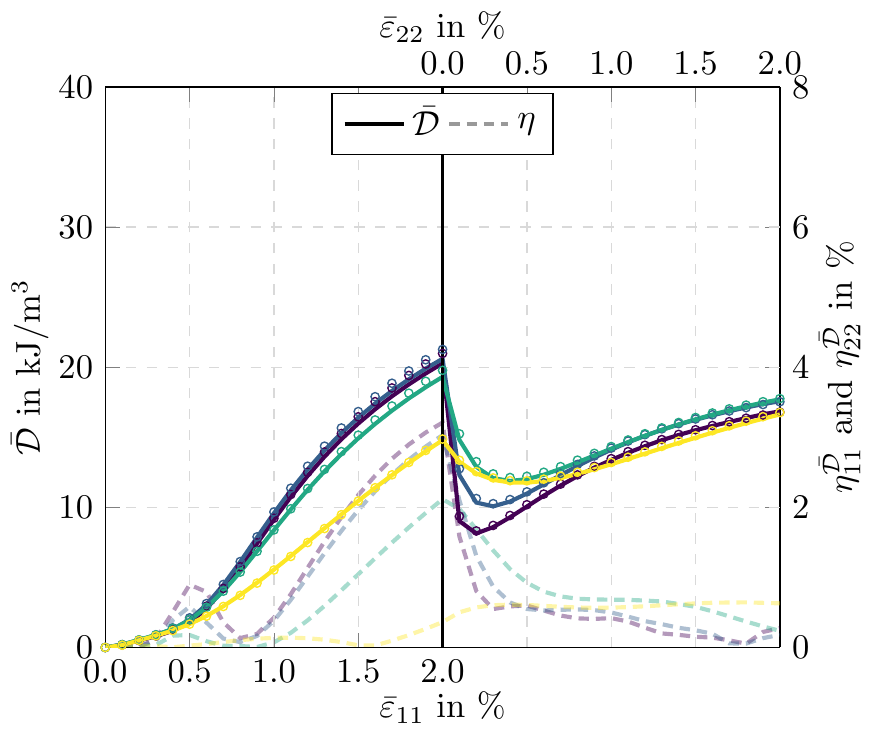}
		\end{subfigure}
	\end{subfigure}
	\caption{Strain-controlled biaxial loading: extension in principal fiber direction followed by an extension perpendicular to the principal fiber direction}
	\label{fig:biaxial_loading}
\end{figure}

In Fig.~\ref{fig:biaxial_loading}, the results for the biaxial loading in the $\fe_1$-$\fe_2$ direction, \review{\ie $(i,j) \equiv (1,2)$,} are illustrated. As we consider a biaxial loading, the effective stress components in the $\fe_1$ and $\fe_2$ are shown in addition to the temperature change and the effective dissipation. \review{Please note that the error measure $\eta^{\bar{\sigma}_{mn}}_{ij}$ denotes the relative error of the $(m,n)$ stress component for a load in the $(i,j)$ direction.} As before, the DMN matches the full-field solutions remarkably well. Relative errors lie below $2\%$ for the effective stress and temperature change and do not exceed $3.5 \%$ for the effective dissipation. The relative errors for all considered load cases are summarized in Tab.~\ref{tab:mat_val_error_monotonic}.

\bibliographystyle{ieeetr}
{\footnotesize\bibliography{literature}}

\end{document}